\documentclass[11pt,a4paper]{article}
\pdfoutput=1
\usepackage{jcappub}
\usepackage{amsmath}
\usepackage{amssymb}
\usepackage{booktabs}
\usepackage{graphicx}
\usepackage{xspace}
\usepackage[table]{xcolor}
\usepackage[normalem]{ulem}
\usepackage{subcaption}
\usepackage{url}
\usepackage{hyperref}
\usepackage{listings}
\definecolor{codegreen}{rgb}{0,0.6,0}
\definecolor{codegray}{rgb}{0.5,0.5,0.5}
\definecolor{codepurple}{rgb}{0.58,0,0.82}
\definecolor{backcolour}{rgb}{0.95,0.95,0.92}
\definecolor{DESYcyan}{RGB}{0,166,235}
\lstdefinestyle{mystyle}{
    backgroundcolor=\color{backcolour},
    commentstyle=\color{codegreen},
    keywordstyle=\color{DESYcyan},
    numberstyle=\tiny\color{codegray},
    stringstyle=\color{codepurple},
    basicstyle=\ttfamily\footnotesize,
    breakatwhitespace=false,
    breaklines=true,
    captionpos=b,
    keepspaces=true,
    numbers=left,
    numbersep=5pt,
    showspaces=false,
    showstringspaces=false,
    showtabs=false,
    tabsize=2
}
\newcommand{\hc}{h_\mathrm{c}}
\newcommand{\hcsq}{h_\mathrm{c}^2}
\newcommand{\rchar}{r_\mathrm{char}}
\newcommand{\nuin}{\nu_\mathrm{inner}}
\newcommand{\nuout}{\nu_\mathrm{outer}}
\newcommand{\tauf}{\tau_\mathrm{f}}
\newcommand{\epsmu}{\epsilon_\mu}

\newif\iffinal\finaltrue
\definecolor{ctcolour}{rgb}{0.75,0.0,0.35}

\AtBeginDocument{\iffinal\else\emergencystretch=3em\fi}
\newcommand{\msol}{M_\odot}
\newcommand{\nHz}{\,\mathrm{nHz}}
\newcommand{\binaryfog}{\texttt{binaryfog}\xspace}
\newcommand{\holodeck}{\texttt{holodeck}\xspace}
\newcommand{\ceffyl}{\texttt{ceffyl}\xspace}
\newcommand{\ptarcade}{\texttt{PTArcade}\xspace}
\newcommand{\enterprise}{\texttt{enterprise}\xspace}

\makeatletter\gdef\@fpheader{}\makeatother
\renewcommand\afterLogoSpace{}
\renewcommand\afterProceedingsSpace{\vskip0pt}
\renewcommand\afterTitleSpace{\vskip22pt}
\renewcommand\afterEmailSpace{\vskip20pt}
\title{\vspace*{-0.3cm}
  Seeing through the fog: astrophysical priors and the evidence for new physics in pulsar timing arrays
  \texorpdfstring{\\[0.2cm]\makebox[\linewidth][c]{\includegraphics[width=4cm]{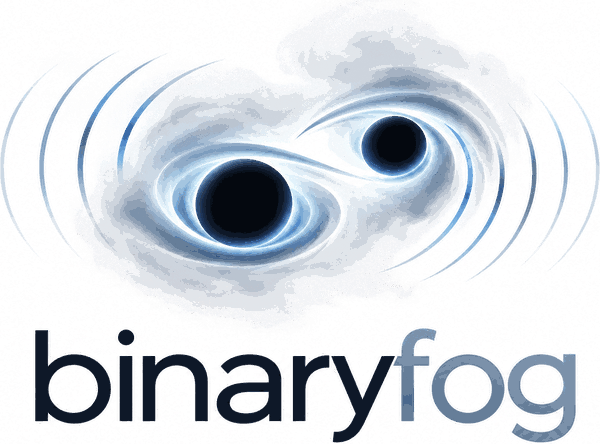}}\\[-0.2cm]}
  }

\author[1,2]{Torsten Bringmann,}
\author[2]{Chris Gordon,}
\author[3]{and Carlo Tasillo}

\affiliation[1]{Department of Physics, University of Oslo, Box 1048, N-0316 Oslo, Norway}
\affiliation[2]{School of Physical and Chemical Sciences, University of Canterbury,\\ Private Bag 4800, Christchurch 8140, New Zealand}
\affiliation[3]{Instituto de F\'isica Corpuscular (IFIC), Universitat de Val\`encia-CSIC,\\ Parc Cient\'ific UV, C/ Catedr\'atico Jos\'e Beltr\'an 2, E-46980 Paterna, Spain}

\emailAdd{torsten.bringmann@fys.uio.no}
\emailAdd{chris.gordon@canterbury.ac.nz}
\emailAdd{carlo.tasillo@ific.uv.es}

\abstract{
Pulsar timing arrays (PTAs) have found evidence for a gravitational wave background at nHz frequencies. Inspiralling supermassive black hole 
binaries (SMBHBs) likely contribute to at least part of the signal, but there may be additional cosmological sources from new physics in the early universe.
Disentangling the two is complicated by the prohibitively slow modelling of SMBHB spectra based on realistic population
models. Here we introduce \binaryfog{}, a surrogate of up-to-date population models 
also used by the NANOGrav collaboration, that returns the statistical distribution of bent spectra 
$\mathcal{O}(10^6)$ faster than
direct computation. We use \binaryfog{} to profile over
the six astrophysical parameters in fits of four extensively discussed new-physics models: a dark sector phase transition, 
cosmic superstrings, inflationary gravitational waves, and domain walls. 
Compared to the widely adopted power-law description of the SMBHB background, 
the preference for a phase transition drops from $2.8\sigma$ to $1.5\sigma$. 
The other three models follow a similar pattern. We also recommend updated astrophysical priors 
based on galaxy surveys published since 2014, which for current PTA data still only give a marginal 
preference of $1.7\sigma$ for a phase transition. 
An interpretation of the signal in terms of SMBHBs alone favours binaries that merge within about 
$2\,$Gyr after their host galaxies have merged, requiring an efficient mechanism to overcome the final parsec of separation.
If binaries need longer than about $6.5\,$Gyr to merge, the preference for a phase transition exceeds 
$3\sigma$ under our recommended updated prior set.
\binaryfog{} can help resolve these questions, as it allows users to study the impact 
of updated astrophysical priors and a realistic SMBHB population model on new-physics searches with PTAs.
}

\begin{document}
\maketitle
\flushbottom

\section{Introduction}
\label{sec:intro}

Pulsar timing arrays (PTAs) have reported evidence for a stochastic gravitational wave (GW) background at nanohertz frequencies, with the 
characteristic angular correlations predicted by Hellings and 
Downs~\cite{Hellings:1983fr,NANOGrav:2023gor,EPTA:2023fyk,Reardon:2023gzh,Xu:2023wog,Miles:2024seg}. A natural explanation 
consists in the 
population of inspiralling supermassive black hole binaries (SMBHBs) left behind by galaxy mergers~\cite{Sesana:2025udx}. Fitting the 
observed GW background amplitude requires black hole masses or merger rates at the high end of what galaxy surveys 
support~\cite{NANOGrav:2023hfp,Sato-Polito:2023gym,Ellis:2023dgf,Chen:2025wel,Mingarelli:2026ykj}. A recent determination of the 
massive end of the galaxy stellar mass function has been reported to remove much of this discrepancy~\cite{Liepold:2024woa}, but a fit that 
includes the environmental hardening of the binaries finds it insufficient~\cite{Matt:2025bao}. The measured spectrum is also flatter than the 
$\hcsq \propto f^{-4/3}$ of binaries driven by gravitational wave emission alone~\cite{Phinney:2001di}; part of the excess may come from the 
priors on the intrinsic red noise of the pulsars~\cite{Dai:2026jjw}, and it is unsettled how binaries traverse the final 
parsec~\cite{Begelman:1980vb,Milosavljevic:2002bn,Khan:2011gi,Vasiliev:2015uzo}.

A cosmological source, such as a phase transition in a dark sector, cosmic strings, domain walls, or blue-tilted inflationary tensor modes, 
could also explain (part of) the 
signal~\cite{NANOGrav:2023hvm,EPTA:2023xxk,Madge:2023dxc,Figueroa:2023zhu,Ellis:2023oxs,Wang:2025wht,Bringmann:2023opz,Balan:2025uke,Bringmann:2026xcx}. 
Fits of such sources so far have left the binaries 
out~\cite{Madge:2023dxc,Figueroa:2023zhu,Bringmann:2023opz,Wang:2025wht,Ellis:2023tsl,Vagnozzi:2023lwo,Franciolini:2023pbf,Bian:2023dnv,Wu:2023hsa}, 
represented them by a power law with amplitude $A$ and spectral index $\gamma$, under a fixed Gaussian prior based on population 
models in which the binaries evolve by gravitational wave emission 
alone~\cite{NANOGrav:2023hvm,NANOGrav:2023hfp,Balan:2025uke,Bringmann:2026xcx,Gouttenoire:2023ftk,Bhaumik:2023wmw,Winkler:2024olr} 
or with $\gamma = 13/3$ fixed~\cite{EPTA:2023xxk,Gouttenoire:2023bqy}, or described them by population parameters fitted to the timing 
data without priors from galaxy surveys~\cite{Ellis:2023oxs}.
We refer to the very common second treatment as 
the $(A,\gamma)$ power law. It discards much of the information the underlying population models contain,
with independent parameters describing the galaxy stellar mass function, the 
relation between black hole and bulge mass, and the hardening of the binaries. 
The background in a given frequency bin, furthermore, is dominated by a  few loud 
sources, so the prediction is a skewed distribution of spectra over realisations of the 
population~\cite{Ali-Haimoud:2026sbk,Lamb:2024gbh,Xue:2024qtx,Agazie:2024jbf}. It is debated whether a pulsar timing array on its own 
can separate SMBHBs from cosmological backgrounds~\cite{Moore:2021ibq,Becsy:2022pnr,Konstandin:2024fyo,Gardiner:2026orz}.

Here we ask how strongly the data prefer -- or constrain -- a cosmological contribution once the binaries are modelled as an actual  
population, and how this preference 
depends on the astrophysical priors. To this end we construct \binaryfog{}, a surrogate built from a library of \holodeck{} 
models~\cite{NANOGrav:2023hfp,holodeck} that returns this distribution in every frequency bin for any point of the six-dimensional 
astrophysical parameter space in $0.2\,$milliseconds instead of three minutes. We use it to profile over those astrophysical parameters in fits of 
four often discussed scenarios of new physics that may contribute to the GW background
-- a dark sector phase transition, cosmic superstrings, inflationary gravitational waves, and domain walls -- 
to the NANOGrav 15\,yr free spectrum re-analysed in 2026~\cite{NANOGrav:2026err}. 
We generally find that the evidence for a phase transition decreases significantly when using the full population model
rather than a simple power law. We point out that this conclusion depends crucially on the adopted priors for the parameters of the population
model, and we discuss in detail the impact of up-to-date astrophysically motivated priors on new-physics interpretations
in the context of the four scenarios mentioned above.

After introducing the population model and the surrogate in section~\ref{sec:background} and the likelihood in section~\ref{sec:likelihood}, 
we test the surrogate and compare the priors in section~\ref{sec:checks}, before turning to our main focus and results, namely
the new-physics fits and their dependence on the free spectrum and on the priors in section~\ref{sec:results}. We conclude in
section~\ref{sec:conclusions}. The appendices collect the new-physics spectra (appendix~\ref{app:models}), the priors and their sources
(appendix~\ref{app:priors}), and a quick-start guide to downloading and using \binaryfog{}  (appendix~\ref{sec:tool}).

\section{A surrogate for GW backgrounds from supermassive black hole binaries}
\label{sec:background}

The standard treatment of the binary background in searches for new physics, the $(A, \gamma)$ power law~\cite{NANOGrav:2023hvm}, 
takes the characteristic strain to be $\hc(f) = A\,(f/f_\mathrm{yr})^{(3-\gamma)/2}$, or
\begin{equation}
h^2\Omega_\mathrm{GW}^\mathrm{bin}(f) = \frac{2\pi^2 f_\mathrm{yr}^2}{3 H_{100}^2}\, A^2
  \left(\frac{f}{f_\mathrm{yr}}\right)^{5-\gamma} ,
  \label{eq:agamma}
\end{equation}
with $f_\mathrm{yr} = 1/\mathrm{yr}$ and $H_{100} = 100\,\mathrm{km\,s^{-1}\,Mpc^{-1}}$. The prior on $(\log_{10} A, \gamma)$ is a 
two-dimensional Gaussian centred at $\log_{10} A = -15.61$ and $\gamma = 4.71$, with standard deviations $0.53$ and $0.35$ and 
covariance $-0.0026$; it derives from power-law fits to population models in which the binaries evolve by gravitational wave emission 
alone~\cite{NANOGrav:2023hvm,NANOGrav:2023hfp}. In these models a binary is taken to emit at the same redshift as its 
galaxy merger. The inspiral is thus effectively treated as instantaneous, while in reality it would take several billion years
without environmental effects. To quote the NANOGrav collaboration~\cite{NANOGrav:2023hfp}, this model therefore
\textit{is not self-consistent, as GW-only evolution is unable to bring binaries to the PTA band within a Hubble time}, but still kept 
for comparison because it \textit{tends to produce the highest GW background amplitudes}. 
This motivates us to consider, instead, a self-consistent population model (and its impact on new-physics searches).

\subsection{Supermassive black hole binary population synthesis}
\label{sec:populations}

We work with the phenomenological semi-analytic models of \holodeck{}, the population synthesis
code used for NANOGrav's own astrophysical interpretation~\cite{NANOGrav:2023hfp,holodeck}, in the
parametrisation of Chen, Sesana, and Conselice~\cite{Chen:2018znx}. All our libraries are built with \holodeck{} in the first version to include 
the correction of the erratum~\cite{NANOGrav:2026err}, see appendix~\ref{app:code} for further details.
In this parametrisation, the differential number of binaries $\mathrm{d}N$ emitting in a logarithmic orbital
frequency interval $\mathrm{d} \ln f_\mathrm{p}$ follows from the comoving number density $n$ and the time a binary spends 
there as~\cite{NANOGrav:2023hfp}
\begin{equation}
  \frac{\mathrm{d}^4 N}{\mathrm{d}\log_{10}\!M \, \mathrm{d}q \, \mathrm{d}z \, \mathrm{d}\ln f_\mathrm{p}}
  = \frac{\mathrm{d}^3 n}{\mathrm{d}\log_{10}\!M \, \mathrm{d}q \, \mathrm{d}z}
    \, \frac{\mathrm{d}V_\mathrm{c}}{\mathrm{d}z} \, \frac{\mathrm{d}t}{\mathrm{d}\ln f_\mathrm{p}} \, \frac{\mathrm{d}z}{\mathrm{d}t} \, ,
  \label{eq:number}
\end{equation}
with $M$ the total binary mass, $q \le 1$ the mass ratio between the two black holes, $z$ the redshift, and $V_\mathrm{c}$ the comoving
volume. The orbital frequency $f_\mathrm{p}$ and the time $t$ are measured in the binary rest frame, so that 
$\mathrm{d}t/\mathrm{d}\ln f_\mathrm{p}$ is the time the binary spends per logarithmic interval of orbital frequency, its residence time. A 
circular binary emits gravitational waves at twice its orbital frequency, observed today at $f = 2f_\mathrm{p}/(1+z)$. The first factor on the 
right-hand side of eq.~\eqref{eq:number} is the comoving number density of binaries per unit mass, mass ratio, and redshift. It follows from 
the galaxy merger rate, which is determined by a galaxy stellar mass function,
a galaxy pair fraction, and a galaxy merger timescale; a
relation between the black hole mass and the stellar bulge mass thus turns the galaxy mergers into mergers of black hole pairs.

The inspiral of each pair can last over a billion years, i.e.~cosmological timescales. \holodeck{} evaluates the number density at the redshift of 
the galaxy merger, integrates the evolution of the binary (described below) from there, and evaluates the residence time and the cosmological 
factors at the later redshift $z$ at which the binary reaches the orbital frequency $f_\mathrm{p}$. Binaries that would reach $f_\mathrm{p}$ 
only in the future hence do not contribute. For more details of the galaxy merger modelling we refer to ref.~\cite{NANOGrav:2023hfp}.

The evolution of the binary through environmental effects is the least certain ingredient of the model. After their host galaxies merge, the black 
hole pairs have to harden from kiloparsec separations to the sub-parsec ones at which gravitational wave emission takes over. In the PTA 
band binaries of $10^8$ to $10^{10}\,\msol$ masses are $0.004$ to $0.1\,$pc apart, and they coalesce at $0.4$ to $40\,\mu$Hz in their rest 
frame, between 
the PTA band and the LISA band~\cite{LISA:2017pwj}. In spherical galaxies, stars orbiting close to the binary (within the so-called loss 
cone) are at some point ejected faster than two-body relaxation refills it.  Hardening therefore stalls at parsec-scale separations, which 
is known as the \textit{final parsec} problem~\cite{Begelman:1980vb,Milosavljevic:2002bn,Merritt:2004gc}.
Proposed solutions include stellar scattering in non-spherical or merging
galaxies~\cite{Khan:2011gi,Vasiliev:2015uzo}, torques from gas~\cite{Armitage:2002uu}, three-body 
interactions~\cite{Ryu:2018yhv,Bonetti:2017lnj}, and the dynamical friction exerted by spikes of self-interacting dark 
matter~\cite{Alonso-Alvarez:2024gdz}. Population synthesis codes do not resolve the problem but parametrise
its outcome~\cite{Ellis:2023dgf}. In this work we use a
broken power law for the rate at which the binary separation $r$ shrinks~\cite{Sesana:2015haa,NANOGrav:2023hfp},
\begin{equation}
    \left.\frac{\mathrm{d}r}{\mathrm{d}t}\right|_\mathrm{env}
  = - A \left(\frac{r}{\rchar}\right)^{1-\nuin}
        \left(1 + \frac{r}{\rchar}\right)^{\nuin - \nuout} \, ,
  \label{eq:hardening}
\end{equation}
with $t$ again the time in the rest frame of the binary. The time spent per logarithmic interval of separation hence 
scales as $r^{\nuin}$ well inside
the break separation $\rchar$ and as $r^{\nuout}$ well outside it. The normalisation $A$ is fixed by requiring that the
total time from an initial separation of $10\,\mathrm{kpc}$ to the innermost stable circular orbit
equals a target lifetime $\tauf$. The hardening of the binary by gravitational wave emission is added to this rate and 
takes over at small separations.

As a default, we use NANOGrav's fiducial \textit{Phenom} model for the galaxy merger rates, see eqs.~(14)--(18) and table~B1 of 
ref.~\cite{NANOGrav:2023hfp}. Its galaxy stellar mass function
\begin{align}
\Psi(m_\star, z) = \ln (10)\,\Psi_0\,\left( \frac{m_\star}{M_\psi} \right)^{\alpha_\psi} \exp\left(-\frac{m_\star}{M_\psi}\right)
\label{eq:gsmf}
\end{align}
is a single Schechter function, providing 
the number of galaxies per comoving $\mathrm{Mpc}^3$ per decade of stellar mass $m_\star$, the total mass 
of the stars in a galaxy, at redshift $z$. It follows a power law of slope $\alpha_\psi = 1 + \alpha_{\psi,0} + \alpha_{\psi,z} z$ at low mass and 
is cut off exponentially above a characteristic mass modelled as $\log_{10}(M_\psi/\msol) = m_{\psi,0} + m_{\psi,z} z$. Its normalisation changes exponentially with redshift,
$\log_{10}(\Psi_0/\mathrm{Mpc^{-3}}) = \psi_0 + \psi_z z$. Only $\psi_0$ and $m_{\psi,0}$ are free parameters, while $\psi_z = -0.60$, 
$m_{\psi,z} = +0.11$, $\alpha_{\psi,0} = -1.21$ and
$\alpha_{\psi,z} = -0.03$ are fixed at the values used in the NANOGrav analysis (table~B1 of ref.~\cite{NANOGrav:2023hfp}), 
see appendix~\ref{app:ngpriors} for further discussion. 
The mass $M_\bullet$ of each black hole of the binary follows the mass of the galactic bulge, 
$M_\mathrm{bulge} = 0.615\,m_\star$, with the bulge fraction $0.615$ taken from measurements in massive 
galaxies~\cite{Lang:2014cta,Bluck:2014ila}, and a log-normal scatter $\epsmu$,
\begin{align}
  \log_{10}(M_\bullet/\msol) = \mu + \alpha_\mu \log_{10}(M_\mathrm{bulge}/10^{11}\msol) + \mathcal{N}(0,\epsmu) \, ,
\end{align}
where $\mu$ is a dimensionless black hole mass normalisation
and the slope is fixed to $\alpha_\mu = 1.10$. 
Black hole masses correlate with properties of their host galaxies, such as the bulge mass, and these correlations are read as evidence that 
black holes and galaxies grow together~\cite{Kormendy:2013dxa}. As recommended by the NANOGrav analysis, we vary six parameters: 
the normalisation $\psi_0$ and the characteristic mass $m_{\psi,0}$ of 
the galaxy stellar mass function, the normalisation $\mu$ and the intrinsic scatter $\epsilon_\mu$ of the black hole
to bulge mass relation, the binary lifetime $\tauf$ in Gyr, and the inner hardening index $\nuin$.
We also follow ref.~\cite{NANOGrav:2023hfp} in fixing the outer index to $\nuout = +2.5$, as motivated 
by dynamical friction against the stars and dark 
matter of the merged host galaxy~\cite{NANOGrav:2023hfp,Kelley:2016gse}. The break separation is held fixed at
$\rchar = 100\,\mathrm{pc}$.\footnote{%
Gardiner et al.~\cite{Gardiner:2025mwf} instead draw $\rchar$ uniformly between $2$ and $20\,\mathrm{pc}$ and set $\nuout = 0$. 
We checked explicitly that profiling 
over a further library of $4096$ \holodeck{} models with $\rchar$ between $2$ and $200\,\mathrm{pc}$ and $\nuout$ between $0$ and $2.5$ 
only marginally changes the allowed new-physics parameter space.
}

We consider four sets of priors on these parameters, as summarised in table~\ref{tab:priors}.  
Flat priors represent a completely uninformed choice. The 
NANOGrav priors~\cite{NANOGrav:2023hfp} are Gaussian in $\psi_0$, $m_{\psi,0}$, $\mu$, and $\epsmu$ and uniform in $\tauf$ and 
$\nuin$. Their centres follow from fits to galaxy survey data and black hole mass 
measurements~\cite{Tomczak:2013bxa,Kormendy:2013dxa,McConnell:2012hz}, but their widths are up to four times larger than the 
uncertainties of these fits. The Tomczak--Kormendy priors take the centres and uncertainties of these fits, and the survey-average priors the centres 
and the spread of new measurements published since then. Section~\ref{sec:galaxypriors} further motivates these specific prior choices, and 
appendix~\ref{app:priors} traces them back to their origin in the literature. When not referring to a specific prior choice in the following, 
results refer to the NANOGrav priors.

\begin{table}[tbp]
\centering
\scriptsize
\setlength{\tabcolsep}{2.5pt}
\begin{tabular}{lcccc}
\toprule
parameter & flat & NANOGrav & Tomczak--Kormendy & \begin{tabular}[b]{@{}c@{}}survey\\ average\end{tabular} \\
\midrule
$\psi_0$ & $\mathcal{U}(-3.5, -1.5)$ & $\mathcal{N}(-2.56, 0.40)$ & $\mathcal{N}(-2.61, 0.10)$ & $\mathcal{N}(-2.20, 0.15)$ \\
$m_{\psi,0}$ & $\mathcal{U}(10.5, 12.5)$ & $\mathcal{N}(10.90, 0.40)$ & $\mathcal{N}(10.96, 0.11)$ & $\mathcal{N}(10.82, 0.11)$ \\
$\mu$ & $\mathcal{U}(7.6, 9.0)$ & $\mathcal{N}(8.60, 0.20)$ & $\mathcal{N}(8.69, 0.05)$ & $\mathcal{N}(8.80, 0.10)$ \\
$\epsmu$ & $\mathcal{U}(0, 0.9)$ & $\mathcal{N}(0.32, 0.15)$ & $\mathcal{N}(0.28, 0.05)$ & $\mathcal{N}(0.40, 0.05)$ \\
$\tauf/\mathrm{Gyr}$ & $\mathcal{U}(0.1, 11)$ & $\mathcal{U}(0.1, 11)$ & $\mathcal{U}(0.1, 11)$ & $\mathcal{U}(0.1, 11)$ \\
$\nuin$ & $\mathcal{U}(-1.5, 0)$ & $\mathcal{U}(-1.5, 0)$ & $\mathcal{U}(-1.5, 0)$ & $\mathcal{U}(-1.5, 0)$ \\
\bottomrule
\end{tabular}
\caption{The six parameters of the population model and the four prior sets that we adopt in this work. $\mathcal{U}$ denotes a 
uniform prior between its two arguments, and $\mathcal{N}$ indicates a Gaussian with stated central value and width. 
$\psi_0$ is stated in units of $\log_{10}\mathrm{Mpc^{-3}}$, 
$m_{\psi,0}$ and $\mu$ in units of $\log_{10}\msol$, $\epsmu$ in dex. The NANOGrav priors are those of the \holodeck{} library of 
ref.~\cite{NANOGrav:2023hfp}; the Tomczak--Kormendy priors take the centres and uncertainties of the fits they are based on, and the 
survey-average priors the centres and spread of measurements published since 2014, see also appendix~\ref{app:priors}.
}
\label{tab:priors}
\end{table}

\subsection{Realisation scatter}
\label{sec:realisations}

The characteristic strain in a redshifted frequency bin is the sum over the population in eq.~\eqref{eq:number} weighted by the squared strain 
$h_\mathrm{s}^2$ of each binary, in the form of
Phinney~\cite{Phinney:2001di} and of Sesana, Vecchio, and Colacino~\cite{Sesana:2008mz}. The population herein is characterised by its 
expected number of binaries in each bin of mass, mass ratio,
redshift, and frequency. The number actually realised in a given universe is modelled as following a Poisson distribution, so the characteristic
strain squared in an observed frequency bin $\Delta f$ is
\begin{equation}
  \hcsq(f) = \sum_{M,q,z} \mathcal{P}\!\left(\frac{\mathrm{d}^4 N}{\mathrm{d}\log_{10}\!M\,\mathrm{d}q\,\mathrm{d}z\,\mathrm{d}\ln f_\mathrm{p}}
  \Delta \log_{10}\!M \, \Delta q \, \Delta z \, \Delta \ln f \right)
  \frac{h_\mathrm{s}^2(f_\mathrm{p})}{\Delta \ln f} \, ,
  \label{eq:realisation}
\end{equation}
where $\mathcal{P}(\lambda)$ is a Poisson variate with mean $\lambda$ and $h_\mathrm{s}$ is the
sky- and polarisation-averaged strain of a single binary, which decreases with its comoving distance $d_\mathrm{c}(z)$ as 
$h_\mathrm{s}^2 \propto d_\mathrm{c}^{-2}$~\cite{NANOGrav:2023hfp}. Keeping the inclination of each binary, which \holodeck{} averages 
over, would broaden the distribution by about ten per cent when many sources share a bin~\cite{Ali-Haimoud:2026sbk}. Averaging 
eq.~\eqref{eq:realisation} over many draws gives a smooth expected strain that, however, no individual realisation follows, since near the 
measured amplitude a few loud binaries typically carry the signal.

The spread between realisations grows with
frequency. We quantify the realisation scatter in each frequency bin as half the difference between the sixteenth and eighty-fourth percentiles 
of the predicted logarithmic strain squared throughout this work. For our fiducial astrophysical model, this half-width increases from 0.03 dex 
in the lowest frequency bin to 0.32 dex in the fourteenth bin. By comparison, the corresponding half-widths of the NANOGrav free-spectrum 
posteriors are 0.40 and 0.81 dex. Thus, the scatter between different realisations of the SMBHB population is smaller than the observational 
uncertainty in both bins. We therefore expect the realisation scatter alone to move the favoured new-physics parameter regions little, given 
the current data.

For the purpose of a likelihood we further need the distribution of $\log_{10}\hcsq$ in each bin.
We adopt a 
one-parameter warp of a normally distributed variable, which we find (cf.~section~\ref{sec:warpchoice}) to describe the distribution of the \holodeck{} realisations better than a log-normal or a map-Airy distribution:
\begin{equation}
  \log_{10}\hcsq(f_i) = m_i + \sigma_i \left( \xi + c \, \xi^2 \right) , \qquad
  \xi \sim \mathcal{N}(0,1) ,
  \label{eq:warp}
\end{equation}
with $m_i$ the median of $\log_{10}\hcsq$ in bin $i$, $\sigma_i$ its half-width between the sixteenth and
eighty-fourth percentiles, and $c$ a skew parameter. The skew is taken to be the same in every frequency bin of a population model but 
depends on the model, i.e.\ on the six astrophysical parameters, through the map spelled out right below in 
section~\ref{sec:surrogate}.\footnote{%
Over the NANOGrav priors its median is $0.20$, with $0.14$ and $0.23$ at the sixteenth and eighty-fourth percentiles. When fitting the skew 
$c$ separately in each bin to the quantiles between $0.5$ and $99.5\%$, which weights the loud tail, we find a slight running from $0.33$ in 
the lowest frequency bin to $0.25$ in the fourteenth bin, which we neglect in this work.
} 
The warp in eq.~\eqref{eq:warp} is held flat below $\xi = -1/(2c)$, where it turns over, which sets a floor on how faint a
realisation can be.

\subsection{A surrogate for the spectral distribution}
\label{sec:surrogate}

Evaluating \holodeck{} with $2000$ realisations takes three minutes on one core. The NANOGrav analysis
therefore interpolates their library with Gaussian 
processes for the median and the standard deviation of $\log_{10}\hcsq$ per bin~\cite{NANOGrav:2023hfp,Taylor:2016ftv}, as do fits of new 
physics affecting the binary evolution~\cite{Tiruvaskar:2025lkq,Tiruvaskar:2025hxl}. Alternative approaches have used
similar neural-network 
emulators~\cite{Bonetti:2023ozm,Tiruvaskar:2026tyl}, a normalising flow over five bins that keeps the correlations between 
them~\cite{Laal:2024trp}, and a direct evaluation of the semi-analytic models~\cite{Iemoto:2026fgk}. 
For realistic full parameter scans, however, we need the shape of the 
distribution in each bin {\it and} an evaluation in a fraction of a millisecond for efficient sampling, and hence 
fit polynomial maps instead.

The median spectrum is described by a  
transition from environment-driven to GW-driven hardening, in the phenomenological turnover form used in pulsar timing analyses~\cite{Sampson:2015ada}:
\begin{equation}
  \hc(f) = A \left(\frac{f}{f_\mathrm{yr}}\right)^{-2/3}
  \left[1 + \left(\frac{f_\mathrm{bend}}{f}\right)^{\kappa}\right]^{-1/2} ,
  \label{eq:bend}
\end{equation}
with $f_\mathrm{yr} = 1\,\mathrm{yr}^{-1}$, an amplitude $A$, a bend frequency $f_\mathrm{bend}$
and a fixed sharpness $\kappa = 1.8$, first used on measured
data in ref.~\cite{NANOGrav:2015aud}.\footnote{%
Ref.~\cite{Sampson:2015ada} quotes $\kappa = 10/3$ for hardening by stellar scattering and $7/3$ for a gas disc, both for a single binary. 
When fitted to each model of the library, $\kappa$ varies between $1.2$ and $6.0$. 
When only considering models where the amplitude is within a factor of ten of the measured one and the four galaxy and black hole parameters are 
within two standard deviations of their measured values, it shows a much smaller spread, between $1.65$ and $1.90$, with a 
median of $1.76$. The fixed value of $\kappa=1.8$ used here reproduces held-out spectra to within $6$--$8\%$.
}

The surrogate maps the six astrophysical parameters $\theta = (\psi_0, m_{\psi,0}, \mu, \epsmu, \tauf, \nuin)$ onto the three quantities  
needed by the likelihood. Each is a polynomial in $u(\theta)$, which maps the ranges of the flat priors linearly onto 
the hypercube $[-1,1]^6$:
\begin{align}
  \log_{10}\hcsq\big|_i(\theta) &= 2\log_{10}\hc\big(f_i; A(\theta), f_\mathrm{bend}(\theta)\big)
    + \sum_{|n|  \le 4} w^{\mathrm{med}}_{n,i}\, u^n \, , \label{eq:surrogate} \\
  \log_{10}\sigma_i(\theta) &= \sum_{|n| \le 4} w^{\sigma}_{n,i}\, u^n \, , \qquad
  c(\theta) = \sum_{|n| \le 2} w^{c}_{n}\, u^n \, .
\end{align}
Here $n = (n_1, \ldots, n_6)$ runs over the multi-indices of total degree $|n| = n_1 + \dots + n_6\leq4$, 
$u^n \equiv \prod_{k=1}^6 u_k^{n_k}$, and $\hc$ in the first term is the bend of
eq.~\eqref{eq:bend}. Rather than fitting $A$ and $f_\mathrm{bend}$ directly, which are
poorly determined for spectra whose turnover falls outside the simulated range, we fit the value and
the slope of $\log_{10}\hcsq$ at a reference frequency of $5\nHz$ and invert
eq.~\eqref{eq:bend} for them in closed form at fixed $\kappa$. 
The total size of the four maps is $54\,$kB:
two polynomials of degree four for the reference value and slope, one per bin for
the median correction and for the width, and one of degree two for the skew, which varies little through the parameter space.
The library holds $2000$ astrophysical models $\theta$, drawn by %
Latin hypercube sampling over the flat prior ranges,
each realised $2000$ times. Models whose spectrum at $5\nHz$ lies more than 
three decades in $\hcsq$ below the measured one are not used. Three quarters of the remaining models 
are used for the fit and one quarter is held out for 
testing.

\section{The pulsar timing likelihoods and parameter inference}
\label{sec:likelihood}

\subsection{Free-spectrum likelihood}
\label{sec:freespec}

We use the free-spectrum posteriors distributed through \ceffyl{}~\cite{Lamb:2023jls} and \ptarcade{}~\cite{Mitridate:2023oar} instead of the 
timing residuals. They are marginalised over the pulsar noise and dispersion measure parameters, which makes the likelihood about a 
thousand times faster than in \enterprise{}~\cite{enterprise}. We restrict ourselves to the $N_\text{f}=14$ lowest frequency bins, as recommended by 
NANOGrav~\cite{NG15Models}, since the higher bins carry almost no information about a signal. In each bin $i$ the free spectrum gives the 
posterior of $\rho_i$, the square root of the timing-residual power, without assuming a spectral shape. A stochastic background 
$h^2 \Omega_\mathrm{GW}$ can be translated to the timing residual power and the characteristic strain $h_\mathrm{c}$ through
\begin{equation}
  h^2 \Omega_\mathrm{GW}(f) = \frac{8\pi^4}{H_{100}^2} \, \rho^2 f^5 T = \frac{2\pi^2}{3 H_{100}^2} f^2 \hcsq(f) \, ,
  \label{eq:conversion}
\end{equation}
where $T$ is the observation time span and $H_{100} \equiv 100\,\mathrm{km\,s^{-1}\,Mpc^{-1}}$, so that the
likelihood of a deterministic spectrum is calculated from the product over bins of the measured density evaluated
at the predicted $\rho_i$. We follow \ceffyl{}'s conventions at the edges of the stored kernel density grids: below the grid, the probability density is held at its value at the lower grid boundary, since sufficiently weak signals are indistinguishable from noise. Above the grid, the model is rejected because the predicted gravitational-wave amplitude exceeds the observational upper limit.

\newpage

\subsection{Marginalisation over realisations, profile likelihood, and statistical preference}
\label{sec:profile}

The binaries do not contribute as a deterministic prediction for the spectrum but through the distribution discussed in 
section~\ref{sec:realisations}. As a result the log-likelihood of a new-physics spectrum $\Omega_i^\mathrm{np}$, depending on the free parameters $\vartheta$ and the model nuisance parameters $\vartheta_\mathrm{nuis}$, together with a population 
model $\theta$ consists of the sum over the $N_\text{f}$ Fourier frequencies of their log-likelihoods. Each Fourier bin's contribution to the log-likelihood is an integral over the binary 
contribution's distribution in that bin, hence
\begin{equation}
  \ln L(\vartheta, \vartheta_\mathrm{nuis}, \theta) = \sum_{i=1}^{N_\text{f}} \ln \int \mathrm{d}\xi \;
  \varphi(\xi) \,
  p\!\left(\rho_i\!\left(\Omega_i^\mathrm{np}(\vartheta, \vartheta_\mathrm{nuis}) + B_i(\xi \mid \theta)\right) \,\middle|\, \mathrm{data}\right) .
  \label{eq:marginal}
\end{equation}
Here $\varphi$ is the standard normal density, and $B_i(\xi \mid \theta)$ the binary contribution in
bin $i$ of the realisation labelled by $\xi$ through eq.~\eqref{eq:warp}, with the median, width and
skew supplied by the surrogate, and $p(\rho_i)$ is the posterior distribution of the free-spectrum analysis. In practice, we evaluate the 
integral as a sum over a fixed grid of $101$ nodes in $\xi$ between
$-4$ and $4$, weighted by $\varphi$.
In the limit of vanishing width of the binary background eq.~\eqref{eq:marginal} reduces to the usual
likelihood of a summed spectrum of two contributions.

For the inference of the $1\sigma$ and $2\sigma$-preferred regions of the new-physics parameter space we use a profile likelihood and 
Wilks' theorem: for a given parameter (combination) $\vartheta$ which we want to constrain, all astrophysical nuisance parameters $\theta$ 
and model nuisance parameters $\vartheta_{\mathrm{nuis}}$ are maximised (profiled) over:
\begin{equation}
  \ln L_\mathrm{prof}(\vartheta) = \max_{\vartheta_\mathrm{nuis}, \, \theta}
  \left[ \ln L(\vartheta, \vartheta_\mathrm{nuis}, \theta) + \ln \pi(\theta) \right] .
  \label{eq:profile}
\end{equation}
We treat the new-physics model parameters as having flat 
priors over their ranges stated in appendix~\ref{app:models}.  
For the SMBHB background parameters $\theta$, we use instead the priors listed in table~\ref{tab:priors}.
The $1\sigma$ and $2\sigma$ regions, which would contain $68.3\%$ and $95.4\%$ of the probability of a Gaussian distribution, then
span the parameter space with $\ln L_\mathrm{prof}$ up to $1.15$ and $3.09$ below the best-fit likelihood 
$\ln L_\mathrm{prof}^\mathrm{max}$ for two scanned model parameters within $\vartheta$, and up to $0.50$ and $2.00$ for one model 
parameter~\cite{Wilks:1938dza}, respectively. To quantify whether new physics is preferred, we use the profile likelihood ratio (or gain) $G$ 
of the best fit over the same fit without it,
\begin{equation}
  G = \ln L_\mathrm{prof}^\mathrm{max} - \max_{\theta} \left[ \ln L(\theta)\big|_{\Omega^\mathrm{np} = 0} + \ln \pi(\theta) \right] .
  \label{eq:gain}
\end{equation}
To express $G$ as a nominal significance, we use the test statistic
$\Lambda = 2G$ and adopt the approximate conversion
$p = \frac{1}{2}P(\chi_k^2 \geq \Lambda)$,
where $\chi_k^2$ denotes a chi-square distribution with $k$
degrees of freedom, and $k$ is the number of free parameters
of the new-physics model. This prescription has also been
used in gamma-ray source searches~\cite{Fermi-LAT:2010feq}.
We quote the corresponding one-sided Gaussian significance
$Z = \Phi^{-1}(1-p)$, where $\Phi$ is the cumulative standard
normal distribution. The conversion is approximate because
the spectral-shape parameters are not identifiable when the
new-physics contribution vanishes, so the exact null distribution
need not follow this prescription~\cite{Gross:2010qma,Protassov:2002sz}. Monte Carlo simulations can be used to go beyond this approximation, see for instance ref.~\cite{CTAO:2024wvb}, which we leave for future studies.
The quoted significances should therefore be regarded as nominal. A significance of $1$, $2$, and $3\sigma$ then requires $G = 0.50$, $2.00$, and $4.50$ 
for $k = 1$, $1.15$, $3.09$, and $5.91$ for $k = 2$, and $1.76$, $4.01$, and $7.08$ for $k = 3$. A contour in a plane uses $k = 2$, and a 
significance derived from $G$ uses the number $k$ of free parameters of the new-physics model, including those profiled over. A model with 
more parameters thus has to gain more in $G$ to reach the same significance, which makes the significances of different models 
comparable. The astrophysical parameters do not enter $k$, since both terms of eq.~\eqref{eq:gain} contain them.

The $(A, \gamma)$ power law can be profiled as well via eq.~\eqref{eq:profile}, with its Gaussian prior as
a penalty, so that both treatments of the binaries differ only in the family of background spectra
and in the prior on it. This prior matters in its own right, since part of $G$ comes from the prior penalty when the astrophysical parameters 
move away from the values galaxy surveys prefer. In section~\ref{sec:priordep} we will get back to this, and discuss  these two 
contributions separately.

\subsection{Noise modelling and available data sets}
\label{sec:data}

The free-spectrum posteriors $p(\rho_i)$ depend on the modelling of the intrinsic pulsar noise, the dispersion measure variations and  
the correlations between pulsars. We use the public posteriors of the NANOGrav 15\,yr data set in their most recent version, 2.0.0 of the data record~\cite{NG15KDE}, re-run after a correction of the sampler~\cite{PTMCMCSampler,NANOGrav:2026err}. This release also includes the spectrum with Hellings--Downs (HD) correlations fitted together with 
a monopole (MP), as e.g.~resulting from clock errors, a dipole (DP), like from errors of the solar system ephemeris, 
and a common uncorrelated red noise 
(CURN). The NANOGrav analysis adopts this HD+MP+DP+CURN spectrum 
as the most conservative choice for their astrophysical interpretation~\cite{NANOGrav:2023hfp,NANOGrav:2026err}. 
Let us remark that, at the time of writing, \ptarcade{}~\cite{PTArcadeData} still downloads the spectra released in 2023 together with NANOGrav's first analyses of the 15\,yr data set, 
from before the correction.\footnote{%
The \ptarcade{} developers confirmed this. Version 1.5 still downloads the old files, and the development branch contains the fix.
}

We use the CURN spectrum of the 2026 release as a baseline for our analysis, 
which assumes the same red noise process in every pulsar without correlations 
between pulsars, with deterministic dispersion measure offsets. The spectrum rests on the auto-correlations of the pulsars, so CURN 
recovers it without the suppression of the Hellings--Downs overlap reduction function~\cite{Lamb:2023jls,Mitridate:2023oar}. NANOGrav's 
new-physics search used the HD spectrum~\cite{NANOGrav:2023hvm}. 
In section~\ref{sec:datasets}, we will explore in detail the impact of these choices on new-physics interpretations
of the PTA data.

\section{The binary background, its priors, and validation of the surrogate}
\label{sec:checks}

Before we use the surrogate for new-physics searches in section~\ref{sec:results}, 
we take a closer look at how the astrophysical parameters individually shape the spectrum (section~\ref{sec:anatomy})
and how this overall translates to the range of spectra compatible with the priors considered in this work
(section~\ref{sec:galaxypriors}). We then perform a number of validation tests for the surrogate,
by comparing the warp of eq.~\eqref{eq:warp} with other distributions used in the literature (section~\ref{sec:warpchoice})
and by reproducing NANOGrav's fit of the population model (section~\ref{sec:validation}).

\subsection{Anatomy of the binary background and comparison to a power law}
\label{sec:anatomy}

\begin{figure}[tbp]
\centering
\includegraphics[width=\textwidth]{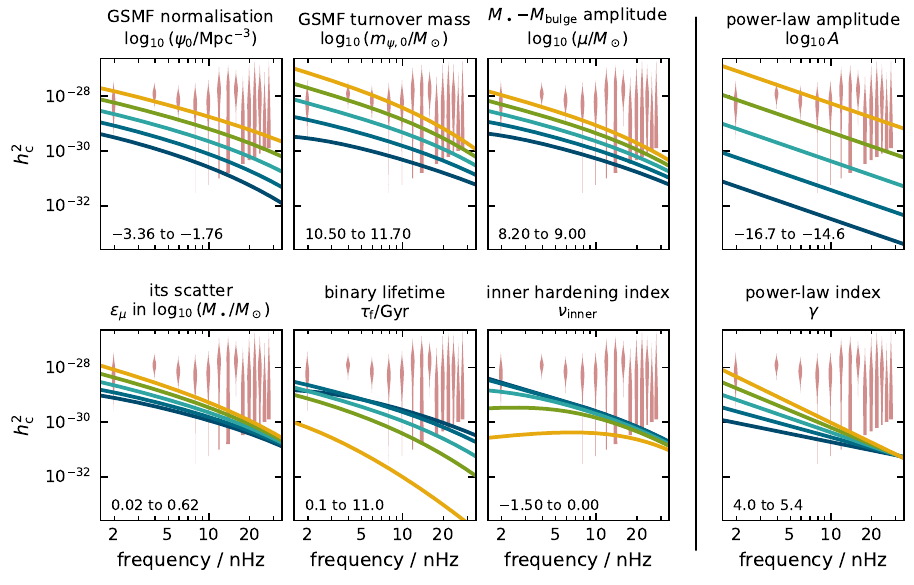}
\caption{The median predicted SMBHB spectra that result when individually varying each of the six parameters of the population 
model (left) and, alternatively, the two parameters of the $(A, \gamma)$ power law (right) across the indicated range (the $2\sigma$ 
range of the NANOGrav priors for the four Gaussian parameters, and the full range of the uniform priors for $\tauf$ and $\nuin$, cf.~table~\ref{tab:priors}). We restrict ourselves to $m_{\psi,0}> 10.5$ here since the $2\sigma$ range of the NANOGrav prior reaches below the lower limit of the uniform prior used to generate the model library.
The five values in each panel are equally spaced, dark blue for the 
smallest and yellow for the largest, with the other parameters at the centres of the NANOGrav priors; for the 
power law the range is the centre of its Gaussian prior $\pm 2\sigma$. For comparison, we also show 
our baseline free spectrum, the CURN spectrum of the 2026 NANOGrav 15\,yr release~\cite{NANOGrav:2026err}.
}
\label{fig:anatomy}
\end{figure}

The first three columns of figure~\ref{fig:anatomy} show 
how varying each of the six parameters individually changes the median spectrum, when holding the other 
parameters fixed at their central values. 
We also include, for comparison, our baseline free spectrum inferred by NANOGrav.
The parameters $\psi_0$, $m_{\psi,0}$, $\mu$, and $\epsmu$ mostly set the amplitude, but $m_{\psi,0}$ and $\epsmu$ also steepen 
the spectrum at high frequencies. Across the ranges shown here the characteristic mass $m_{\psi,0}$ moves $\hcsq$ at $2\nHz$ by 
more than two orders of magnitude, $\psi_0$ and $\mu$ by about one and a half, and $\epsmu$ by one, while the slope in the lowest 
five bins changes by less than one unit. These four parameters are clearly
strongly degenerate and hence need galaxy survey priors to be disentangled. The binary lifetime $\tauf$ and the inner hardening index 
$\nuin$ instead bend the spectrum at low frequencies: across their priors they change the slope in the lowest five bins by more than one and 
a half units~\cite{NANOGrav:2023hfp}, and with it $\hcsq$ at $2\nHz$ by one and a half to two orders of magnitude. A good fit to the data 
requires a short lifetime and a shallow inner index. 

The final column of figure~\ref{fig:anatomy} shows, instead, how varying the parameters of the much simpler, commonly
used power-law description affects the predicted GW background. Clearly, this simpler prescription can imitate the amplitude 
parameters of the population model, but not the bend induced by the hardening parameters. The full model therefore 
has in particular more freedom to fit the relatively high second frequency bin, implying less need for an additional 
cosmological contribution.

\subsection{Galaxy survey priors}
\label{sec:galaxypriors}

\begin{figure}[tbp]
\centering
\includegraphics[width=\textwidth]{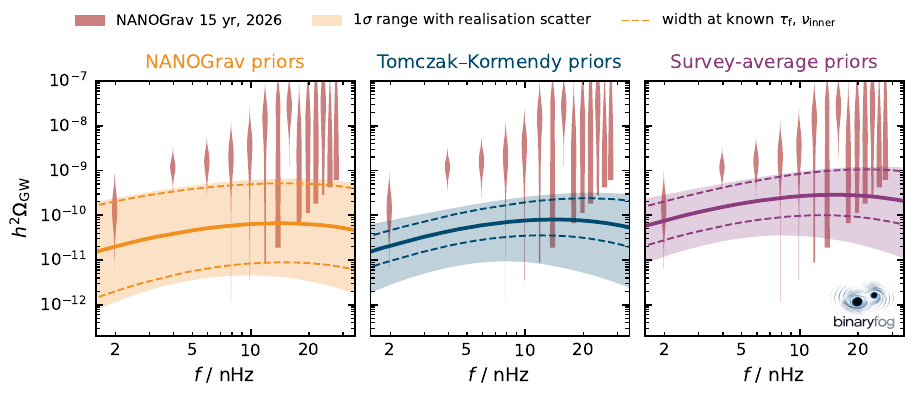}
\caption{The gravitational wave background of the binaries predicted under the NANOGrav (left), Tomczak--Kormendy (middle) and survey-average 
priors (right) of table~\ref{tab:priors}, before any fit to the timing data, from \binaryfog{} with $40000$ draws from each prior set and one 
realisation of the background per draw. Solid lines show the median, shaded bands the $1\sigma$ range over the prior and the realisation 
scatter. The dashed lines show the $1\sigma$ range due to the remaining
four astrophysical parameters and realisation scatter when
$\tauf$ and $\nuin$ are held fixed. This range is averaged
over their uniform priors and centred on the median of the
full distribution. For the Tomczak--Kormendy and survey-average priors these two hardening 
parameters set most of the width of the band. The violins are the NANOGrav 15\,yr CURN free spectrum. Under all three prior sets the 
median prediction lies below the measured spectrum, in the lowest bin by $0.8\,$dex in $h^2\Omega_\mathrm{GW}$ for the NANOGrav and 
Tomczak--Kormendy priors and by $0.2\,$dex for the survey-average priors.
}
\label{fig:priorsets}
\end{figure}

Figure~\ref{fig:priorsets} shows the background that the three Gaussian prior sets, cf.~table~\ref{tab:priors}, 
predict before any fit to the timing data. The NANOGrav and 
Tomczak--Kormendy priors predict a median spectrum below the measured one, but the survey-average priors come noticeably
closer to it. The width of the 
prediction comes mostly from the uniform priors on the two hardening parameters, except under the NANOGrav priors, whose wide 
Gaussian priors of the other parameters contribute a similar amount.

The NANOGrav priors are centred on fits of the galaxy stellar mass function to the data of Tomczak et al.~\cite{Tomczak:2013bxa} and on 
the black hole to bulge mass relations of Kormendy and Ho~\cite{Kormendy:2013dxa} and of McConnell and Ma~\cite{McConnell:2012hz}, 
but their widths are about four times the uncertainties of these fits, without a stated justification (appendix~\ref{app:ngpriors}). Since these 
widths strongly affect the predicted amplitude of the binary background, and hence the preference for new physics, the Tomczak--Kormendy 
priors take the centres and uncertainties of the fits themselves (appendix~\ref{app:tkpriors}). We do not claim that the astrophysical 
parameters are known this precisely: fits to different galaxy catalogues differ by more than these 
uncertainties~\cite{Leja:2019lfo,Sato-Polito:2025dqw}, and so do the bulge-mass fits of Kormendy and Ho and of McConnell and Ma. 
Whether such narrow priors are compatible with the measured amplitude is 
debated~\cite{Sato-Polito:2023gym,Lapi:2025wxt,Ellis:2024wdh,Liepold:2024woa,Sato-Polito:2025dqw}. They basically illustrate what the data say about 
new physics if the astrophysical inputs are as precise as the measurements they are based on. The survey-average priors take the mean and the 
spread of six determinations of the galaxy stellar mass function published since 2014, and of black hole mass relations brought to a common 
initial mass function (appendix~\ref{app:surveypriors}). Their widths are similar to those of the Tomczak--Kormendy priors, but their centres 
predict a louder background, by a factor of about two in $\hc$ (figure~\ref{fig:priorsets}).

\subsection{Distribution of the realisation scatter and accuracy of the surrogate}
\label{sec:warpchoice}

The distribution of eq.~\eqref{eq:realisation} is asymmetric in $\log_{10}\hcsq$, with a heavy tail towards loud signals. A finite number of 
sources steepens the spectrum and scatters it~\cite{Sesana:2008mz,Roebber:2015iva}, and ignoring this scatter biases the inferred 
amplitude~\cite{Ravi:2012bz}. When many sources share a bin the distribution approaches a universal limit~\cite{Ellis:2023owy}, which 
Ali-Ha\"imoud identified as a reflected map-Airy distribution%
~\cite{Ali-Haimoud:2026sbk}. In the opposite limit of finite source counts the t-process of 
refs.~\cite{Sardesai:2024rdd,Moreschi:2026ere} offers a good fit, as it is skewed towards large amplitudes.
The distribution has been computed for various population models~\cite{Sato-Polito:2024lew}, its loud 
tail traced to the chance of a nearby source~\cite{Raidal:2026ezm}, its higher moments derived with a low-redshift 
regularisation~\cite{Lamb:2024gbh,Hisamatsu:2026qvd}, and wave interference added~\cite{Xue:2024qtx}. Simplified versions of the 
integral over realisations in eq.~\eqref{eq:marginal} are used in 
refs.~\cite{Tiruvaskar:2025lkq,Tiruvaskar:2025hxl,Yu:2026rba,Falxa:2026mhh,Falxa:2025qxr,Sardesai:2024rdd,Ellis:2023dgf,Ellis:2023oxs}. 

Here we compare the warp with two other distributions, each matched to the median and the width of the realisations in each bin
(the latter defined here as the half-distance between the sixteenth and eighty-fourth percentiles~\cite{Ellis:2023owy}): 
the log-normal, 
common in the literature~\cite{NANOGrav:2023hvm,Moore:2021ibq,Becsy:2022pnr,Konstandin:2024fyo}, and the map-Airy 
distribution~\cite{Ali-Haimoud:2026sbk}. Against $2000$ realisations of each of $150$ \holodeck{} models held out of the fit of 
section~\ref{sec:surrogate}, the Kolmogorov--Smirnov distance averaged over models and bins is $0.044$ for the warp, $0.073$ for the
log-normal, and $0.094$ for the map-Airy distribution. This ordering is opposite to the one of ref.~\cite{Ali-Haimoud:2026sbk}, where the 
map-Airy law fits better than a log-normal distribution at every amplitude.\footnote{%
The map-Airy law can be matched to the same percentiles only where the realisations are narrower than about $0.55$ in $\log_{10}\hcsq$, 
and the distance is computed only there. We note that this distribution %
is derived for bins that hold many sources, while the effective source count read off the 
measured width falls from $620$ in the lowest bin through $13$ in the fourth to $1.2$ in the fourteenth, so that above about $8\nHz$ the 
map-Airy law 
is used outside the regime it is derived in. In the three lowest bins at the centre of the NANOGrav priors it fits better than the log-normal and 
the warp in eq.~\eqref{eq:warp}.
}

\begin{figure}[tbp]
\centering
\includegraphics[width=\textwidth]{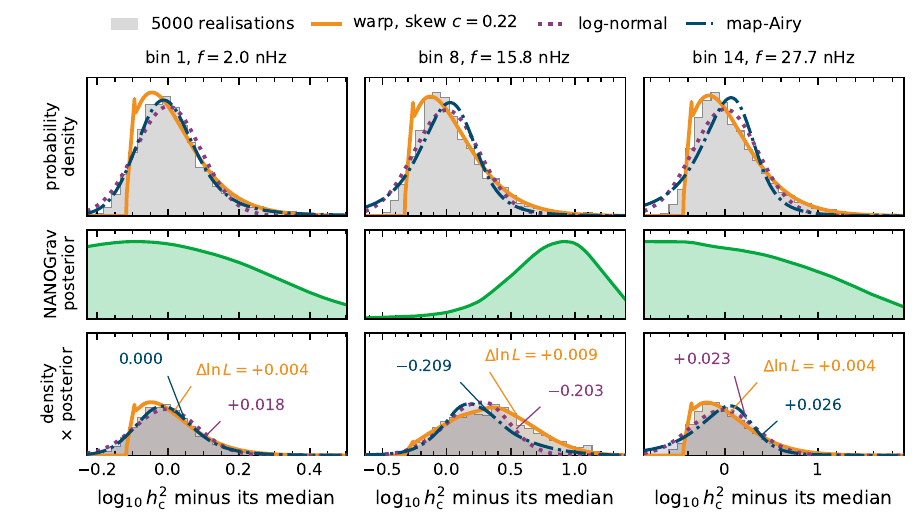}
\caption{%
{\it Top panel:} The distribution of $\log_{10}\hcsq$ over $5000$ realisations of the \holodeck{} model at the best fit of the binaries alone (NANOGrav 
15\,yr common uncorrelated red noise spectrum, NANOGrav priors), in the lowest, the eighth, and the highest of the fourteen bins (grey), 
against a log-normal (purple, dotted), the warp of eq.~\eqref{eq:warp} with the skew $c = 0.22$ of the surrogate at this point (orange, solid) 
and the map-Airy distribution (dark blue, dash-dotted), each matched to the same median and width. The spike of the warp is the probability 
it piles up at its floor (section~\ref{sec:realisations}). {\it Middle panel:} The posterior of the amplitude of the bin in the NANOGrav CURN free spectrum, 
as a function of the binary amplitude. {\it Bottom panel:} The product of the two top panels, the integrand of eq.~\eqref{eq:marginal},
normalised such that the area under the histogram equals one. The annotated numbers 
denote $\Delta\ln L$, the difference of the logarithm of the area under the histogram and the respective parametrisation of the same colour, 
corresponding to the difference in the likelihood when using the respective distributions.
The warp of eq.~\eqref{eq:warp} is the 
only one which consistently reproduces the likelihood of the realisations on the (sub-)per cent level across the bins.}
\label{fig:realisations}
\end{figure}

In order to resolve the tails of the distributions well, we generate $5000$ realisations at the best fit of the binaries alone to the CURN 
spectrum under the NANOGrav priors. In figure~\ref{fig:realisations}, we  show the resulting histograms together with the three analytical 
parametrisations described above. The width in $\log_{10}\hcsq$ grows from $0.09$ in the first bin to $0.35$ in the fourteenth, and the 
distribution becomes increasingly skewed towards loud realisations. In the fourteenth bin, the mean sits a factor of $2.5$ above the 
median~\cite{Sesana:2008mz}. In that bin the Kolmogorov--Smirnov distance, most sensitive to the bulk, is $0.048$ for the warp, $0.076$ 
for the log-normal, and $0.110$ for the map-Airy law. The 1-Wasserstein distance, the mean distance between quantiles and sensitive to the 
tails, is $0.016$, $0.079$, and $0.130$ in $\log_{10}\hcsq$. The warp gets the loud tail right, and the other two are too little skewed.

The warp is flat below $\xi = -1/(2c)$ and misdescribes the faintest realisations. Replacing this cutoff by a constant floor moves the best-fit 
$\ln L$ for a phase transition by $3 \times 10^{-6}$, since such faint realisations lie far below the measured spectrum. 
Figure~\ref{fig:realisations} shows where the shape of the adopted parametrisation matters in practice. 
In the lowest and the highest bin, the posterior of the free spectrum 
hardly varies across the binary distribution, and all three parametrisations result in the same log-likelihood for this bin 
to within $0.03$. %
In the eighth bin the 
posterior rises steeply towards louder signals and the integrand is carried by the loud tail. The warp follows the realisations there ($\Delta\ln 
L = +0.009$), while the log-normal and the map-Airy law put too little probability into the tail ($\Delta\ln L = -0.20$ and $-0.21$). Summed 
over all fourteen bins, the warp overestimates $\ln L$ of the binaries alone by $0.13$, and the log-normal and the map-Airy law 
underestimate it by $0.07$ and $0.20$. At the best fit with a phase transition (see section~\ref{sec:pt}) all three agree with the realisations to 
$0.002$.

On models held out of the fit the surrogate reproduces the median spectrum to $0.048$ in $\log_{10}\hcsq$, the width to four per cent and 
the skew to $0.05$, which is small compared to the width of the measured free spectrum in every bin.  
The error on the held-out models stops decreasing once about half of the library is used, and splitting the realisations of each model into 
halves changes the median, the width, and the skew two to eight times less than these errors. Its errors shift $\ln L$ of a held-out 
model by $0.20$ through the median, $0.05$ through the width, and $0.03$ through the skew, in the median over models, and by $-1.2$ to 
$+1.2$ in total for the middle two thirds of the models. While this sounds like a dramatic bias on the following analyses, we remark that a 
contour is drawn relative to the maximum of its own scan, and the gain $G$ is a difference of two maxima. A shift of $\ln L$ common to all 
points of a scan therefore cancels in both, and only the part of the error that varies across a scan can move a contour or a preference. 
We tested this directly: with the background taken from \holodeck{} instead of the surrogate, the profile likelihood of the phase 
transition changes by $0.003$ inside the $1\sigma$ region.
The shape of the distribution therefore mostly matters in those bins where the measured posterior changes across the spread 
of the realisations. 

\subsection{Validation against the NANOGrav posterior}
\label{sec:validation}

\begin{figure}[tbp]
\centering
\vspace*{-1cm}
\includegraphics[width=0.9\textwidth]{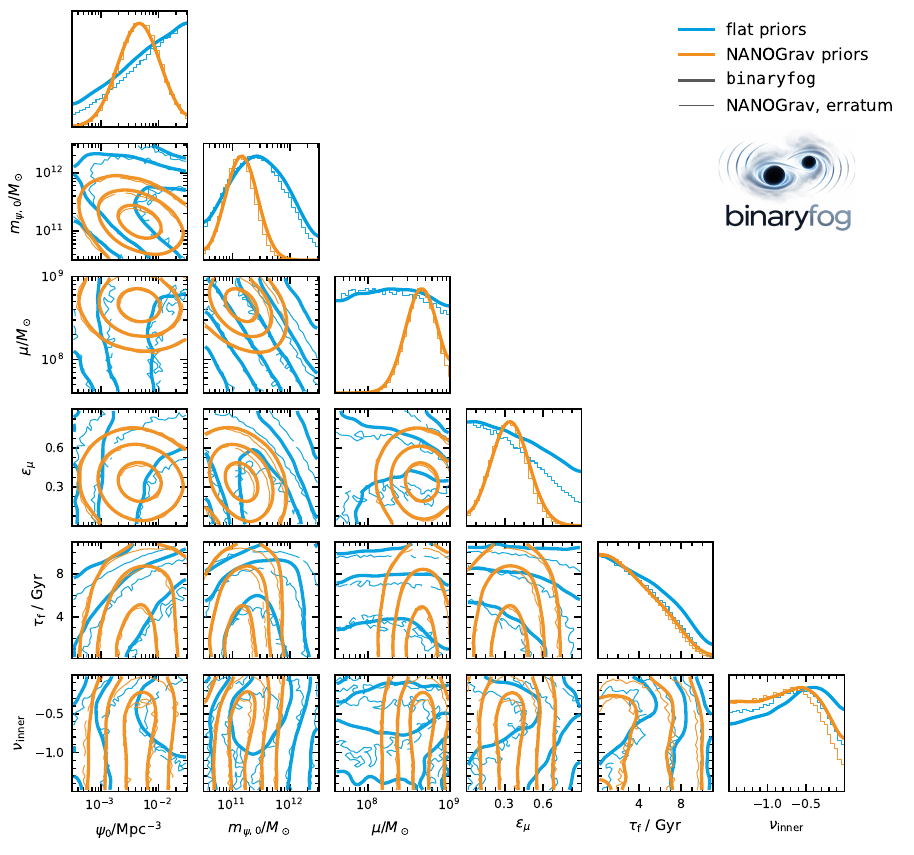}
\caption{Posteriors for the six astrophysical parameters obtained with \binaryfog (thick) against those extracted from
the vector graphics of figure~9 of refs.~\cite{NANOGrav:2023hfp,NANOGrav:2026err} (thin),
for flat
priors over the ranges in table~\ref{tab:priors} (blue) and the NANOGrav priors (orange). Our fits in this figure and the 
published ones both use the 2026 HD+MP+DP+CURN 
free spectrum. The published contours derive from an unsmoothed histogram of 34 bins; pieces 
spanning fewer than three bins are 
not drawn here. Contours enclose
the $1$, $2$, and $3\sigma$ credible regions; the diagonal shows the one-dimensional marginalised posteriors.
}
\label{fig:validation}
\end{figure}

Figure~9 in refs.~\cite{NANOGrav:2023hfp,NANOGrav:2026err} 
shows NANOGrav's posterior of the six 
parameters $\theta$ from the five lowest bins of the HD+MP+DP+CURN spectrum, under flat and under NANOGrav priors, with Gaussian 
processes for the median and the standard deviation of $\log_{10}\hcsq$ in each bin. We repeat this fit with our surrogate and the warp of 
section~\ref{sec:realisations}, on the 2026 re-run of the same spectrum~\cite{NG15KDE}. The two pipelines share no code except 
\holodeck{}. We sample with \texttt{emcee}~\cite{Foreman-Mackey:2012ig}, with $64$ walkers for $30000$ steps, discarding the first $5000$ 
and thinning by ten, which leaves $160000$ samples per prior set, more than a hundred autocorrelation times.

Figure~\ref{fig:validation} compares our posteriors with the published contours, extracted from the vector graphics of the published figure. 
We observe that under the NANOGrav priors the two sets of contours lie on top of each other in every panel. 
For flat priors the published contours are noisier, since they are drawn from an unsmoothed histogram, and we identify 
visible differences in particular in the row of the inner hardening index $\nuin$, where our posterior prefers shallower values. 
To quantify these differences, for every point of each  
published contour we take the distance to the closest point of ours, with each axis scaled to the flat prior range, and the 
median over points and panels. Under the NANOGrav priors this distance has a median of 
$1.3\%$ at $1\sigma$ ($1.6\%$ at $2\sigma$), about the 
width of the lines, and every median agrees with the published one to within $0.17$ posterior standard deviations. Under flat priors it is 
$6.4\%$ and $6.7\%$, mostly because the published contours are derived from an unsmoothed histogram. Drawn the same way, ours give 
$2.2\%$ and $3.2\%$, against $2.0\%$ and $1.8\%$ between two halves of our own chain. 
The excess over this noise comes from the five panels involving $\nuin$, which differ by $3.1\%$ at $1\sigma$ while the other ten differ by $2.0\%$.
The remaining difference we associate dominantly with a shift of the inner 
hardening index by a quarter of a posterior standard deviation, with a median of $-0.66$ against $-0.75$.\footnote{%
The shift survives additional tests like 
removing the truncation of the priors at the ranges of the flat priors, refitting the surrogate on other training splits, setting the skew to 
zero and changing the random seed. An emulator built in the same way as NANOGrav's, with Gaussian processes trained with the code of 
\holodeck{}~\cite{holodeck} on our library and a Gaussian likelihood, moves our median a third of the way, to $-0.67$ and $-0.69$ in two 
chains. The inner hardening index is set almost entirely by the lowest bin, where one posterior standard deviation in $\nuin$ moves the median of 
$\log_{10}\hcsq$ by $0.54\,$dex, so a difference of $0.05\,$dex between two emulators there, the size of our surrogate's error, shifts it by a 
tenth of a standard deviation. The remaining shift may come from their library or from details of their Gaussian-process training, which are 
not public.
} 
Despite these slight discrepancies in the $\nuin$ posterior distribution, we consider the surrogate validated.

\section{Impact of the astrophysical background on new-physics searches}
\label{sec:results}

In this section, we turn to the main focus of this work, namely the impact of the SMBHB modelling on new-physics searches
with PTAs. We stress that using the surrogate developed within \binaryfog{} is paramount for this purpose: 
the surrogate evaluates the distribution of the binary background in $0.18\,$milliseconds on one core, against the $180\,$seconds 
\holodeck{} needs for the same prediction. A profile likelihood over the six astrophysical parameters and the parameters of a new-physics 
model needs of order $10^6$ spectra per scan, i.e.~a few minutes with the surrogate and about six CPU years with \holodeck{}. 
In other words, the surrogate profiles over $\theta$ in the time \holodeck{} needs for the realisations of a single $\theta$. 
We start by identifying which parameter regions of a dark sector phase transition are preferred by the data in section~\ref{sec:pt},
by treating the astrophysical parameters as nuisance parameters, and then perform the same exercise for cosmic superstrings, 
inflationary gravitational waves and domain walls in section~\ref{sec:examples}. We compare this to results obtained based on
other versions of the free spectrum in section~\ref{sec:datasets}, and to variations in astrophysical priors
in section~\ref{sec:priordep}.

\subsection{Dark sector phase transition}
\label{sec:pt}

A first-order phase transition in a dark sector is among the best-motivated early universe
sources of a nanohertz background: a hidden sector that breaks a symmetry at temperatures of MeV
to GeV puts the peak of its spectrum in the pulsar timing band, and such transitions have been
fitted to the PTA signal since the first hints of evidence for a GW background were 
announced~\cite{NANOGrav:2023hvm,EPTA:2023xxk,Bringmann:2023opz,Madge:2023dxc,Winkler:2024olr,Bringmann:2026xcx}.
We parametrise the phase transition by its strength $\alpha$, its inverse duration $\beta/H$, and the reheating temperature
$T_\mathrm{reh}$, with the sound-wave spectrum of ref.~\cite{Bringmann:2026xcx}. Appendix~\ref{app:models}
describes the further modelling choices and the spectral template that we used.

\begin{figure}[tbp]
\centering
\includegraphics[width=\textwidth]{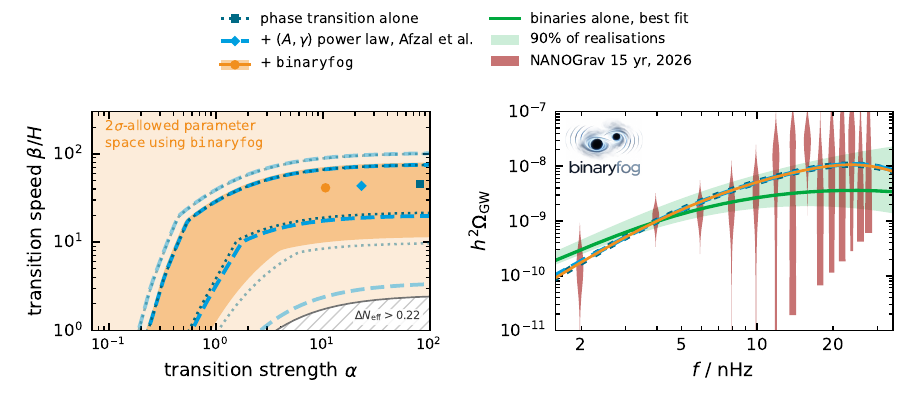}
\caption{{\it Left panel:} The allowed region in the plane of phase transition strength $\alpha$ and inverse duration $\beta/H$, with 
$T_\mathrm{reh}$ being profiled over. With the binaries treated as a population with realisation scatter 
under the 
NANOGrav priors (orange), the preferred regions
are significantly larger than with the commonly adopted $(A, \gamma)$ power law~\cite{NANOGrav:2023hvm} (dashed blue)  
as well as with no binaries included in the fit (dotted blue); 
in all cases, darker (lighter) colours refer to $1\sigma$ ($2\sigma$) confidence regions. The symbols mark the best fit of each treatment,
whose spectrum is shown on the right; the maxima lie on a ridge along which $\ln L$ changes by less than $10^{-4}$.
The \binaryfog{} $2\sigma$ region covers the whole panel outside the hatched region, in which the gravitational waves of the phase transition 
alone add more than $0.22$ to $N_\mathrm{eff}$~\cite{Yeh:2022heq}. {\it Right panel:} The best-fit spectra for a phase transition alone, a phase transition combined with the $(A,\gamma)$ power-law background, and a phase transition combined with the \binaryfog{} population model, compared with the NANOGrav 15\,yr CURN spectrum. Also shown is the best fit of the binaries alone (green), with the band containing 90\% of its realisations.
}
\label{fig:pt}
\end{figure}

The left panel of figure~\ref{fig:pt} shows the profile likelihood in the plane of $\alpha$ and $\beta/H$, with $T_\mathrm{reh}$ profiled over 
and evaluated without binaries, with the $(A,\gamma)$ power law and with \binaryfog{}, respectively. 
Above $\alpha \simeq 10$ the phase transition happens practically in vacuum and its 
spectrum no longer depends on $\alpha$, so the $1\sigma$ region becomes a horizontal slab. Its range in $\beta/H$ is $22$ to $76$ without 
binaries, $20$ to $76$ with the power law and $12$ to $81$ with \binaryfog{}, the widening mostly happening as a result of 
the binary lifetime and the inner hardening index, which bend the binary spectrum at low frequencies. 
There is a similar broadening of the $1\sigma$ band at smaller values of $\alpha$.
Slow phase transitions in that region are therefore far less constrained than a power-law treatment suggests
-- which translates to less severe tuning requirements for a weakly coupled dark sector to produce such transitions~\cite{Bringmann:2026xcx}. 
The $2\sigma$ region widens even further: with \binaryfog{}, it covers the whole plane outside the hatched region, which is 
excluded from cosmology because the gravitational waves of the phase transition alone would 
add more than $\Delta N_\mathrm{eff} = 0.22$ to the radiation density~\cite{Yeh:2022heq}. The gain $G = 2.9$ is below the $2\sigma$ 
threshold of $3.09$, and at every point the reheating temperature can move the signal out of the PTA frequency band. 

In the right panel of figure~\ref{fig:pt}, we show the three best-fit spectra. They are almost 
indistinguishable, since the phase transition dominates the signal in all cases. At the 
\binaryfog{} best fit the four Gaussian parameters sit at the centres of the NANOGrav priors and $\tauf$ at the lower end of its range 
(see section~\ref{sec:priordep}). The maximum likelihood is reached along a ridge from $(\alpha, \beta/H) = (55, 46)$ with 
$T_\mathrm{reh} = 21\,$MeV to weaker and 
slower phase transitions at higher reheating temperatures, along which $\ln L$ changes by less than $10^{-4}$. The fit of the binaries alone 
(green), on the other hand, pulls the galaxy stellar mass function away from the centres of the NANOGrav priors and misses the maximum of 
the posterior in the first bin.

For slow phase transitions, with $\beta/H\lesssim10$, the peak of the induced GW spectrum moves to about $100\nHz$ (at 
$T_\mathrm{reh} \simeq 1\,$GeV). 
The binaries then contribute mostly to the low-frequency part of the observed spectrum, in particular the lowest bin at $2\nHz$, and the 
steeper spectrum of the phase transition fits the upper bins. The binaries alone cannot fit the apparent turnover at $3$ to 
$6\nHz$ (green curve, see also 
ref.~\cite{Agazie:2026tui}), which is why a steeper additional contribution is preferred. This holds in particular for the 
simple power-law description, but to a somewhat lesser degree for the bent spectra expected in population models. Both fits place the binary 
lifetime $\tauf$ at the lower end of its range, $0.1\,$Gyr.
In section~\ref{sec:priordep} we discuss how the preference for a phase transition depends on the astrophysical priors
and population models.

Both the favoured region and the preference depend on the spectral template that is used for the phase transition. 
The best fits lie along a ridge that reaches $\alpha \gg 1$, 
while the sound-wave template of appendix~\ref{app:models} rests on simulations of phase transitions with $\alpha$ up to about 
$0.1$~\cite{Caprini:2024hue}. If the signal is to be interpreted in terms of a strong phase transition, the relevant template is instead 
the one of 
ref.~\cite{Lewicki:2025hxg}, from simulations of strongly supercooled phase transitions that include the expansion of the universe during the 
phase transition. For dissipative shells and $\alpha \to \infty$ it no longer depends on $\alpha$, so the question of how strongly the data 
prefer a {\it strong} phase transition has a quantitative answer that does not rest on a template outside its range. Scanning again
$\beta/H$ with $T_\mathrm{reh}$ profiled over for this template, 
the preference for a strong phase transition is $1.9\sigma$ under the 
NANOGrav priors and $3.2\sigma$ with the $(A,\gamma)$ power law. Its $1\sigma$ interval spans the region 
from $\beta/H = 5.8$ to $14.9$ and from 
$6.2$ to $14.3$, respectively. 

\subsection{Cosmic superstrings, inflationary gravitational waves and domain walls}
\label{sec:examples}

We repeat the study of phase transitions for the three further cosmological sources 
that we consider in this work (see appendix~\ref{app:models}). We chose them as representatives of the many 
new-physics models that have been discussed in the literature in this context, 
because their spectra feature three very distinct shapes. The spectrum of cosmic superstring loops is close to a power law 
across the entire frequency band. 
Inflationary gravitational waves follow a power law of free blue tilt, $\Omega_\mathrm{GW} \propto f^{n_\mathrm{t}}$, which 
turns over to $f^{n_\mathrm{t}-2}$ above the frequency of the mode that re-entered the horizon at reheating. Domain walls annihilating into a 
dark sector give a peaked spectrum, rising as $f^3$ and falling as $f^{-b}$ with $b \le 1$ above a peak that can lie inside the band. For the 
strings we scan the tension $G\mu$ and profile over the intercommutation probability $P$. For inflation we scan the tensor index 
$n_\mathrm{t}$ and profile over the tensor-to-scalar ratio $r$ and the reheating temperature $T_\mathrm{rh}$, since $r$ and $n_\mathrm{t}$ 
are strongly degenerate in the band (cf.~figure~5 of ref.~\cite{NANOGrav:2023hvm}). For the domain walls we scan the annihilation 
temperature $T_*$ and profile over $b$, the sharpness $c$ of the peak and $\Delta N_\mathrm{eff}$, which measures the energy in the 
walls. The six astrophysical parameters are profiled over as before.

\begin{figure}[t]
\centering
\vspace*{-1cm}
\includegraphics[width=\textwidth]{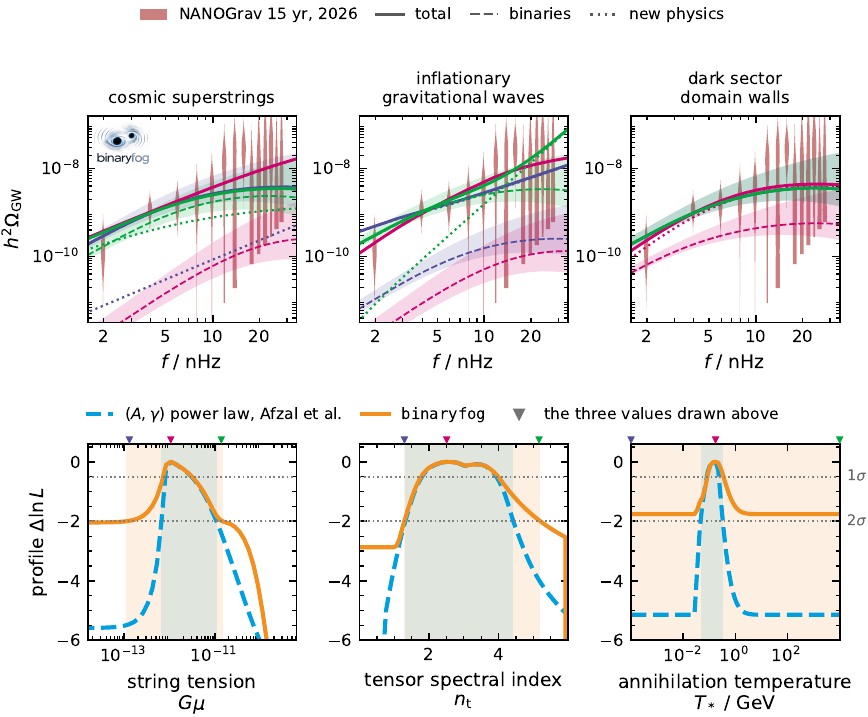}
\caption{Comparison of the impact of a population-based astrophysical background instead of a power law on the preferred parameter 
regions for three distinct new-physics models. The solid curves in the top panels indicate the total signal for a given model parameter 
as marked by a triangle in the lower panels; their dashed and dotted counterparts indicate the astrophysical and the new-physics contributions, 
with the shaded bands holding 90\% of the realisations the respective population would produce. Magenta curves belong to the best fit, 
purple (green) curves to the lower (upper) end of the $2\sigma$-favoured region in parameter space;
these are the ends and the best fit of the \binaryfog{} interval. In the lower panels, we plot the profile likelihood ratio, with the 
dotted horizontal lines indicating the $1\sigma$ and $2\sigma$ thresholds. The shaded bands represent the $2\sigma$ intervals 
resulting from the NANOGrav 15\,yr common uncorrelated red noise spectrum, when using the power-law (light blue) or
\binaryfog{} prescription (orange) for the SMBHB contribution, respectively. 
}
\label{fig:examples}
\end{figure}

Figure~\ref{fig:examples} shows the gravitational wave signals predicted by the three models. Since the cosmic string spectrum 
stays close to a power law, the $(A, \gamma)$ power law imitates it about as well as the full population model. 
The $2\sigma$ interval of $\log_{10} G\mu$ 
nevertheless grows by $75\%$ when profiling with \binaryfog, from $1.2$ to $2.1$ decades, almost entirely at the low-tension end, which 
moves down by a factor of $6$ as the binaries carry more of the signal. This lower end rests on a small margin, since the gain $G = 2.04$ 
lies only marginally above the one-parameter threshold of $2.00$, and most of the widening disappears when $P$ is allowed below 
$10^{-3}$. 
At the best fit point we find an intercommutation probability of $P = 1.3 \times 10^{-3}$, close to the lower end of its prior, 
where the signal is strongest. 
In the figure, we still show results based on this prior, for better comparability with ref.~\cite{NANOGrav:2023hvm},
but note that 
the preferred parameter space opens up when decreasing the lower prior boundary of $P$, allowing arbitrarily small
values of $G\mu$ at $2\sigma$.

For inflationary gravitational waves, the tensor index is bounded at $2\sigma$ between $n_\mathrm{t} = 1.32$ and $4.42$ with the power 
law and between $1.29$ and $5.20$ with \binaryfog{}, a widening of the interval by $26\%$ at the steep end. 
Such a highly blue-tilted inflationary spectrum contributes only in the upper bins and 
leaves the bend at low frequencies to the binaries. At the best fit, $n_\mathrm{t} = 2.5$ and $T_\mathrm{rh} = 0.34\,$GeV, and the binary 
contribution is negligible. Within the $2\sigma$ region 
the tensor-to-scalar ratio falls as 
$n_\mathrm{t}$ grows, from $r \approx 10^{-4}$ at the lower end to $r \approx 10^{-35}$ at its upper end.

The annihilation temperature $T_*$ of the domain walls is affected most among the new-physics scenarios we consider here: 
with the power law it is pinned to a window a tenth of its prior range wide, 
while with \binaryfog{} it remains unconstrained at $2\sigma$. 
The domain wall spectrum turns over inside the frequency band in a way similar to a population with efficient 
environmental hardening. Since its amplitude prior is limited by $\Delta N_\mathrm{eff} \le 0.39$ the 
two contributions thus become interchangeable beyond the $1\sigma$ band. At the best fit point, 
$\Delta N_\mathrm{eff} = 0.39$ and $b = 0.50$ sit 
at the ends of their prior ranges, where 
the signal is largest and falls most slowly above the peak, with $c = 2.47$. 
Lowering the upper bound from $\Delta N_\mathrm{eff} \le 0.39$ (as used by NANOGrav) to $\Delta N_\mathrm{eff} \le 0.22$
(the same bound as we apply to the 
phase transition), domain walls are no longer preferred; concretely, we find $G = 1.6$ for the power law and $0.6$ for \binaryfog{}
in that case. 

In comparison, the preference with the power law is $2.9\sigma$ for cosmic strings, $2.8\sigma$ for inflation, and 
$2.1\sigma$ for domain walls, which drops to $1.5\sigma$, $1.5\sigma$, and $0.7\sigma$ with \binaryfog{}. 
We observe that the preference drops by a similar amount for every model -- which is directly related to the main phenomenological
property of the SMBHB population spectrum stressed several times above, namely the flexibility of bending the spectrum
compared to a simple power law.
Phase transition and inflationary contributions, in particular, generally reach the same gain 
since they both fit an excess (on top of the binaries) in the upper bins.

Let us conclude by mentioning that, for the $(A,\gamma)$ power law, our $1\sigma$ intervals for the three models 
shown in figure~\ref{fig:examples} are very similar to NANOGrav's $1\sigma$ credible 
intervals~\cite{NANOGrav:2023hvm}: for cosmic superstrings we find $\log_{10} G\mu$ from $-12.09$ to 
$-11.52$ against their $-12.09$ to $-11.49$, for inflation, with $r$ as the scanned parameter as in ref.~\cite{NANOGrav:2023hvm}, 
$\log_{10} r$ from $-23.9$ to $-8.2$ against $-23.0$ to $-7.2$, and for domain walls $\log_{10}(T_*/\mathrm{GeV})$ from $-1.09$ to $-0.64$ 
against $-1.27$ to $-0.67$. Differences of this size are expected between a profile likelihood analysis 
and Bayesian credible intervals, and between 
our spectrum and the 2023 Hellings--Downs spectrum they used, which predates the correction of ref.~\cite{NANOGrav:2026err}.

\subsection{Comparison of different NANOGrav free spectra}
\label{sec:datasets}

\begin{figure}[tbp]
\centering
\includegraphics[width=\textwidth]{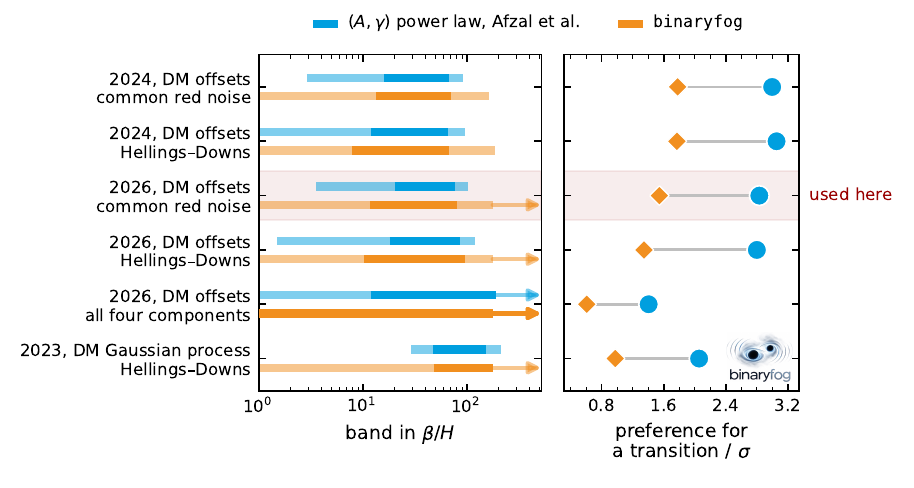}
\caption{Six versions of the NANOGrav 15\,yr free spectrum, each fitted with the binaries as the
$(A, \gamma)$ power law (blue) and as the \binaryfog{} background (orange). {\it Left panel:} The band in $\beta/H$ for a
strong phase transition, in the limit $\alpha \to \infty$, solid for $1\sigma$ and pale for $2\sigma$. An arrow marks a band that continues past
the end of the scanned range. {\it Right panel:} The preference for a phase transition, i.e.~the gain $G$ of
eq.~\eqref{eq:gain} converted as described in section~\ref{sec:profile}. 
The shaded row is the default spectrum used in the rest of this paper.
}
\label{fig:datasets}
\end{figure}

The results presented above are all based on the same 
free spectrum, CURN of the 2026 release with deterministic dispersion measure offsets. 
Here we repeat the phase transition fit with five additional versions, and summarise the results in figure~\ref{fig:datasets}: 
the CURN and HD spectra of the 2024 \ptarcade{} 
record~\cite{PTArcadeData} and of the 2026 release~\cite{NG15KDE}, all with deterministic dispersion measure offsets, the 2026 
HD+MP+DP+CURN spectrum, the most conservative of the available noise models, and an HD spectrum from NANOGrav's 2023 
record~\cite{NG15KDEv1} in which the dispersion measure variations are modelled as a Gaussian process to capture the many effects that 
may delay the radio pulses of a given pulsar. 
Overall, the release date and the correlation pattern are found to matter less than the noise model and the set of components. 

The choice between HD and CURN only marginally affects the preference and the band in $\beta/H$ in either release, 
given the current state of the data. In every version the preference with \binaryfog{} is about $1\sigma$ lower than with the 
power law, and for HD+MP+DP+CURN only an extremely mild preference of $0.6\sigma$ remains. 
With the Gaussian-process model of the dispersion measure variations the phase 
transition has to be weaker, at larger $\beta/H$, and is less preferred.\footnote{%
The Gaussian-process product is the 2023 one, since the published record carries no re-run of it. A re-run after the correction of the sampler, 
which W.~Lamb kindly provided ahead of its release, agrees with it to within $0.09$ in $\log_{10}\rho$ in every bin and gives 
$2.2\sigma$ with the $(A, \gamma)$ power law and $1.0\sigma$ with \binaryfog{}, against $2.1\sigma$ and $1.0\sigma$, with the band 
under \binaryfog{} from $\beta/H = 71$ to $215$.
} 
With \binaryfog{} the $2\sigma$ band in $\beta/H$ is comparable across the 2026 
products. The preference therefore depends on the modelling of the dispersion measure variations as strongly as on the binary background.

Let us mention that the 2023 posteriors carry a numerical artefact for amplitudes 
far below the measured signal: their density grids run down to 
$\log_{10}(\rho_i/\mathrm{s}) = -15.1$, but the Markov chains never visited values below about $-9$, 
resulting in a numerical floor with isolated spikes up to $31$ in $\ln L$ above it. 
The tails of the distribution thus cannot be used far from the measured values, as already pointed out in 
refs.~\cite{Bringmann:2023opz,Bringmann:2026xcx}.
The 2026 posteriors, in contrast, only keep the range that is actually populated by the chains. 

All the posteriors explored here 
were obtained with NANOGrav's log-uniform priors on the amplitude and the slope of the intrinsic red noise of each 
pulsar. Dai et al.~\cite{Dai:2026jjw} showed that, since the auto-correlations of a pulsar only measure the sum of the background and its 
intrinsic red noise, the prior volume
grows with the background amplitude and pulls the inferred amplitude up: at a fixed slope 
of $13/3$ by $0.06$ to $0.2$ in $\log_{10} A$ compared with other choices of the noise priors, and with a free slope by $0.3$ in 
$\log_{10} A$. With those other pulsar-intrinsic red noise priors the Bayes factors for new-physics sources against binaries fall to order unity. 
Their free spectra are not public. To estimate the size of the effect on our results, we lowered $\log_{10}\rho_i$ of the common uncorrelated 
red noise spectrum by either $0.1$ (or $0.2$) in all bins. The preference for a phase transition then falls from $2.8\sigma$ to $2.7\sigma$ (or 
$2.6\sigma$) with the $(A,\gamma)$ power law, and from $1.5\sigma$ to $1.4\sigma$ (or $1.2\sigma$) under the NANOGrav priors. Under 
the Tomczak--Kormendy priors it drops from $3.3\sigma$ to $2.7\sigma$ (or $2.2\sigma$), and from $1.7\sigma$ to $1.3\sigma$ (or 
$1.0\sigma$) under the survey-average priors (section~\ref{sec:priordep}). The best fit with a phase transition is unchanged by the shift, since the 
transition takes up the difference, while the fit of the binaries alone improves, and with it the tension between the binaries and the 
astrophysical priors. The steeper spectrum that the alternative noise priors also give is not included in this estimate; it would further 
favour the binaries.

\subsection{Dependence on the astrophysical priors}
\label{sec:priordep}

The preferences for a phase transition stated in sections~\ref{sec:pt} and~\ref{sec:examples} depend strongly on the astrophysical priors. With flat 
priors as the most extreme example, $99\%$ of the scanned plane of $\alpha$ and $\beta/H$ 
would lie inside the $1\sigma$ region and the preference for a phase 
transition drop to $0.2\sigma$, since the binaries can then be tuned to reproduce the measured spectrum. 
For the three Gaussian prior sets 
that do take into account the information contained in galaxy surveys, 
table~\ref{tab:priordep} summarises the preference for all four new-physics models, with the $(A,\gamma)$ power law and with 
\binaryfog{}, on the three free spectra of section~\ref{sec:data}. With CURN and HD free spectra the 
Tomczak--Kormendy priors raise the preference for new physics  
slightly above the respective power-law values and well above the respective values under the NANOGrav 
priors, while the $2\sigma$ intervals of the model parameters change only slightly. 
Under our recommended survey-average priors, the preferences drop again, to values similar to those found
when adopting the NANOGrav priors.
Within an HD+MP+DP+CURN fit the preferences drop to between $0.2\sigma$ 
and $1.6\sigma$, because the additional degrees of freedom in the overlap reduction function widen the uncertainty of the 
measured spectrum sufficiently for even a faint binary background to fit the data.

\begin{table}[tbp]
\centering
\footnotesize
\setlength{\tabcolsep}{3pt}
\begin{tabular}{llcccc}
\toprule
 & & $(A,\gamma)$ & \multicolumn{3}{c}{\binaryfog{}} \\
\cmidrule(l){4-6}
model & free spectrum & power law & NANOGrav & Tomczak--Kormendy & \begin{tabular}[b]{@{}c@{}}survey\\ average\end{tabular} \\
\midrule
phase transition ($k = 3$) & CURN & $2.8\sigma$ & $1.5\sigma$ & $3.3\sigma$ & $1.7\sigma$ \\
 & HD & $2.8\sigma$ & $1.3\sigma$ & $2.9\sigma$ & $1.5\sigma$ \\
 & HD+MP+DP+CURN & $1.4\sigma$ & $0.6\sigma$ & $1.2\sigma$ & $0.4\sigma$ \\
\midrule
cosmic superstrings ($k = 2$) & CURN & $2.9\sigma$ & $1.5\sigma$ & $3.4\sigma$ & $1.8\sigma$ \\
 & HD & $2.9\sigma$ & $1.3\sigma$ & $3.0\sigma$ & $1.5\sigma$ \\
 & HD+MP+DP+CURN & $1.6\sigma$ & $0.8\sigma$ & $1.4\sigma$ & $0.5\sigma$ \\
\midrule
inflation ($k = 3$) & CURN & $2.8\sigma$ & $1.5\sigma$ & $3.3\sigma$ & $1.7\sigma$ \\
 & HD & $2.8\sigma$ & $1.4\sigma$ & $2.9\sigma$ & $1.5\sigma$ \\
 & HD+MP+DP+CURN & $1.4\sigma$ & $0.6\sigma$ & $1.2\sigma$ & $0.4\sigma$ \\
\midrule
domain walls ($k = 4$) & CURN & $2.1\sigma$ & $0.7\sigma$ & $2.6\sigma$ & $1.1\sigma$ \\
 & HD & $2.3\sigma$ & $0.7\sigma$ & $2.4\sigma$ & $1.0\sigma$ \\
 & HD+MP+DP+CURN & $1.1\sigma$ & $0.4\sigma$ & $0.9\sigma$ & $0.2\sigma$ \\
\bottomrule
\end{tabular}
\caption{Preference for each new-physics model over the binaries alone, expressed as one-sided significance from the gain $G$ in eq.~\eqref{eq:gain}; 
$k$ is the number of free parameters of the model. The binaries are modelled by the $(A, \gamma)$ power law under its Gaussian prior, 
or by \binaryfog{} under the NANOGrav, the Tomczak--Kormendy, or the survey-average priors of table~\ref{tab:priors}. The three free spectra are 
those of the 2026 NANOGrav\,15\,yr release with deterministic dispersion measure offsets: the common uncorrelated red noise (CURN) used 
as reference choice throughout this paper, the Hellings--Downs (HD) spectrum, and the Hellings--Downs spectrum fitted together with a 
monopole (MP), a dipole (DP), and CURN. }
\label{tab:priordep}
\end{table}

The preferences stated in table~\ref{tab:priordep} are hence a result of two effects: the prior penalty in fits with only the binaries, and the improved
goodness-of-fit when considering both contributions. Under the NANOGrav priors the fit \textit{without} a phase transition signal puts the binary lifetime at 
$0.1\,$Gyr, the fastest hardening of NANOGrav's library~\cite{NANOGrav:2023hfp}, and moves the normalisation and the characteristic 
mass of the galaxy stellar mass function by $1.3$ and $1.0$ standard deviations from their prior centres. Of the gain $G = 2.9$ in the
\textit{combined} fit, a gain of $1.6$ results from the improved goodness-of-fit and the remaining $1.3$ come from the prior penalty, against $3.5$ and $3.0$ when using the $(A, \gamma)$ power law.
Under the Tomczak--Kormendy priors the fit of the binaries alone moves $m_{\psi,0}$ by $1.9$, $\psi_0$ by $1.2$ and $\mu$ by $0.8$ 
standard deviations, its $\ln L$ falls from $-1.38$ to $-6.51$, and the gain rises to $G = 8.0$ ($3.3\sigma$), of which $5.0$ come 
from the data and $3.0$ from the prior penalty: these priors predict too faint a background, which any additional power, 
astrophysical or cosmological, would relieve. The survey-average priors predict a louder background (see figure~\ref{fig:priorsets}), 
which the binaries alone reach close to the prior centres, so that the penalty disappears despite the narrow widths and all 
preferences stay below $2\sigma$.

\begin{figure}[tbp]
    \centering
\includegraphics[width=\linewidth]{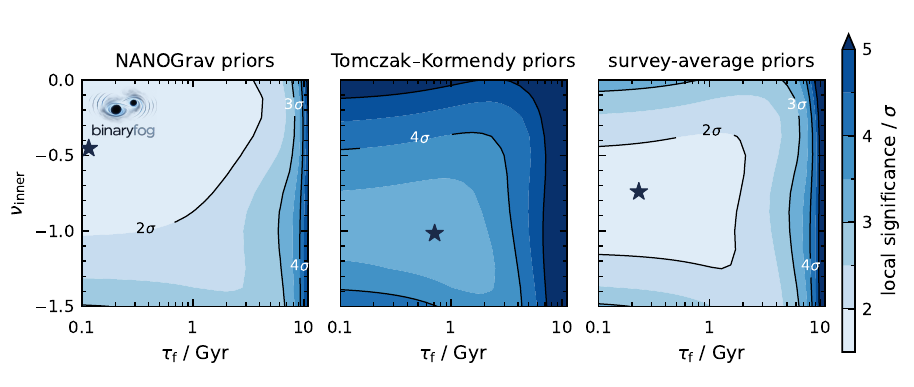}
    \caption{Preference for a dark sector phase transition over the binaries alone on the CURN spectrum, at fixed 
    binary lifetime $\tauf$ and inner hardening index $\nuin$, with the other four astrophysical parameters profiled under 
    the NANOGrav (left), Tomczak--Kormendy (middle) and survey-average (right) priors of table~\ref{tab:priors}. 
    The colour gives the local one-sided significance from the gain $G$ of eq.~\eqref{eq:gain} in steps of $0.5\sigma$, 
    and the black lines mark $2$, $3$, $4$ and $5\sigma$. The stars mark the best fit of the binaries alone with all six 
    parameters free, where the preference is $1.5\sigma$, $3.3\sigma$ and $1.7\sigma$, respectively.}
    \label{fig:hardening}
\end{figure}

In all of the considered prior sets of table~\ref{tab:priors}, the hardening parameters $\tauf$ and $\nuin$ are modelled by
uniform distributions. Figure~\ref{fig:hardening} shows the
preference for a 
phase transition with $\tauf$ and $\nuin$ fixed and the other four astrophysical parameters profiled over.
In each of the panels, representing the three different prior choices, %
the preference is dominated by how well the binaries alone can fit the data. The data prefer fast hardening
in order to match the required signal amplitude,
$\tauf = 0.1$, $0.7$, and $0.2\,$Gyr 
at the best fit points (cf.~the posterior of $\tauf$ in figure~\ref{fig:validation}), and moderate turnovers of the GW spectrum due to the hardening process
having a mild slope $\nuin = -0.45$, $-1.0$, and $-0.75$ %
(see also figure~\ref{fig:anatomy}). At the centres of the
NANOGrav priors the background at 
$\tauf = 11\,$Gyr is more than an order of magnitude fainter than at $1\,$Gyr. Hence, while a lower bound on $\tauf$ of $1\,$Gyr leaves the 
preference almost unchanged ($1.6\sigma$, $3.3\sigma$ and $1.8\sigma$), raising the lower bound to $3$ and $6\,$Gyr leads to an increased preference of $1.8\sigma$ and 
$2.3\sigma$, respectively, under the NANOGrav priors. Under our recommended survey-average priors, the preference for a phase transition exceeds the $2\sigma$ and $3\sigma$ thresholds for hardening timescales larger than $2.0\,$Gyr and $6.5\,$Gyr, respectively.

The above discussion illustrates that the preference for a new-physics contribution
to the PTA data depends as much on where the astrophysical priors are centred as on their widths;
a more robust future determination of such a signal component will thus require a significant gain in 
{\it both} accuracy and precision from galaxy surveys and merger simulations.
The other conclusion is that prior boundary effects on the hardening parameters 
have a particularly significant impact on the preference for new physics. This motivates further
studies as to how SMBHBs overcome the final parsec~\cite{Tiruvaskar:2025lkq, Alonso-Alvarez:2024gdz},
and in particular how small timescales $\tauf$ can realistically be achieved.

\subsection{Discussion of assumptions and caveats}
\label{sec:limits}

One of the basic assumptions in the population model used here is that the binaries' orbits are circular. 
Eccentricity shifts power to higher harmonics and hardens binaries faster, 
which suppresses the signal at the bottom of the band, as environmental hardening does, and in some models enhances it nearer the 
top~\cite{Kelley:2017lek}. Raidal et al.~\cite{Raidal:2024odr} find the two effects to be
near-degenerate in the spectrum, although the degeneracy is 
broken by the spectral fluctuations and the correlations between frequency bins, which our free-spectrum likelihood drops. They obtain a 
mean eccentricity of $0.83 \pm 0.05$ at $2\nHz$ for evolution by gravitational wave emission alone, while with environmental effects circular 
binaries remain only mildly disfavoured. The eccentricity distribution is degenerate with the binary mass function rather than with the stellar 
density of the hosts, since the turnover frequency depends only weakly on the latter~\cite{Chen:2016zyo}. Our library already spans 
turnovers from below to above the PTA frequency band, but strongly eccentric binaries with fast hardening and large merger rates could give flat and loud 
spectra outside it. Including them would need a change to \holodeck{}.

Another choice is that we treat the background as stochastic. We thus 
do not separate the loudest binaries as resolvable sources, which a joint search can 
do~\cite{Goncharov:2026cdh} and which biases a spectral fit otherwise~\cite{Becsy:2023qul}. The same loud binaries make the background 
anisotropic~\cite{Gardiner:2023zzr,Konstandin:2024fyo}. The realisation statistics themselves cannot be measured by a pulsar timing 
array~\cite{Cecchini:2026cru}, so they have to come from a population model. Given the lack of evidence for continuous wave 
signals~\cite{NANOGrav:2023pdq} and only modest chances of observing other signs of discrete sources at the current quality of PTA 
data~\cite{Konstandin:2024fyo}, we hence do not expect any relevant biases from treating the GW background as stochastic.

Like \holodeck{}, we moreover assume that every binary emits at an 
exact multiple of the inverse observation time~\cite{Lamb:2025niq}, neglect the correlations of the realisations between bins, and use one 
parametrisation of the population. Others exist, for instance a double Schechter mass function with an Illustris merger 
rate~\cite{Leja:2019lfo,Rodriguez-Gomez:2015aua}, used with an evolving black hole mass relation in ref.~\cite{Matt:2025bao}.
While the amount of additional small-mass galaxies as described by a double Schechter mass function only has a minor effect on the PTA 
signal (cf.~figure~\ref{fig:surveyfits} in the appendix), an evolving black hole mass relation may affect  our results by increasing 
the signal source counts at large redshift, and hence lead to somewhat stronger binary signals. As nHz GWs themselves are 
currently one of the most precise probes of this relation~\cite{Leja:2019lfo, Matt:2025bao}, the latter caveat should rather be understood as a 
source of uncertainty, similar to the unknown details of binary hardening, which requires more detailed studies.

The free-spectrum likelihood treats the frequency bins as independent, in the same way
as NANOGrav's astrophysical analysis does. Normalising-flow 
likelihoods go beyond this~\cite{Laal:2024trp,Wright:2026lwu}, which mostly 
matters for eccentric populations~\cite{Raidal:2024odr}. Taking into account 
these correlations is beyond the scope of this work. In fact, it does not seem required 
at the current quality level of pulsar timing data, since the bins of the free 
spectrum appear to be close to uncorrelated, which allowed the factorised likelihood of \ceffyl{} in the first place~\cite{Lamb:2023jls}. 
Within a given bin, 
eq.~\eqref{eq:marginal} treats the timing residuals as Gaussian for each realisation and averages over the realisations. This is accurate for a 
single pulsar~\cite{Xue:2024qtx}, and for many pulsars it requires the effective number of sources in a bin to be large compared to a third of 
the number of pulsars~\cite{Ali-Haimoud:2026sbk}. For the $67$ pulsars of NANOGrav and the source counts of 
section~\ref{sec:warpchoice} this holds only in the three lowest bins, so the treatment of the higher bins is the least controlled step of our 
likelihood (just as of NANOGrav's).

\newpage
\section{Conclusions}
\label{sec:conclusions}

Supermassive black hole binaries (SMBHBs) are the expected source of the nHz gravitational wave background,
but there are several well-motivated new-physics scenarios that would add, possibly subdominant, cosmological contributions.
In this work we have explored how the modelling of the astrophysical SMBHB background
affects any conclusions about those new-physics contributions.
We also presented a new tool  \binaryfog{} that we developed 
specifically for this purpose, allowing us to compute the astrophysical background 
based on state-of-the-art models with unprecedented speed, without compromising on accuracy.

We stress that runtime was a bottleneck in previous analyses searching for new physics in pulsar timing array data,
which therefore largely left out the SMBHB contribution entirely or roughly approximated it by a simple phenomenological power law.
The actual prediction from population models, however, is a distribution of bent spectra that depends on at least six uncertain 
astrophysical parameters. 
\binaryfog{} returns the median,  width, and skew of the resulting distribution in each frequency bin in $0.2\,$milliseconds, a million 
times faster than \holodeck{}, which previous analyses had to rely on. 
It reproduces held-out \holodeck{} models to $0.048$ in $\log_{10}\hcsq$, its warped Gaussian describes the 
realisations better than the log-normal or map-Airy distributions that have previously been promoted in this context, 
and it fully reproduces NANOGrav's posterior of the population parameters. It is this 
speed-up that makes it possible to profile over the six astrophysical parameters in any new-physics fit.

We find that under the same priors as in the new-physics analysis by NANOGrav~\cite{NANOGrav:2023hvm}, 
the more realistic description of the binary population adopted here
lowers the preference for a dark sector phase transition to $1.5\sigma$, compared to $2.8\sigma$ when assuming a simple 
$(A,\gamma)$ power law. 
The preference for signal contributions from cosmic superstrings, inflationary gravitational waves, or domain walls 
drops in a very similar way. %

We have discussed that the astrophysical parameters which most affect how well the \mbox{SMBHBs} can fit the measured spectrum are the 
binary lifetime and the inner hardening index,  which bends the spectrum at low frequencies. Changing the astrophysical priors,
in particular for these parameters, therefore strongly affects the numbers quoted in the previous paragraph.
The NANOGrav priors are up to four times wider than the fits they are based on, and when instead using
the widths of those fits the preference for a phase transition rises to $3.3\sigma$.
We also presented updated astrophysical priors in this work, based on newer measurements published since 2014 (for a detailed 
discussion, see appendix~\ref{app:surveypriors}). With these priors, which appear to be closer to representing the 
current state of knowledge, the preference for a phase transition drops to a very mild $1.7\sigma$. 
Lowering the spectrum by $0.1$ to $0.2$ in $\log_{10}\rho$ (the square root of the timing-residual power), as the 
intrinsic pulsar noise priors may require~\cite{Dai:2026jjw}, further decreases the preference to $1.0\sigma$. This clearly demonstrates that 
any preference for new physics from pulsar timing arrays must be quoted along with the astrophysical priors (and the free spectrum 
it rests on). 

In assessing these findings, it is important to realise that in a fit of the binaries alone the data favour binaries that 
merge within about $2\,$Gyr after their host galaxies have merged. Including a phase transition in the fit, 
on the other hand, the data hardly 
constrain the binary lifetime or the inner hardening index (section~\ref{sec:priordep}). If binaries turn out to require 
longer than about 
$6.5\,$Gyr to merge, %
the preferred binary 
background becomes fainter and the preference for a phase transition exceeds $3\sigma$ under our
recommended updated prior set. Whether the gravitational wave background is explained by the 
binaries alone or with an additional new-physics contribution thus remains tightly coupled to a resolution of the final 
parsec problem.

In the future, better determinations of the galaxy stellar mass function, of the black hole to bulge mass relation, and of how binaries harden 
will be decisive in settling whether the data describing the nHz gravitational wave background prefer a cosmological 
contribution. We have designed \binaryfog{} to help with this task. 
It takes  as input any new-physics spectrum as a Python function, along with any choice of astrophysical priors, and 
profiles over the binary population in minutes on a laptop, so that global fits of new-physics models to pulsar timing data can incorporate 
a realistic astrophysical background. \binaryfog{} is straightforward to use, cf.~appendix~\ref{sec:tool}, 
and available on GitHub~\cite{binaryfog}.

\bigskip

\acknowledgments We thank William Lamb, Kayhan Gültekin and David Wright for discussions of \holodeck{}, \ptarcade{}, and \ceffyl{},
and Shreyas Tiruvaskar for sharing his version of \holodeck{} with us.
We are grateful to the NANOGrav collaboration for releasing the model files, the free-spectrum data
products, and the corrected versions of both.
CT gratefully acknowledges support from the
Alexander von Humboldt Foundation, through a Feodor Lynen
Research Fellowship, as well as from the Spanish National Grant
PID2025-171343NB-C55 and from Generalitat Valenciana through the grant CIPROM/22/69.
TB is supported through a FRIPRO grant of the Norwegian Research Council
(project ID 353561 `DarkTurns'), and thanks the University of Canterbury for
hospitality and support in the form of a Visiting Erskine Fellowship. CG acknowledges support from the Marsden Fund Council
from Government funding administered by the Royal Society
Te Apārangi (grants MFP-UOA2131 and MFP-UOA2531). The computations for this work
were performed on the local SOM clusters at the Instituto de F\'isica Corpuscular (IFIC). The
authors used Claude Opus 5.5 for assistance with the Python code. We have checked and take full responsibility for the results.

\appendix

\section{Gravitational wave spectra of the new-physics models}
\label{app:models}

We study four different cosmological contributions to the nanohertz signal. This appendix briefly summarises the spectral shape for 
each of them, and how it is determined. 
For the inflationary, domain wall, and cosmic superstring contributions, we use the implementations and prior 
ranges provided in the online repository~\cite{NG15Models} belonging to the NANOGrav 15\,yr search for new-physics 
signals~\cite{NANOGrav:2023hvm}, and for the phase transition interpretation we use the 
\texttt{TransitionListener}~\cite{Matuszak:2026xsz} implementation (see also ref.~\cite{Bringmann:2026xcx}). %
We refer to 
these works for further information on the individual contributions and an in-depth discussion of the underlying model assumptions.

\subsection{Dark sector phase transition}
\label{app:pt}

A first-order phase transition in a dark sector proceeds by
the nucleation of bubbles, which then expand and source fluid motion in the dark
plasma. Here we assume the rather generic case of bubbles which reach a terminal velocity (e.g.~due to the presence of gauge bosons 
obtaining a mass in the transition, resulting in velocity-dependent friction terms~\cite{Gouttenoire:2021kjv}), so sound waves dominate the 
gravitational wave production. We keep that contribution alone and set the
efficiencies of wall collisions and of turbulence to zero. The template is based on simulations~\cite{Caprini:2024hue,Jinno:2022mie}, 
used exactly as in ref.~\cite{Bringmann:2026xcx},
\begin{align}
\begin{split}
  h^2\Omega_\mathrm{GW}^\mathrm{PT}(f)
    &= \mathcal{R}h^2\, A_\mathrm{sw}\, K^2\, \mathcal{Y}_\mathrm{sw}\, RH_*\, \tilde{S}(f) \, , %
  \label{eq:ptspec} \\
  \tilde S(f) &= N \left(\frac{f}{f_2}\right)^{3}
    \left[1 + \left(\frac{f}{f_1}\right)^{2}\right]^{-1}
    \left[1 + \left(\frac{f}{f_2}\right)^{4}\right]^{-1} , \\ %
  N &= \frac{2\sqrt{2}}{\pi}\left[1 + \frac{f_2^2}{f_1^2}
       + \sqrt{2}\,\frac{f_2}{f_1}\right] ,
    \end{split}
\end{align}
a doubly broken power law rising as $f^3$ below $f_1$, as $f$ between the breaks, and falling as
$f^{-3}$ above $f_2$. The two break frequencies are given by
\begin{align}
  f_1 \simeq 0.2\,\frac{a_* H_*}{RH_*} \, , \qquad
  f_2 \simeq 0.5\,\Delta_\mathrm{w}^{-1}\,\frac{a_* H_*}{RH_*} \, , \qquad
  \Delta_\mathrm{w} = \frac{|v_\mathrm{w} - c_\mathrm{s}|}{\max(v_\mathrm{w}, c_\mathrm{s})} \, ,
\end{align}
where $RH_*$ is the mean bubble separation at percolation in units of the Hubble radius and
$a_* H_*$ the Hubble rate at percolation redshifted to today. Three further factors set the amplitude: the
normalisation $A_\mathrm{sw} \simeq 0.11$ measured in lattice simulations, the fraction $K$ of the
released energy that ends up in bulk fluid motion, and the factor $\mathcal{Y}_\mathrm{sw}$ by which a
finite sound wave lifetime suppresses the signal,
\begin{align}
  K = 0.6\,\kappa_\mathrm{sw}(\alpha, v_\mathrm{w})\,\frac{\alpha}{1+\alpha} \, , \qquad
  \mathcal{Y}_\mathrm{sw} = \min\!\left(1, \tau_\mathrm{sh} H_*\right) , \qquad
  \tau_\mathrm{sh} H_* \simeq \frac{RH_*}{\sqrt{3K/4}} \, ,
\end{align}
with the efficiency factor $\kappa_\mathrm{sw}$ taken from ref.~\cite{Espinosa:2010hh} and the factor $0.6$
found in the Higgsless simulations of ref.~\cite{Jinno:2022mie}. The amplitude of the spectrum is redshifted by a factor
\begin{align}
  \mathcal{R}h^2 = \Omega_\gamma h^2
    \left(\frac{h_{0}}{h_{\mathrm{reh}}}\right)^{4/3}
    \frac{g_\mathrm{reh}}{g_\gamma} \, ,
    \end{align}
with $\Omega_\gamma h^2 = 2.473 \times 10^{-5}$, $g_\gamma = 2$, and $h_{0} = 3.93$, where
$h$ counts entropy degrees of freedom and $g$ energy ones. We take
$v_\mathrm{w} = 1$, $c_\mathrm{s} = 1/\sqrt{3}$ and two light dark sector degrees of freedom beyond
the Standard Model bath, so that $g_\mathrm{reh} = g_\mathrm{SM}(T_\mathrm{reh}) + 2$ and likewise
for $h_{\mathrm{reh}}$. The three parameters scanned in section~\ref{sec:pt} are the transition
strength $\alpha$, the inverse duration $\beta/H$, and the reheating temperature $T_\mathrm{reh}$,
which fixes $a_* H_*$ and hence where the spectral breaks appear today. We profile the reheating temperature between $1\,$MeV and
$10\,$GeV%
, where the lower bound is motivated by bounds from big bang nucleosynthesis~\cite{deSalas:2015glj}. The inverse duration enters through 
the bubble separation,
\begin{align}
  \frac{\beta}{H} = \left(\frac{8\pi}{1 - P_\mathrm{f}(T_*)}\right)^{1/3}
    \frac{\max(c_\mathrm{s}, v_\mathrm{w})}{RH_*} \, ,
\end{align}
evaluated at the percolation criterion $1 - P_\mathrm{f} = 0.29$, with $P_\mathrm{f}$ the fraction of space still in the false vacuum. We keep 
the factor $0.29$ due to the argument presented in ref.~\cite{Matuszak:2026xsz}, noting that it is often omitted in the literature, see for 
instance ref.~\cite{Caprini:2024hue}.

\subsection{Dark sector domain walls}
\label{app:dw}

A broken discrete symmetry leaves a
network of domain walls whose energy density grows relative to the surrounding radiation bath until the network annihilates at a temperature 
$T_*$. The bath the network annihilates into is here assumed to be a secluded dark sector, instead of the photon bath. We take the model file
\texttt{dw\_ds} of ref.~\cite{NG15Models} unchanged, where
\begin{align}
\begin{split}
  h^2\Omega_\mathrm{GW}^\mathrm{dw}(f) &= h^2\,\frac{\pi\,\epsilon_\mathrm{gw}}{960}\,\alpha_\mathrm{dw}^2\,
    g_*\!\left(\frac{h_{\mathrm{eq}}}{h_{*}}\right)^{4/3}
    \frac{T_0^4}{M_\mathrm{pl}^2 H_0^2}\; S\!\left(\frac{f}{f_0}\right) , \quad \text{with} \quad
    \alpha_\mathrm{dw} = \frac{\Delta N_\mathrm{eff} h_{*}^{4/3}}{13.6\; g_*\, }
   \, , \\
   S(x) &= \frac{(3+b)^{c}}{\left(b\,x^{-3/c} + 3\,x^{b/c}\right)^{c}}
   \, , \quad
  f_0 = \left(\frac{h_{\mathrm{eq}}}{h_{*}}\right)^{1/3}
    \left(\frac{\pi^2 g_*}{90}\right)^{1/2} \frac{T_0\,T_*}{M_\mathrm{pl}} \, .
\end{split}
\end{align}
Here $\epsilon_\mathrm{gw} = 0.7$ is the efficiency with which the annihilating network radiates,
as measured in the simulations of ref.~\cite{Hiramatsu:2013qaa},
$\alpha_\mathrm{dw}$ the fraction of the energy density stored in the walls at annihilation,
$\Delta N_\mathrm{eff}$ the relativistic degrees of freedom that the decay products contribute, $g_*$ and
$h_{*}$ the energy and entropy degrees of freedom at $T_*$, $h_{\mathrm{eq}}$ the latter at
matter-radiation equality, $T_0$ the present photon temperature, $M_\mathrm{pl}$ the reduced Planck
mass and $H_0$ the Hubble rate today. The spectrum rises as $f^3$ below $f_0$ and falls as $f^{-b}$
above it, with $c$ controlling how sharply it turns over. The four free model parameters are $-4 < \log_{10} (T_* / \text{GeV}) < 4$,
$-3 < \log_{10} \Delta N_\mathrm{eff}< \log_{10} 0.39$, $0.5 < b < 1$ and $0.3 < c<3$, in the parametrisation of ref.~\cite{Ferreira:2022zzo}. 
We keep $T_*$ as a scanned parameter and profile over the three others. The prior on
$\Delta N_\mathrm{eff}$ stops at $0.39$, the bound from nucleosynthesis alone adopted in ref.~\cite{Ferreira:2022zzo}. It is weaker than the 
joint bound $0.22$ from nucleosynthesis and the cosmic microwave background (CMB)~\cite{Yeh:2022heq} that we apply to the phase transition, because the CMB bound 
holds only if the decay products of the walls remain relativistic until recombination.

\subsection{Cosmic superstrings}
\label{app:strings}

Loops chopped off a cosmic string network oscillate and radiate gravitational waves. The resulting stochastic
background can be computed via the sum over loop harmonics integrated over cosmic history. We take the model file
\texttt{super} of ref.~\cite{NG15Models}, which computes the signal from field-theory cosmic superstrings and then normalises it to the 
intercommutation probability $P$~\cite{Jackson:2004zg}, such that it corresponds to the superstring prediction~\cite{NANOGrav:2023hvm},
\begin{align}
\begin{split}
h^2\Omega_\mathrm{GW}^\mathrm{super}(f) &= \frac{h^2}{P}\,\Omega_\mathrm{GW}^\mathrm{cs}(f) \, , \qquad \text{where} \qquad
  \Omega_\mathrm{GW}^\mathrm{cs}(f) = \frac{8\pi}{3 H_0^2}\,(G\mu)^2
    \sum_{k=1}^{k_\mathrm{max}} P_k\,\mathcal{I}_k(f) \, ,\\
  P_k &= \frac{\Gamma}{\zeta(q)}\,\frac{1}{k^{q}} \, , \qquad
  \mathcal{I}_k(f) = \frac{2k}{f} \int_{t_\mathrm{ini}}^{t_0}\!\mathrm{d}t\,
    \left(\frac{a(t)}{a(t_0)}\right)^{5}
    n_\ell\!\left(\frac{2k}{f}\frac{a(t)}{a(t_0)},\, t\right) \, .
  \end{split}
\end{align}
The string tension $G\mu$ is given in units of Newton's constant, $P_k$ is the power radiated into the $k$-th
harmonic of a loop, normalised such that $\sum_k P_k = \Gamma = 50$ as found in
simulations~\cite{Blanco-Pillado:2017oxo}, and $q = 4/3$ for emission dominated by cusps. The loop
number density $n_\ell(\ell,t)$ follows from the velocity-dependent one-scale model with the
loop-chopping parameter and loop size fixed to their simulated
values~\cite{Blanco-Pillado:2013qja}, and the sum over harmonics is truncated at $k_\mathrm{max} = 10^5$.
The two parameters are $-14 < \log_{10} G\mu < -6$ and $-3 < \log_{10} P < 0$, and we profile over $P$.

\subsection{Inflationary gravitational waves}
\label{app:igw}

Inflationary tensor fluctuations are generally expected to follow, to leading approximation, 
a power-law shape $\mathcal{P}_T$ in frequency. Converted to our units and redshifted 
through the cosmic expansion history, the spectrum (\texttt{igw} of ref.~\cite{NG15Models}) reads
\begin{align}
\begin{split}
  h^2\Omega_\mathrm{GW}^\mathrm{igw}(f) &= h^2\,\frac{\Omega_\mathrm{rad}}{24}\,
    \frac{g(f)}{g_{0}}
    \left(\frac{h_{0}}{h(f)}\right)^{4/3}
    \mathcal{P}_T(f)\; T(f) \, , \qquad
  \mathcal{P}_T(f) = r\,A_\text{s} \left(\frac{f}{f_\mathrm{cmb}}\right)^{n_\text{t}} , \\
  T(f) &= \left[1 - 0.22\left(\frac{f}{f_\mathrm{rh}}\right)^{3/2}
                 + 0.65\left(\frac{f}{f_\mathrm{rh}}\right)^{2}\right]^{-1} ,
\end{split}
\end{align}
with transfer function $T(f)$ as fitted in refs.~\cite{Kuroyanagi:2014nba,Kuroyanagi:2020sfw}.
Here $\Omega_\mathrm{rad}$ is the present radiation density, $A_\text{s}$ the measured scalar amplitude at the CMB
pivot scale $f_\mathrm{cmb}$, and $f_\mathrm{rh}$ the frequency of a mode that was of horizon size at
reheating, fixed by the reheating temperature $T_\mathrm{rh}$. The degrees of freedom functions $g$ and $h$ are evaluated
at the temperature at which the corresponding mode re-entered the Hubble horizon. Modes that re-entered the 
horizon before reheating are suppressed by
the transfer function, which falls as $f^{-2}$ above $f_\mathrm{rh}$. The three free parameters
are the tensor-to-scalar ratio $-40 < \log_{10} r < 0$, the (tensorial) spectral tilt  $0 < n_\text{t} < 6$ and the reheating temperature 
$-3 < \log_{10} ( T_\mathrm{rh} / \text{GeV}) < 3$. A blue tilt,
$n_\text{t} > 0$, is needed for the signal to reach the nanohertz band at all, and the three parameters are
strongly degenerate, which is why we scan $n_\mathrm{t}$ and profile over $r$ and $T_\mathrm{rh}$ (section~\ref{sec:examples}).

\section{Priors of the population model}
\label{app:priors}

Table~\ref{tab:priors} in section~\ref{sec:populations} lists the six parameters of the population model
with their uniform prior ranges and the three Gaussian prior sets. This appendix traces 
the three Gaussian prior sets -- excluding the completely uninformed, flat one -- to the measurements they are based on.

\subsection{NANOGrav priors}
\label{app:ngpriors}

The priors on the galaxy stellar mass function used in the NANOGrav analysis~\cite{NANOGrav:2023hfp} are stated as being based on fits to the data of 
Tomczak et al.~\cite{Tomczak:2013bxa} (table~B1 of ref.~\cite{NANOGrav:2023hfp}, footnotes a and d), without explicit reference as to
how their widths were chosen. A comment in  the
parameter space of the \holodeck{} library indicates that they derive from fits to Tomczak et al.\ ``with 
4x stdev values''.\footnote{%
File \texttt{param\_spaces\_classic.py} of \holodeck{}~\cite{holodeck}; the fits are in the notebook 
\texttt{notebooks/devs/sams/parameter-selection.ipynb}.
} 
That notebook fits straight lines in redshift to the parameters of the single Schechter fits, eq.~\eqref{eq:gsmf}, of the total galaxy sample in 
table~3 of ref.~\cite{Tomczak:2013bxa} (page 12) and finds $\psi_0 = -2.57 \pm 0.11$ and $m_{\psi,0} = 10.94 \pm 0.11$, with slopes 
$\psi_z = -0.59$ and $m_{\psi,z} = +0.12$ and the low-mass slope $\alpha_{\psi,0} = -1.21$, close to the fixed values of 
section~\ref{sec:populations}. We reproduced these fits. Two of the rows the notebook takes as the total sample, at $1.5 < z < 2.0$ and 
$2.0 < z < 2.5$, erroneously belong to the star-forming sample of the same table; with the rows of the total sample the fit gives 
$\psi_0 = -2.61 \pm 0.10$ and $m_{\psi,0} = 10.96 \pm 0.11$. The NANOGrav priors $\mathcal{N}(-2.56, 0.40)$ and 
$\mathcal{N}(10.90, 0.40)$ hence appear to have about four times the width of the actual fits.

The NANOGrav priors on the normalisation and scatter of the black hole to bulge mass relation, $\mu = 8.60 \pm 0.20$ and 
$\epsmu = 0.32 \pm 0.15$, lie between two measurements. Kormendy and Ho~\cite{Kormendy:2013dxa} find 
$M_\bullet/10^9\msol = 0.49^{+0.06}_{-0.05}\,(M_\mathrm{bulge}/10^{11}\msol)^{1.16 \pm 0.08}$ with an intrinsic scatter of 
$0.29\,$dex (their eq.~10, page 571 of the journal version), i.e.\ $\mu = 8.69 \pm 0.05$. They quote no uncertainty on that scatter. For the 
relations with the infrared bulge luminosity and with the velocity dispersion they find $\epsilon_\mu = 0.31 \pm 0.02$ and 
$0.29 \pm 0.03\,$dex (page 566). McConnell and Ma~\cite{McConnell:2012hz} find $\mu = 8.46 \pm 0.08$ with slope $1.05$ and scatter
 $\epsilon_\mu = 0.34\,$dex (their table~2). The $M$--$\sigma$ and $M$--$L$ relations of G\"ultekin et al.~\cite{Gultekin:2009qn}, 
 cited by NANOGrav as well, which relate the black hole mass to the bulge velocity dispersion and luminosity, respectively, and do 
 not involve the bulge mass, give scatters of 
 $\epsilon_\mu = 0.44 \pm 0.06$ and $\epsilon_\mu = 0.38 \pm 0.09\,$dex. The width $0.20$ on $\mu$ thus appears to span the difference 
 between the two bulge-mass fits, and the width $0.15$ on 
 $\epsmu$ the spread of the quoted scatters (although this is not the reasoning provided in ref.~\cite{NANOGrav:2023hfp}). 
 \holodeck{} also contains 
 a narrower prior set based on Kormendy and Ho, with $\mu = 8.69 \pm 0.05$, slope $1.17 \pm 0.08$ and $\epsmu = 0.28 \pm 0.05$, where 
 a comment notes that no uncertainty on the scatter is given, and $0.28\,$dex is the scatter that Kormendy and Ho find for the relation with the 
 velocity dispersion (their eq.~5). Also the widths of the NANOGrav priors on $\mu$ and $\epsmu$ are hence about four times those of this 
 narrower set. While we could not identify a clear justification of this discrepancy, this difference has a strong effect on the predicted 
 amplitude of the binary background and hence on the preference for new physics (as demonstrated in the main text).

\subsection{Tomczak--Kormendy priors}
\label{app:tkpriors}

The Tomczak--Kormendy priors take the fits described in appendix~\ref{app:ngpriors} themselves. For the galaxy stellar mass function they take the fit to the total sample of Tomczak 
et al.~\cite{Tomczak:2013bxa}, $\psi_0 \sim \mathcal{N}(-2.61, 0.10)$ and $m_{\psi,0} \sim \mathcal{N}(10.96, 0.11)$, for the black hole to bulge mass relation the 
narrower \holodeck{} set, $\mu \sim \mathcal{N}(8.69, 0.05)$ and $\epsmu \sim \mathcal{N}(0.28, 0.05)$, and for $\tauf$ and $\nuin$ the 
same uniform priors as before (table~\ref{tab:priors}). Like the NANOGrav priors, they combine the stellar masses of Tomczak et al.\ 
with the relation of Kormendy and Ho, whose bulge masses are calibrated to masses inferred from the orbital velocities of stars and are 
therefore larger by a factor of $1.34$ for the same galaxy (see section~6.6.1 in ref.~\cite{Kormendy:2013dxa}). Inserting the stellar masses of Tomczak et al.\ 
into this relation underestimates the actual black hole masses by about $0.15\,$dex, just like a smaller $\mu$ would, and reduces the predicted 
$h^2\Omega_\mathrm{GW}$ by about a factor of $1.7$.

\subsection{Survey-average priors}
\label{app:surveypriors}

What we refer to as survey-average priors throughout our analysis is 
based on measurements of the galaxy stellar mass function and of the black hole to bulge mass relation published 
since 2014. Since there is no individual reference which fits these recent measurements to the parametrisation of 
section~\ref{sec:populations}, we perform here a simple average of the centres and spreads of such measurements, which is sufficient for 
our purpose of illustrating the effect of the priors on the preference for new physics. The following discussion is not meant to be exhaustive, 
but to give a representative overview of the recent literature.

Galaxy surveys publish the mass function in redshift bins as fits of a single Schechter function in the form of eq.~\eqref{eq:gsmf}, or of a 
double Schechter function, the sum of two Schechter functions with the same characteristic mass and different low-mass slopes, which also 
follows the rise in the number of galaxies below about $10^{10}\,\msol$~\cite{Baldry:2011fj,Tomczak:2013bxa}.
Above the characteristic mass the mass function falls off exponentially. 
In surveys, there is a statistical bias known as the Eddington bias~\cite{Eddington:1913}: more galaxies scatter up from the many lower 
masses than down from the few higher mass ones, and the measured mass function lies above the true one at the massive end. A survey 
corrects for it by convolving its model of the mass function with the error on the stellar mass of a single galaxy, $0.1\,$dex in
Wright et al.~\cite{Wright:2018gsm} and $0.15\,$dex in McLeod et al.~\cite{McLeod:2021gsm}, before comparing it with the counts,
so that the published parameters aim to describe the true mass function.

We use the six available determinations that are corrected in this way, all of the total galaxy population, star-forming and 
quiescent~\cite{Davidzon:2017rib,Weaver:2023cos,McLeod:2021gsm,Leja:2019lfo,Wright:2018gsm,Driver:2022vyh}. Davidzon et 
al.\ (COSMOS2015, table~1) and Weaver et al.\ (COSMOS2020, table~C.1) use deep near-infrared imaging of the two square degrees of 
the COSMOS field and reach $z \simeq 6$ and $7.5$. McLeod et al.\ (table~5) combine ground-based near-infrared surveys with imaging of 
the Hubble Space Telescope. Leja et al.\ (appendix~B) use galaxies of the COSMOS and 3D-HST fields and give a mass function 
continuous in redshift for $0.2 < z < 3$. Their masses come from star formation histories without a fixed functional form, and their number 
densities are higher than those of the other surveys (their section~6.1). Wright et al.\ (table~A3) combine the GAMA, G10-COSMOS and 
3D-HST surveys and also give a mass function continuous in redshift. Driver et al.\ (GAMA DR4, table~7) give the local mass function at 
$z < 0.1$ from spectroscopic redshifts over $250$ square degrees. The first three share galaxies of the COSMOS field and are not fully 
independent. The mass functions of Tomczak et al., on which the NANOGrav priors rest, are not corrected for the Eddington bias and are not 
among the six.

\begin{figure}[tbp]
\centering
\includegraphics[width=\textwidth]{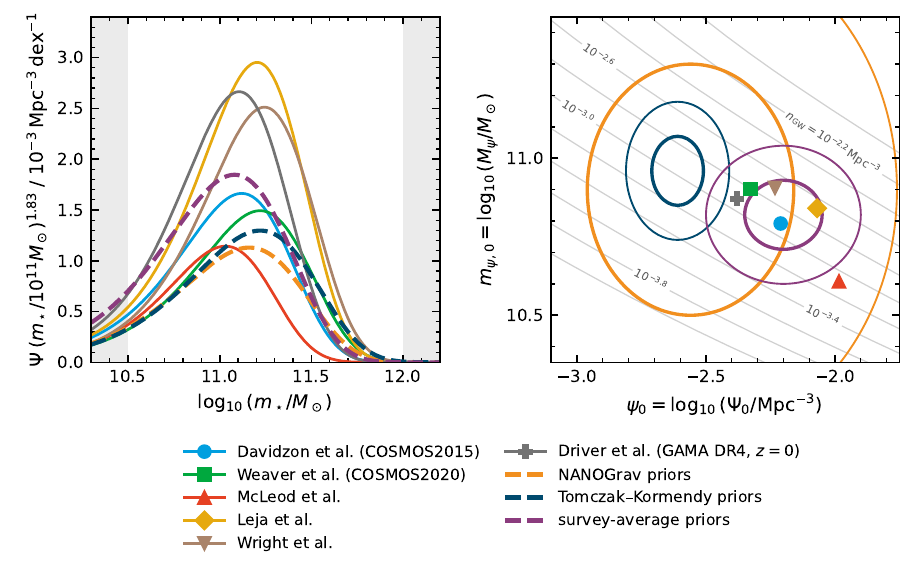}
\caption{{\it Left panel:} The six galaxy stellar mass functions from refs.~\cite{Davidzon:2017rib,Weaver:2023cos,McLeod:2021gsm,Leja:2019lfo,Wright:2018gsm,Driver:2022vyh}, multiplied by 
$(m_\star/10^{11}\msol)^{1.83}$, the contribution of each decade of stellar mass to 
$\hcsq$, in the redshift bin of each survey that contains $z = 0.5$ ($z = 0$ for GAMA DR4), and the single Schechter function of 
eq.~\eqref{eq:gsmf} at $z = 0.5$ at the centres of the three Gaussian prior sets (dashed). The area under each curve between the grey 
bands is the $\hcsq$-weighted number density $n_\mathrm{GW}$. {\it Right panel:} 
The values of $\psi_0$ and $m_{\psi,0}$ fitted to each survey 
(coloured symbols), with the $1\sigma$ (thick) and $2\sigma$ (thin) ellipses of the three prior sets, drawn without correlation as they enter 
the fits. The $2\sigma$ ellipse of the NANOGrav priors extends beyond the panel. Along each grey line $n_\mathrm{GW}$ is constant
(as indicated by the labels, the lines are $0.2\,$dex apart).}
\label{fig:surveyfits}
\end{figure}

The binary background weights galaxies by their mass. For binaries that shrink by gravitational wave emission, the characteristic strain 
squared, $\hcsq$, at a given frequency is proportional to the number density of binaries times their chirp mass to the power 
$5/3$~\cite{Phinney:2001di}, which Liepold and Ma use to translate the local mass function into an amplitude of the 
background~\cite{Liepold:2024woa}. In the population model that we use the galaxy pair fraction and the galaxy merger time do not 
depend on the stellar mass (table~B1 of ref.~\cite{NANOGrav:2023hfp}), the mass of a black hole grows as 
$M_\bullet \propto M_\mathrm{bulge}^{1.10} \propto m_\star^{1.10}$, and at a fixed mass ratio the chirp mass is proportional to $M_\bullet$. 
The galaxies of one decade of stellar mass therefore contribute to $\hcsq$ in proportion to 
$\Psi\, m_\star^{5/3 \times 1.10} = \Psi\, m_\star^{1.83}$. At the prior centres this product peaks at $m_\star \simeq 10^{11.1}\,\msol$, and 
about $90\%$ of its integral comes from $10^{10.5}$ to $10^{12}\,\msol$ (figure~\ref{fig:surveyfits}, left).

We fit the single Schechter function of eq.~\eqref{eq:gsmf} to each survey. Its low-mass slope and redshift evolution are fixed at the values 
of section~\ref{sec:populations}, so only $\psi_0$ and $m_{\psi,0}$ are free. With these two parameters it cannot follow the double 
Schechter curves of the surveys at all masses, and the fit has to choose where to match them. We minimise the squared difference in 
$\log_{10}\Psi$ between the two on a grid of masses over $10.5 < \log_{10}(m_\star/\msol) < 12$, in all redshift bins of a survey with 
$z \le 3$ at once, and weight each mass by its contribution to $\hcsq$, $\Psi\, m_\star^{1.83}$. The fit then reproduces the published curves 
to about $0.1\,$dex where this weight peaks. The mean and standard deviation of the six fits give $\psi_0 = -2.20 \pm 0.15$ and 
$m_{\psi,0} = 10.82 \pm 0.11$. 
We use this result to motivate our choice of survey-average priors in table~\ref{tab:priors}, as visualised in the right panel of figure~\ref{fig:surveyfits}.
This spread exceeds the statistical error of each survey, $0.02$ to $0.1\,$dex on the normalisations (e.g.\ 
COSMOS2020 table~C.1), and comes mainly from the modelling of the stellar populations (ref.~\cite{Leja:2019lfo}, section~6.1).

The background does not depend on $\psi_0$ and $m_{\psi,0}$ separately but on how many massive galaxies they predict. We measure 
this by the $\hcsq$-weighted number density of galaxies,
\begin{equation}
  n_\mathrm{GW} = \int_{10.5}^{12} \Psi(m_\star, z = 0.5)\, \left(\frac{m_\star}{10^{11}\,\msol}\right)^{1.83} \mathrm{d}\log_{10}(m_\star/\msol) \, ,
  \label{eq:ngw}
\end{equation}
evaluated at $z = 0.5$, between the redshifts at which the contributions to $\hcsq$ peak in the models of Liepold and Ma, $z \simeq 0.3$, 
and of NANOGrav, $z \simeq 1$~\cite{Liepold:2024woa,NANOGrav:2023hfp}. In the right panel of figure~\ref{fig:surveyfits}, we show how the surveys 
spread along the lines of constant $n_\mathrm{GW}$, since a higher normalisation of the mass function is compensated by a lower 
characteristic mass. They agree on $\log_{10} n_\mathrm{GW}$ to $0.15\,$dex but fix $\psi_0$ and $m_{\psi,0}$ only to $0.15$ and $0.11$, 
with a correlation coefficient of $-0.77$ that the Gaussian priors neglect. The survey-average centre
lies $0.20\,$dex above the NANOGrav centre in 
$\log_{10} n_\mathrm{GW}$. The NANOGrav centre is a fit of straight lines in redshift to the Schechter parameters of their table~3 
(appendix~\ref{app:ngpriors}). Our weighted fit of the same tables gives $\log_{10} n_\mathrm{GW}$ between $0.12$ and $0.14\,$dex 
above the NANOGrav centre, so most of the $0.20\,$dex come from the fit, and the newer surveys add $0.06$ to $0.08\,$dex.

For the relation between black hole and bulge mass we centre the prior between two fits whose bulge masses are on the stellar-population 
scale of the galaxy surveys rather than on a dynamical one. Ref.~\cite{Sahu:2019bhm} finds $\mu = 8.77$ (table~5), and 
eq.~10 of ref.~\cite{Kormendy:2013dxa} corresponds to $\mu = 8.84$ without the $0.126\,$dex by which its bulge masses were raised to a 
dynamical scale (before eq.~8). We take their mean, $\mu = 8.80$. For the width we take $0.10$, corresponding to  
the spread of four fits: adding to those of refs.~\cite{Sahu:2019bhm,Kormendy:2013dxa}, we take the fits of 
refs.~\cite{McConnell:2012hz,Saglia:2016bhm}, whose bulge masses we lower by the same $0.126\,$dex. The intrinsic scatters 
$\epsmu = 0.40\,$dex~\cite{Sahu:2019bhm} and $\epsmu = 0.42 \pm 0.04\,$dex~\cite{Saglia:2016bhm} motivate a prior of $\epsmu = 0.40 \pm 0.05$. 
In short, the survey-average priors lie in between the other two sets in how tightly they constrain $\mu$: the width of $0.10$ on $\mu$ 
lies between the $0.20$ of the NANOGrav priors and the $0.05$ of the Tomczak--Kormendy priors. On the stellar-population mass scale of the galaxy surveys, however, 
black holes are heavier at a given bulge mass, so that the centre of the prior on $\mu$ lies $0.20\,$dex above that of the NANOGrav 
priors and $0.11\,$dex above that of the Tomczak--Kormendy priors. Together with the larger number of massive galaxies in the newer mass 
functions, our updated survey-average priors predict the loudest SMBHB background of all prior sets considered here.

\section{The \binaryfog{} code}\label{app:code}
\label{sec:tool}

The machinery used to derive the results of this paper is released as \binaryfog{}~\cite{binaryfog} together with this paper and can be found at
\begin{center}
\url{https://github.com/tasicarl/binaryfog}
\end{center}
The libraries of this first \binaryfog{} release
are built with \holodeck{} at the current version $1.7.1$~\cite{holodeck}, including the corrections described in
ref.~\cite{NANOGrav:2026err}. The \holodeck{} libraries, which took 
$130$ CPU hours to generate, and the scripts that produce every figure and number in this paper accompany the publication as an 
archived release with a digital object identifier \url{https://doi.org/10.5281/zenodo.24294394}. \binaryfog{} downloads the measured free spectra~\cite{PTArcadeData,NG15KDE,NG15KDEv1} from their public 
records on first use.

In the code, a new-physics model is supplied as a simple Python function which takes a frequency array as input and returns 
$h^2\Omega_\mathrm{GW}$ evaluated at these frequencies. Nothing about the package is
specific to the models studied here, any of the publicly available model files of the NANOGrav new-physics
search~\cite{NG15Models} can be handed to it unchanged, and any free-spectrum product in the same format can replace the measured spectra. A typical scan (see \texttt{examples/peaked\_source.py} in the repository) looks like
\begin{lstlisting}[language=python]
import numpy as np
import binaryfog as bf

def peaked(f_Hz, log10_amp, log10_fpeak):
    x = np.asarray(f_Hz) / 10.0 ** log10_fpeak
    return 10.0 ** log10_amp * x ** 3 / (1.0 + x ** 7)


data = bf.FreeSpectrum.load("ng15_cp_v2")
fpeak = np.linspace(-8.5, -7.0, 16)
like = bf.Likelihood(data, bf.Background())
scan = bf.ProfileScan(like, peaked, scan={"log10_fpeak": fpeak},
                      nuisance={"log10_amp": (-12, -5, "linear")},
                      astro_prior="nanograv", restarts=3)

table = scan.run(jobs=8)
inside = bf.region(table, 2, ndim=1)
\end{lstlisting}
and returns the profiled likelihood $\ln L_\mathrm{prof}(f_\mathrm{peak})$ together with the astrophysical parameters that achieved it,
in this case for a 
simple broken power law $h^2 \Omega_\mathrm{GW}(f) = A (f/f_\mathrm{peak})^3/(1 + (f/f_\mathrm{peak})^7)$ in the \texttt{pandas} \texttt{table}. From this, the $2\sigma$ region is obtained in the last line of the above code. The
function \texttt{bf.preference(table, like)} computes the fit without new physics and returns the gain $G$ of eq.~\eqref{eq:gain}. Further code documentation can be found online, at the repository address stated above.
We encourage users to get in touch with us\footnote{\href{mailto:carlo.tasillo@ific.uv.es}{carlo.tasillo@ific.uv.es}} 
regarding any questions or issues related to the code.

\bibliographystyle{JHEP_improved}
\bibliography{bibliography}

\providecommand{\href}[2]{#2}\begingroup\raggedright\begin{thebibliography}{100}

\bibitem{Hellings:1983fr}
R.~W. Hellings and G.~S. Downs, \href{http://dx.doi.org/10.1086/183954}{{\it
  {Upper limits on the isotropic gravitational radiation background from pulsar
  timing analysis}}, } {\em Astrophys. J. Lett.} {\bf 265} (1983) L39--L42.

\bibitem{NANOGrav:2023gor}
{\bf NANOGrav Collaboration}, G.~Agazie et~al.,
  \href{http://dx.doi.org/10.3847/2041-8213/acdac6}{{\it {The NANOGrav 15 yr
  Data Set: Evidence for a Gravitational-wave Background}}, } {\em Astrophys.
  J. Lett.} {\bf 951} (2023), no.~1 L8,
  [\href{http://arxiv.org/abs/2306.16213}{{\tt 2306.16213}}].

\bibitem{EPTA:2023fyk}
{\bf EPTA, InPTA Collaboration}, J.~Antoniadis et~al.,
  \href{http://dx.doi.org/10.1051/0004-6361/202346844}{{\it {The second data
  release from the European Pulsar Timing Array - III. Search for gravitational
  wave signals}}, } {\em Astron. Astrophys.} {\bf 678} (2023) A50,
  [\href{http://arxiv.org/abs/2306.16214}{{\tt 2306.16214}}].

\bibitem{Reardon:2023gzh}
D.~J. Reardon et~al., \href{http://dx.doi.org/10.3847/2041-8213/acdd02}{{\it
  {Search for an Isotropic Gravitational-wave Background with the Parkes Pulsar
  Timing Array}}, } {\em Astrophys. J. Lett.} {\bf 951} (2023), no.~1 L6,
  [\href{http://arxiv.org/abs/2306.16215}{{\tt 2306.16215}}].

\bibitem{Xu:2023wog}
H.~Xu et~al., \href{http://dx.doi.org/10.1088/1674-4527/acdfa5}{{\it {Searching
  for the Nano-Hertz Stochastic Gravitational Wave Background with the Chinese
  Pulsar Timing Array Data Release I}}, } {\em Res. Astron. Astrophys.} {\bf
  23} (2023), no.~7 075024, [\href{http://arxiv.org/abs/2306.16216}{{\tt
  2306.16216}}].

\bibitem{Miles:2024seg}
M.~T. Miles et~al., \href{http://dx.doi.org/10.1093/mnras/stae2571}{{\it {The
  MeerKAT Pulsar Timing Array: the first search for gravitational waves with
  the MeerKAT radio telescope}}, } {\em Mon. Not. Roy. Astron. Soc.} {\bf 536}
  (2024), no.~2 1489--1500, [\href{http://arxiv.org/abs/2412.01153}{{\tt
  2412.01153}}].

\bibitem{Sesana:2025udx}
A.~Sesana and D.~G. Figueroa,
  \href{http://dx.doi.org/10.1146/annurev-astro-122325-025132}{{\it {Nanohertz
  Gravitational Waves}}, } {\em Ann. Rev. Astron. Astrophys.} {\bf 64} (2026),
  no.~1 537--586, [\href{http://arxiv.org/abs/2512.18822}{{\tt 2512.18822}}].

\bibitem{NANOGrav:2023hfp}
{\bf NANOGrav Collaboration}, G.~Agazie et~al.,
  \href{http://dx.doi.org/10.3847/2041-8213/ace18b}{{\it {The NANOGrav 15 yr
  Data Set: Constraints on Supermassive Black Hole Binaries from the
  Gravitational-wave Background}}, } {\em Astrophys. J. Lett.} {\bf 952}
  (2023), no.~2 L37, [\href{http://arxiv.org/abs/2306.16220}{{\tt
  2306.16220}}]. [Erratum: Astrophys.J.Lett. 1006, L67 (2026)].

\bibitem{Sato-Polito:2023gym}
G.~Sato-Polito, M.~Zaldarriaga, and E.~Quataert,
  \href{http://dx.doi.org/10.1103/PhysRevD.110.063020}{{\it {Where are the
  supermassive black holes measured by PTAs?}}, } {\em Phys. Rev. D} {\bf 110}
  (2024), no.~6 063020, [\href{http://arxiv.org/abs/2312.06756}{{\tt
  2312.06756}}].

\bibitem{Ellis:2023dgf}
J.~Ellis, M.~Fairbairn, G.~H{\"u}tsi, J.~Raidal, J.~Urrutia, et~al.,
  \href{http://dx.doi.org/10.1103/PhysRevD.109.L021302}{{\it {Gravitational
  waves from supermassive black hole binaries in light of the NANOGrav 15-year
  data}}, } {\em Phys. Rev. D} {\bf 109} (2024), no.~2 L021302,
  [\href{http://arxiv.org/abs/2306.17021}{{\tt 2306.17021}}].

\bibitem{Chen:2025wel}
N.~Chen, T.~Di~Matteo, Y.~Zhou, L.~Z. Kelley, L.~Blecha, et~al.,
  \href{http://dx.doi.org/10.3847/2041-8213/adefe2}{{\it {The
  Gravitational-wave Background from Massive Black Holes in the ASTRID
  Simulation}}, } {\em Astrophys. J. Lett.} {\bf 991} (2025), no.~1 L19,
  [\href{http://arxiv.org/abs/2502.01024}{{\tt 2502.01024}}].

\bibitem{Mingarelli:2026ykj}
C.~M.~F. Mingarelli, {\it {Energetic Ceilings and Benchmarks of Astrophysical
  Gravitational-Wave Backgrounds}},
  \href{http://arxiv.org/abs/2601.18859}{{\tt 2601.18859}}.

\bibitem{Liepold:2024woa}
E.~R. Liepold and C.-P. Ma,
  \href{http://dx.doi.org/10.3847/2041-8213/ad66b8}{{\it {Big Galaxies and Big
  Black Holes: The Massive Ends of the Local Stellar and Black Hole Mass
  Functions and the Implications for Nanohertz Gravitational Waves}}, } {\em
  Astrophys. J. Lett.} {\bf 971} (2024), no.~2 L29,
  [\href{http://arxiv.org/abs/2407.14595}{{\tt 2407.14595}}]. [Erratum:
  Astrophys.J.Lett. 1004, L20 (2026)].

\bibitem{Matt:2025bao}
C.~Matt et~al., \href{http://dx.doi.org/10.3847/1538-4357/ae2480}{{\it
  {Inferring M$_{BH}${\textendash}M$_{bulge}$ Evolution from the
  Gravitational-wave Background}}, } {\em Astrophys. J.} {\bf 997} (2026),
  no.~2 188, [\href{http://arxiv.org/abs/2508.18126}{{\tt 2508.18126}}].

\bibitem{Phinney:2001di}
E.~S. Phinney, {\it {A Practical theorem on gravitational wave backgrounds}},
  \href{http://arxiv.org/abs/astro-ph/0108028}{{\tt astro-ph/0108028}}.

\bibitem{Dai:2026jjw}
B.~Dai, M.~Zaldarriaga, and Y.~Ali-Ha{\"\i}moud, {\it {Prior volume effect
  inflates the gravitational-wave background in pulsar timing arrays}},
  \href{http://arxiv.org/abs/2610.28739}{{\tt 2610.28739}}.

\bibitem{Begelman:1980vb}
M.~C. Begelman, R.~D. Blandford, and M.~J. Rees,
  \href{http://dx.doi.org/10.1038/287307a0}{{\it {Massive black hole binaries
  in active galactic nuclei}}, } {\em Nature} {\bf 287} (1980) 307--309.

\bibitem{Milosavljevic:2002bn}
M.~Milosavljevic and D.~Merritt, \href{http://dx.doi.org/10.1086/378086}{{\it
  {Long term evolution of massive black hole binaries}}, } {\em Astrophys. J.}
  {\bf 596} (2003) 860, [\href{http://arxiv.org/abs/astro-ph/0212459}{{\tt
  astro-ph/0212459}}].

\bibitem{Khan:2011gi}
F.~M. Khan, A.~Just, and D.~Merritt,
  \href{http://dx.doi.org/10.1088/0004-637X/732/2/89}{{\it {Efficient Merger of
  Binary Supermassive Black Holes in Merging Galaxies}}, } {\em Astrophys. J.}
  {\bf 732} (2011) 89, [\href{http://arxiv.org/abs/1103.0272}{{\tt
  1103.0272}}].

\bibitem{Vasiliev:2015uzo}
E.~Vasiliev, F.~Antonini, and D.~Merritt,
  \href{http://dx.doi.org/10.1088/0004-637X/810/1/49}{{\it {The Final-parsec
  Problem in the Collisionless Limit}}, } {\em Astrophys. J.} {\bf 810} (2015),
  no.~1 49, [\href{http://arxiv.org/abs/1505.05480}{{\tt 1505.05480}}].

\bibitem{NANOGrav:2023hvm}
{\bf NANOGrav Collaboration}, A.~Afzal et~al.,
  \href{http://dx.doi.org/10.3847/2041-8213/acdc91}{{\it {The NANOGrav 15 yr
  Data Set: Search for Signals from New Physics}}, } {\em Astrophys. J. Lett.}
  {\bf 951} (2023), no.~1 L11, [\href{http://arxiv.org/abs/2306.16219}{{\tt
  2306.16219}}]. [Erratum: Astrophys.J.Lett. 971, L27 (2024)].

\bibitem{EPTA:2023xxk}
{\bf EPTA, InPTA Collaboration}, J.~Antoniadis et~al.,
  \href{http://dx.doi.org/10.1051/0004-6361/202347433}{{\it {The second data
  release from the European Pulsar Timing Array - IV. Implications for massive
  black holes, dark matter, and the early Universe}}, } {\em Astron.
  Astrophys.} {\bf 685} (2024) A94,
  [\href{http://arxiv.org/abs/2306.16227}{{\tt 2306.16227}}].

\bibitem{Madge:2023dxc}
E.~Madge, E.~Morgante, C.~Puchades-Ib{\'a}{\~n}ez, N.~Ramberg, W.~Ratzinger,
  et~al., \href{http://dx.doi.org/10.1007/JHEP10(2023)171}{{\it {Primordial
  gravitational waves in the nano-Hertz regime and PTA data {\textemdash}
  towards solving the GW inverse problem}}, } {\em JHEP} {\bf 10} (2023) 171,
  [\href{http://arxiv.org/abs/2306.14856}{{\tt 2306.14856}}].

\bibitem{Figueroa:2023zhu}
D.~G. Figueroa, M.~Pieroni, A.~Ricciardone, and P.~Simakachorn,
  \href{http://dx.doi.org/10.1103/PhysRevLett.132.171002}{{\it {Cosmological
  Background Interpretation of Pulsar Timing Array Data}}, } {\em Phys. Rev.
  Lett.} {\bf 132} (2024), no.~17 171002,
  [\href{http://arxiv.org/abs/2307.02399}{{\tt 2307.02399}}].

\bibitem{Ellis:2023oxs}
J.~Ellis, M.~Fairbairn, G.~Franciolini, G.~H{\"u}tsi, A.~Iovino, et~al.,
  \href{http://dx.doi.org/10.1103/PhysRevD.109.023522}{{\it {What is the source
  of the PTA GW signal?}}, } {\em Phys. Rev. D} {\bf 109} (2024), no.~2 023522,
  [\href{http://arxiv.org/abs/2308.08546}{{\tt 2308.08546}}].

\bibitem{Wang:2025wht}
M.~Wang, Z.~Zhang, and H.~Xu, {\it {Discriminating Between Models of the
  Nanohertz Gravitational-Wave Background with Pulsar Timing Arrays}},
  \href{http://arxiv.org/abs/2510.22713}{{\tt 2510.22713}}.

\bibitem{Bringmann:2023opz}
T.~Bringmann, P.~F. Depta, T.~Konstandin, K.~Schmidt-Hoberg, and C.~Tasillo,
  \href{http://dx.doi.org/10.1088/1475-7516/2023/11/053}{{\it {Does NANOGrav
  observe a dark sector phase transition?}}, } {\em JCAP} {\bf 11} (2023) 053,
  [\href{http://arxiv.org/abs/2306.09411}{{\tt 2306.09411}}].

\bibitem{Balan:2025uke}
S.~Balan, T.~Bringmann, F.~Kahlhoefer, J.~Matuszak, and C.~Tasillo,
  \href{http://dx.doi.org/10.1088/1475-7516/2025/08/062}{{\it {Sub-GeV dark
  matter and nano-Hertz gravitational waves from a classically conformal dark
  sector}}, } {\em JCAP} {\bf 08} (2025) 062,
  [\href{http://arxiv.org/abs/2502.19478}{{\tt 2502.19478}}].

\bibitem{Bringmann:2026xcx}
T.~Bringmann, T.~Konstandin, J.~Matuszak, K.~Schmidt-Hoberg, and C.~Tasillo,
  {\it {Tuning the violins: dark sector phase transition models for the PTA
  signal}},  \href{http://arxiv.org/abs/2602.09092}{{\tt 2602.09092}}.

\bibitem{Ellis:2023tsl}
J.~Ellis, M.~Lewicki, C.~Lin, and V.~Vaskonen,
  \href{http://dx.doi.org/10.1103/PhysRevD.108.103511}{{\it {Cosmic
  superstrings revisited in light of NANOGrav 15-year data}}, } {\em Phys. Rev.
  D} {\bf 108} (2023), no.~10 103511,
  [\href{http://arxiv.org/abs/2306.17147}{{\tt 2306.17147}}].

\bibitem{Vagnozzi:2023lwo}
S.~Vagnozzi, \href{http://dx.doi.org/10.1016/j.jheap.2023.07.001}{{\it
  {Inflationary interpretation of the stochastic gravitational wave background
  signal detected by pulsar timing array experiments}}, } {\em JHEAp} {\bf 39}
  (2023) 81--98, [\href{http://arxiv.org/abs/2306.16912}{{\tt 2306.16912}}].

\bibitem{Franciolini:2023pbf}
G.~Franciolini, A.~Iovino, Jr., V.~Vaskonen, and H.~Veerm{\"a}e,
  \href{http://dx.doi.org/10.1103/PhysRevLett.131.201401}{{\it {Recent
  Gravitational Wave Observation by Pulsar Timing Arrays and Primordial Black
  Holes: The Importance of Non-Gaussianities}}, } {\em Phys. Rev. Lett.} {\bf
  131} (2023), no.~20 201401, [\href{http://arxiv.org/abs/2306.17149}{{\tt
  2306.17149}}].

\bibitem{Bian:2023dnv}
L.~Bian, S.~Ge, J.~Shu, B.~Wang, X.-Y. Yang, et~al.,
  \href{http://dx.doi.org/10.1103/PhysRevD.109.L101301}{{\it {Gravitational
  wave sources for pulsar timing arrays}}, } {\em Phys. Rev. D} {\bf 109}
  (2024), no.~10 L101301, [\href{http://arxiv.org/abs/2307.02376}{{\tt
  2307.02376}}].

\bibitem{Wu:2023hsa}
Y.-M. Wu, Z.-C. Chen, and Q.-G. Huang,
  \href{http://dx.doi.org/10.1007/s11433-023-2298-7}{{\it {Cosmological
  interpretation for the stochastic signal in pulsar timing arrays}}, } {\em
  Sci. China Phys. Mech. Astron.} {\bf 67} (2024), no.~4 240412,
  [\href{http://arxiv.org/abs/2307.03141}{{\tt 2307.03141}}].

\bibitem{Gouttenoire:2023ftk}
Y.~Gouttenoire and E.~Vitagliano,
  \href{http://dx.doi.org/10.1103/PhysRevD.110.L061306}{{\it {Domain wall
  interpretation of the PTA signal confronting black hole overproduction}}, }
  {\em Phys. Rev. D} {\bf 110} (2024), no.~6 L061306,
  [\href{http://arxiv.org/abs/2306.17841}{{\tt 2306.17841}}].

\bibitem{Bhaumik:2023wmw}
N.~Bhaumik, R.~K. Jain, and M.~Lewicki,
  \href{http://dx.doi.org/10.1103/PhysRevD.108.123532}{{\it {Ultralow mass
  primordial black holes in the early Universe can explain the pulsar timing
  array signal}}, } {\em Phys. Rev. D} {\bf 108} (2023), no.~12 123532,
  [\href{http://arxiv.org/abs/2308.07912}{{\tt 2308.07912}}].

\bibitem{Winkler:2024olr}
M.~W. Winkler and K.~Freese,
  \href{http://dx.doi.org/10.1103/PhysRevD.111.083509}{{\it {Origin of the
  stochastic gravitational wave background: First-order phase transition versus
  black hole mergers}}, } {\em Phys. Rev. D} {\bf 111} (2025), no.~8 083509,
  [\href{http://arxiv.org/abs/2401.13729}{{\tt 2401.13729}}].

\bibitem{Gouttenoire:2023bqy}
Y.~Gouttenoire, \href{http://dx.doi.org/10.1103/PhysRevLett.131.171404}{{\it
  {First-Order Phase Transition Interpretation of Pulsar Timing Array Signal Is
  Consistent with Solar-Mass Black Holes}}, } {\em Phys. Rev. Lett.} {\bf 131}
  (2023), no.~17 171404, [\href{http://arxiv.org/abs/2307.04239}{{\tt
  2307.04239}}].

\bibitem{Ali-Haimoud:2026sbk}
Y.~Ali-Ha{\"\i}moud, {\it {A practical theorem on gravitational-wave background
  statistics}},  \href{http://arxiv.org/abs/2604.19701}{{\tt 2604.19701}}.

\bibitem{Lamb:2024gbh}
W.~G. Lamb and S.~R. Taylor,
  \href{http://dx.doi.org/10.3847/2041-8213/ad654a}{{\it {Spectral Variance in
  a Stochastic Gravitational-wave Background from a Binary Population}}, } {\em
  Astrophys. J. Lett.} {\bf 971} (2024), no.~1 L10,
  [\href{http://arxiv.org/abs/2407.06270}{{\tt 2407.06270}}].

\bibitem{Xue:2024qtx}
X.~Xue, Z.~Pan, and L.~Dai,
  \href{http://dx.doi.org/10.1103/PhysRevD.111.043022}{{\it {Non-Gaussian
  statistics of nanohertz stochastic gravitational waves}}, } {\em Phys. Rev.
  D} {\bf 111} (2025), no.~4 043022,
  [\href{http://arxiv.org/abs/2409.19516}{{\tt 2409.19516}}].

\bibitem{Agazie:2024jbf}
{\bf NANOGrav Collaboration}, G.~Agazie et~al.,
  \href{http://dx.doi.org/10.3847/1538-4357/ad93d5}{{\it {The NANOGrav 15 yr
  Data Set: Looking for Signs of Discreteness in the Gravitational-wave
  Background}}, } {\em Astrophys. J.} {\bf 978} (2025), no.~1 31,
  [\href{http://arxiv.org/abs/2404.07020}{{\tt 2404.07020}}].

\bibitem{Moore:2021ibq}
C.~J. Moore and A.~Vecchio,
  \href{http://dx.doi.org/10.1038/s41550-021-01489-8}{{\it {Ultra-low-frequency
  gravitational waves from cosmological and astrophysical processes}}, } {\em
  Nature Astron.} {\bf 5} (2021), no.~12 1268--1274,
  [\href{http://arxiv.org/abs/2104.15130}{{\tt 2104.15130}}].

\bibitem{Becsy:2022pnr}
B.~B{\'e}csy, N.~J. Cornish, and L.~Z. Kelley,
  \href{http://dx.doi.org/10.3847/1538-4357/aca1b2}{{\it {Exploring Realistic
  Nanohertz Gravitational-wave Backgrounds}}, } {\em Astrophys. J.} {\bf 941}
  (2022), no.~2 119, [\href{http://arxiv.org/abs/2207.01607}{{\tt
  2207.01607}}].

\bibitem{Konstandin:2024fyo}
T.~Konstandin, A.-M. Lemke, A.~Mitridate, and E.~Perboni,
  \href{http://dx.doi.org/10.1088/1475-7516/2025/04/059}{{\it {The impact of
  cosmic variance on PTAs anisotropy searches}}, } {\em JCAP} {\bf 04} (2025)
  059, [\href{http://arxiv.org/abs/2408.07741}{{\tt 2408.07741}}].

\bibitem{Gardiner:2026orz}
E.~C. Gardiner et~al., {\it {Constraints on Supermassive Black Hole Binaries
  from the Lack of Resolvable Sources in the NANOGrav 15-yr Dataset}},
  \href{http://arxiv.org/abs/2609.32306}{{\tt 2609.32306}}.

\bibitem{holodeck}
{The NANOGrav Collaboration}, {\it {holodeck: massive black hole binary
  population synthesis}},  2026.
\newblock software, version 1.7.1; libraries built at commit 44fe308,
  \url{https://github.com/nanograv/holodeck}.

\bibitem{NANOGrav:2026err}
{\bf NANOGrav Collaboration}, G.~Agazie et~al.,
  \href{http://dx.doi.org/10.3847/2041-8213/ae8764}{{\it {Erratum: The NANOGrav
  15 yr Data Set: Constraints on Supermassive Black Hole Binaries from the
  Gravitational-wave Background}}, } {\em Astrophys. J. Lett.} {\bf 1006}
  (2026), no.~2 L67.

\bibitem{Chen:2018znx}
S.~Chen, A.~Sesana, and C.~J. Conselice,
  \href{http://dx.doi.org/10.1093/mnras/stz1722}{{\it {Constraining
  astrophysical observables of Galaxy and Supermassive Black Hole Binary
  Mergers using Pulsar Timing Arrays}}, } {\em Mon. Not. Roy. Astron. Soc.}
  {\bf 488} (2019), no.~1 401--418,
  [\href{http://arxiv.org/abs/1810.04184}{{\tt 1810.04184}}].

\bibitem{LISA:2017pwj}
{\bf LISA Collaboration}, P.~Amaro-Seoane et~al., {\it {Laser Interferometer
  Space Antenna}},  \href{http://arxiv.org/abs/1702.00786}{{\tt 1702.00786}}.

\bibitem{Merritt:2004gc}
D.~Merritt and M.~Milosavljevic,
  \href{http://dx.doi.org/10.12942/lrr-2005-8}{{\it {Massive black hole binary
  evolution}}, } {\em Living Rev. Rel.} {\bf 8} (2005) 8,
  [\href{http://arxiv.org/abs/astro-ph/0410364}{{\tt astro-ph/0410364}}].

\bibitem{Armitage:2002uu}
P.~J. Armitage and P.~Natarajan, \href{http://dx.doi.org/10.1086/339770}{{\it
  {Accretion during the merger of supermassive black holes}}, } {\em Astrophys.
  J. Lett.} {\bf 567} (2002) L9--L12,
  [\href{http://arxiv.org/abs/astro-ph/0201318}{{\tt astro-ph/0201318}}].

\bibitem{Ryu:2018yhv}
T.~Ryu, R.~Perna, Z.~Haiman, J.~P. Ostriker, and N.~C. Stone,
  \href{http://dx.doi.org/10.1093/mnras/stx2524}{{\it {Interactions between
  multiple supermassive black holes in galactic nuclei: a solution to the final
  parsec problem}}, } {\em Mon. Not. Roy. Astron. Soc.} {\bf 473} (2018), no.~3
  3410--3433, [\href{http://arxiv.org/abs/1709.06501}{{\tt 1709.06501}}].

\bibitem{Bonetti:2017lnj}
M.~Bonetti, A.~Sesana, E.~Barausse, and F.~Haardt,
  \href{http://dx.doi.org/10.1093/mnras/sty874}{{\it {Post-Newtonian evolution
  of massive black hole triplets in galactic nuclei {\textendash} III. A robust
  lower limit to the nHz stochastic background of gravitational waves}}, } {\em
  Mon. Not. Roy. Astron. Soc.} {\bf 477} (2018), no.~2 2599--2612,
  [\href{http://arxiv.org/abs/1709.06095}{{\tt 1709.06095}}].

\bibitem{Alonso-Alvarez:2024gdz}
G.~Alonso-{\'A}lvarez, J.~M. Cline, and C.~Dewar,
  \href{http://dx.doi.org/10.1103/PhysRevLett.133.021401}{{\it
  {Self-Interacting Dark Matter Solves the Final Parsec Problem of Supermassive
  Black Hole Mergers}}, } {\em Phys. Rev. Lett.} {\bf 133} (2024), no.~2
  021401, [\href{http://arxiv.org/abs/2401.14450}{{\tt 2401.14450}}].

\bibitem{Sesana:2015haa}
A.~Sesana and F.~M. Khan, \href{http://dx.doi.org/10.1093/mnrasl/slv131}{{\it
  {Scattering experiments meet N-body {\textendash} I. A practical recipe for
  the evolution of massive black hole binaries in stellar environments}}, }
  {\em Mon. Not. Roy. Astron. Soc.} {\bf 454} (2015), no.~1 L66--L70,
  [\href{http://arxiv.org/abs/1505.02062}{{\tt 1505.02062}}].

\bibitem{Lang:2014cta}
P.~Lang et~al., \href{http://dx.doi.org/10.1088/0004-637X/788/1/11}{{\it {Bulge
  Growth and Quenching since $z =$ 2.5 in CANDELS/3D-HST}}, } {\em Astrophys.
  J.} {\bf 788} (2014) 11, [\href{http://arxiv.org/abs/1402.0866}{{\tt
  1402.0866}}].

\bibitem{Bluck:2014ila}
A.~F.~L. Bluck, J.~T. Mendel, S.~L. Ellison, J.~Moreno, L.~Simard, et~al.,
  \href{http://dx.doi.org/10.1093/mnras/stu594}{{\it {Bulge mass is king: The
  dominant role of the bulge in determining the fraction of passive galaxies in
  the Sloan Digital Sky Survey}}, } {\em Mon. Not. Roy. Astron. Soc.} {\bf 441}
  (2014), no.~1 599--629, [\href{http://arxiv.org/abs/1403.5269}{{\tt
  1403.5269}}].

\bibitem{Kormendy:2013dxa}
J.~Kormendy and L.~C. Ho,
  \href{http://dx.doi.org/10.1146/annurev-astro-082708-101811}{{\it
  {Coevolution (Or Not) of Supermassive Black Holes and Host Galaxies}}, } {\em
  Ann. Rev. Astron. Astrophys.} {\bf 51} (2013) 511--653,
  [\href{http://arxiv.org/abs/1304.7762}{{\tt 1304.7762}}].

\bibitem{Kelley:2016gse}
L.~Z. Kelley, L.~Blecha, and L.~Hernquist,
  \href{http://dx.doi.org/10.1093/mnras/stw2452}{{\it {Massive Black Hole
  Binary Mergers in Dynamical Galactic Environments}}, } {\em Mon. Not. Roy.
  Astron. Soc.} {\bf 464} (2017), no.~3 3131--3157,
  [\href{http://arxiv.org/abs/1606.01900}{{\tt 1606.01900}}].

\bibitem{Gardiner:2025mwf}
E.~C. Gardiner, B.~B{\'e}csy, L.~Z. Kelley, and N.~J. Cornish,
  \href{http://dx.doi.org/10.3847/1538-4357/ade4c2}{{\it {Characterizing
  Continuous Gravitational Waves from Supermassive Black Hole Binaries in
  Realistic Pulsar Timing Array Data}}, } {\em Astrophys. J.} {\bf 988} (2025),
  no.~2 222, [\href{http://arxiv.org/abs/2502.16016}{{\tt 2502.16016}}].

\bibitem{Tomczak:2013bxa}
A.~R. Tomczak et~al., \href{http://dx.doi.org/10.1088/0004-637X/783/2/85}{{\it
  {Galaxy Stellar Mass Functions from ZFOURGE/CANDELS: An Excess of Low-mass
  Galaxies since $z = 2$ and the Rapid Buildup of Quiescent Galaxies}}, } {\em
  Astrophys. J.} {\bf 783} (2014) 85,
  [\href{http://arxiv.org/abs/1309.5972}{{\tt 1309.5972}}].

\bibitem{McConnell:2012hz}
N.~J. McConnell and C.-P. Ma,
  \href{http://dx.doi.org/10.1088/0004-637X/764/2/184}{{\it {Revisiting the
  Scaling Relations of Black Hole Masses and Host Galaxy Properties}}, } {\em
  Astrophys. J.} {\bf 764} (2013) 184,
  [\href{http://arxiv.org/abs/1211.2816}{{\tt 1211.2816}}].

\bibitem{Sesana:2008mz}
A.~Sesana, A.~Vecchio, and C.~N. Colacino,
  \href{http://dx.doi.org/10.1111/j.1365-2966.2008.13682.x}{{\it {The
  stochastic gravitational-wave background from massive black hole binary
  systems: implications for observations with Pulsar Timing Arrays}}, } {\em
  Mon. Not. Roy. Astron. Soc.} {\bf 390} (2008) 192,
  [\href{http://arxiv.org/abs/0804.4476}{{\tt 0804.4476}}].

\bibitem{Taylor:2016ftv}
S.~R. Taylor, J.~Simon, and L.~Sampson,
  \href{http://dx.doi.org/10.1103/PhysRevLett.118.181102}{{\it {Constraints On
  The Dynamical Environments Of Supermassive Black-hole Binaries Using
  Pulsar-timing Arrays}}, } {\em Phys. Rev. Lett.} {\bf 118} (2017), no.~18
  181102, [\href{http://arxiv.org/abs/1612.02817}{{\tt 1612.02817}}].

\bibitem{Tiruvaskar:2025lkq}
S.~Tiruvaskar and C.~Gordon, \href{http://dx.doi.org/10.1103/2hqm-qv99}{{\it
  {Self-interacting dark-matter spikes and the final-parsec problem: Bayesian
  constraints from the NANOGrav 15-year gravitational-wave background}}, } {\em
  Phys. Rev. D} {\bf 113} (2026), no.~4 043501,
  [\href{http://arxiv.org/abs/2506.18153}{{\tt 2506.18153}}].

\bibitem{Tiruvaskar:2025hxl}
S.~Tiruvaskar, R.~Boey, R.~Easther, and C.~Gordon,
  \href{http://dx.doi.org/10.1103/4bw6-2tt8}{{\it {Ultralight dark matter
  constraints from nano-Hertz gravitational waves}}, } {\em Phys. Rev. D} {\bf
  113} (2026), no.~6 063541, [\href{http://arxiv.org/abs/2512.15292}{{\tt
  2512.15292}}].

\bibitem{Bonetti:2023ozm}
M.~Bonetti, A.~Franchini, B.~G. Galuzzi, and A.~Sesana,
  \href{http://dx.doi.org/10.1051/0004-6361/202348433}{{\it {Neural networks
  unveiling the properties of gravitational wave background from supermassive
  black hole binaries}}, } {\em Astron. Astrophys.} {\bf 687} (2024) A42,
  [\href{http://arxiv.org/abs/2311.04276}{{\tt 2311.04276}}].

\bibitem{Tiruvaskar:2026tyl}
S.~Tiruvaskar and C.~Gordon, {\it {Replacing Gaussian Processes with Neural
  Networks in Pulsar Timing Array Inference of the Gravitational-Wave
  Background}},  \href{http://arxiv.org/abs/2604.04340}{{\tt 2604.04340}}.

\bibitem{Laal:2024trp}
N.~Laal et~al., \href{http://dx.doi.org/10.3847/1538-4357/adb4ef}{{\it {Deep
  Neural Emulation of the Supermassive Black Hole Binary Population}}, } {\em
  Astrophys. J.} {\bf 982} (2025), no.~1 55,
  [\href{http://arxiv.org/abs/2411.10519}{{\tt 2411.10519}}].

\bibitem{Iemoto:2026fgk}
K.~Iemoto, B.~Goncharov, G.~Sato-Polito, and X.~Bi, {\it {Bridging the
  Population Synthesis of Supermassive Binary Black Holes and the Gravitational
  Wave Background}},  \href{http://arxiv.org/abs/2609.07703}{{\tt 2609.07703}}.

\bibitem{Sampson:2015ada}
L.~Sampson, N.~J. Cornish, and S.~T. McWilliams,
  \href{http://dx.doi.org/10.1103/PhysRevD.91.084055}{{\it {Constraining the
  Solution to the Last Parsec Problem with Pulsar Timing}}, } {\em Phys. Rev.
  D} {\bf 91} (2015), no.~8 084055,
  [\href{http://arxiv.org/abs/1503.02662}{{\tt 1503.02662}}].

\bibitem{NANOGrav:2015aud}
{\bf NANOGrav Collaboration}, Z.~Arzoumanian et~al.,
  \href{http://dx.doi.org/10.3847/0004-637X/821/1/13}{{\it {The NANOGrav
  Nine-year Data Set: Limits on the Isotropic Stochastic Gravitational Wave
  Background}}, } {\em Astrophys. J.} {\bf 821} (2016), no.~1 13,
  [\href{http://arxiv.org/abs/1508.03024}{{\tt 1508.03024}}].

\bibitem{Lamb:2023jls}
W.~G. Lamb, S.~R. Taylor, and R.~van Haasteren,
  \href{http://dx.doi.org/10.1103/PhysRevD.108.103019}{{\it {Rapid refitting
  techniques for Bayesian spectral characterization of the gravitational wave
  background using pulsar timing arrays}}, } {\em Phys. Rev. D} {\bf 108}
  (2023), no.~10 103019, [\href{http://arxiv.org/abs/2303.15442}{{\tt
  2303.15442}}].

\bibitem{Mitridate:2023oar}
A.~Mitridate, D.~Wright, R.~von Eckardstein, T.~Schr{\"o}der, J.~Nay, et~al.,
  {\it {PTArcade}},  \href{http://arxiv.org/abs/2306.16377}{{\tt 2306.16377}}.

\bibitem{enterprise}
J.~A. Ellis, M.~Vallisneri, S.~R. Taylor, and P.~T. Baker,
  \href{http://dx.doi.org/10.5281/zenodo.4059815}{{\it {ENTERPRISE: Enhanced
  Numerical Toolbox Enabling a Robust PulsaR Inference SuitE}}, } 2020.
\newblock software, \url{https://github.com/nanograv/enterprise}.

\bibitem{NG15Models}
{The NANOGrav Collaboration},
  \href{http://dx.doi.org/10.5281/zenodo.8092761}{{\it {The NANOGrav 15-year
  search for new physics signals -- model files}}, } 2023.

\bibitem{Wilks:1938dza}
S.~S. Wilks, \href{http://dx.doi.org/10.1214/aoms/1177732360}{{\it {The
  Large-Sample Distribution of the Likelihood Ratio for Testing Composite
  Hypotheses}}, } {\em Annals Math. Statist.} {\bf 9} (1938), no.~1 60--62.

\bibitem{Fermi-LAT:2010feq}
{\bf Fermi-LAT Collaboration}, A.~A. Abdo et~al.,
  \href{http://dx.doi.org/10.1088/0067-0049/188/2/405}{{\it {Fermi Large Area
  Telescope First Source Catalog}}, } {\em Astrophys. J. Suppl.} {\bf 188}
  (2010) 405--436, [\href{http://arxiv.org/abs/1002.2280}{{\tt 1002.2280}}].

\bibitem{Gross:2010qma}
E.~Gross and O.~Vitells,
  \href{http://dx.doi.org/10.1140/epjc/s10052-010-1470-8}{{\it {Trial factors
  for the look elsewhere effect in high energy physics}}, } {\em Eur. Phys. J.
  C} {\bf 70} (2010) 525--530, [\href{http://arxiv.org/abs/1005.1891}{{\tt
  1005.1891}}].

\bibitem{Protassov:2002sz}
R.~Protassov, D.~A. van Dyk, A.~Connors, V.~L. Kashyap, and A.~Siemiginowska,
  \href{http://dx.doi.org/10.1086/339856}{{\it {Statistics: handle with care,
  detecting multiple model components with the likelihood ratio test}}, } {\em
  Astrophys. J.} {\bf 571} (2002) 545,
  [\href{http://arxiv.org/abs/astro-ph/0201547}{{\tt astro-ph/0201547}}].

\bibitem{CTAO:2024wvb}
{\bf CTAO Collaboration}, S.~Abe et~al.,
  \href{http://dx.doi.org/10.1088/1475-7516/2024/07/047}{{\it {Dark matter line
  searches with the Cherenkov Telescope Array}}, } {\em JCAP} {\bf 07} (2024)
  047, [\href{http://arxiv.org/abs/2403.04857}{{\tt 2403.04857}}].

\bibitem{NG15KDE}
{The NANOGrav Collaboration},
  \href{http://dx.doi.org/10.5281/zenodo.21844115}{{\it {KDE representations of
  the gravitational wave background free spectra present in the NANOGrav
  15-year dataset, version 2.0.0}}, } 2026.

\bibitem{PTMCMCSampler}
J.~Ellis and R.~van Haasteren,
  \href{http://dx.doi.org/10.5281/zenodo.1037579}{{\it {jellis18/PTMCMCSampler:
  Official Release}}, } 2017.

\bibitem{PTArcadeData}
A.~Mitridate and D.~Wright,
  \href{http://dx.doi.org/10.5281/zenodo.10495907}{{\it {PTArcade -- data}}, }
  2024.

\bibitem{Leja:2019lfo}
J.~Leja, J.~S. Speagle, B.~D. Johnson, C.~Conroy, P.~van Dokkum, et~al.,
  \href{http://dx.doi.org/10.3847/1538-4357/ab7e27}{{\it {A New Census of the
  $0.2 < z < 3.0$ Universe. I. The Stellar Mass Function}}, } {\em Astrophys.
  J.} {\bf 893} (2020) 111, [\href{http://arxiv.org/abs/1910.04168}{{\tt
  1910.04168}}].

\bibitem{Sato-Polito:2025dqw}
G.~Sato-Polito and M.~Zaldarriaga,
  \href{http://dx.doi.org/10.1103/vswd-swb3}{{\it {Uncertainties in the
  supermassive black hole abundance and implications for the GW background}}, }
  {\em Phys. Rev. D} {\bf 114} (2026), no.~6 063052,
  [\href{http://arxiv.org/abs/2509.08041}{{\tt 2509.08041}}].

\bibitem{Lapi:2025wxt}
A.~Lapi, F.~Shankar, M.~Bosi, D.~Roberts, H.~Fu, et~al.,
  \href{http://dx.doi.org/10.1088/1475-7516/2026/02/001}{{\it {Semi-empirical
  framework of supermassive black hole evolution: highlighting a possible
  tension between demographics and gravitational wave background}}, } {\em
  JCAP} {\bf 02} (2026) 001, [\href{http://arxiv.org/abs/2507.15436}{{\tt
  2507.15436}}].

\bibitem{Ellis:2024wdh}
J.~Ellis, M.~Fairbairn, G.~H{\"u}tsi, J.~Urrutia, V.~Vaskonen, et~al.,
  \href{http://dx.doi.org/10.1051/0004-6361/202450846}{{\it {Consistency of
  JWST black hole observations with NANOGrav gravitational wave measurements}},
  } {\em Astron. Astrophys.} {\bf 691} (2024) A270,
  [\href{http://arxiv.org/abs/2403.19650}{{\tt 2403.19650}}].

\bibitem{Roebber:2015iva}
E.~Roebber, G.~Holder, D.~E. Holz, and M.~Warren,
  \href{http://dx.doi.org/10.3847/0004-637X/819/2/163}{{\it {Cosmic variance in
  the nanohertz gravitational wave background}}, } {\em Astrophys. J.} {\bf
  819} (2016) 163, [\href{http://arxiv.org/abs/1508.07336}{{\tt 1508.07336}}].

\bibitem{Ravi:2012bz}
V.~Ravi, J.~S.~B. Wyithe, G.~Hobbs, R.~M. Shannon, R.~N. Manchester, et~al.,
  \href{http://dx.doi.org/10.1088/0004-637X/761/2/84}{{\it {Does a `stochastic'
  background of gravitational waves exist in the pulsar timing band?}}, } {\em
  Astrophys. J.} {\bf 761} (2012) 84,
  [\href{http://arxiv.org/abs/1210.3854}{{\tt 1210.3854}}].

\bibitem{Ellis:2023owy}
J.~Ellis, M.~Fairbairn, G.~H{\"u}tsi, M.~Raidal, J.~Urrutia, et~al.,
  \href{http://dx.doi.org/10.1051/0004-6361/202346268}{{\it {Prospects for
  future binary black hole gravitational wave studies in light of PTA
  measurements}}, } {\em Astron. Astrophys.} {\bf 676} (2023) A38,
  [\href{http://arxiv.org/abs/2301.13854}{{\tt 2301.13854}}].

\bibitem{Sardesai:2024rdd}
S.~C. Sardesai, J.~Simon, and S.~J. Vigeland,
  \href{http://dx.doi.org/10.3847/1538-4357/ad8a60}{{\it {Characterizing the
  Nanohertz Gravitational Wave Background Using a t-process Power Spectral
  Density}}, } {\em Astrophys. J.} {\bf 976} (2024), no.~2 212,
  [\href{http://arxiv.org/abs/2408.10139}{{\tt 2408.10139}}].

\bibitem{Moreschi:2026ere}
B.~E. Moreschi, K.~A. Gersbach, G.~Shaifullah, S.~R. Taylor, and A.~Sesana,
  {\it {Modeling the nanohertz gravitational wave background with the
  t-process: Improved frequentist analysis of gravitational-wave power}},
  \href{http://arxiv.org/abs/2610.08906}{{\tt 2610.08906}}.

\bibitem{Sato-Polito:2024lew}
G.~Sato-Polito and M.~Zaldarriaga,
  \href{http://dx.doi.org/10.1103/PhysRevD.111.023043}{{\it {Distribution of
  the gravitational-wave background from supermassive black holes}}, } {\em
  Phys. Rev. D} {\bf 111} (2025), no.~2 023043,
  [\href{http://arxiv.org/abs/2406.17010}{{\tt 2406.17010}}].

\bibitem{Raidal:2026ezm}
J.~Raidal, J.~Urrutia, V.~Vaskonen, and H.~Veerm{\"a}e,
  \href{http://dx.doi.org/10.1103/kjg1-mwpx}{{\it {Heavy tailed non-Gaussianity
  of the supermassive black hole gravitational wave background}}, } {\em Phys.
  Rev. D} {\bf 114} (2026), no.~4 043550,
  [\href{http://arxiv.org/abs/2604.08506}{{\tt 2604.08506}}].

\bibitem{Hisamatsu:2026qvd}
H.~Hisamatsu and K.~Kyutoku, \href{http://dx.doi.org/10.1103/kttj-kdyw}{{\it
  {Higher-order statistics of the stochastic gravitational wave background from
  supermassive black hole binaries}}, } {\em Phys. Rev. D} {\bf 114} (2026),
  no.~6 063041, [\href{http://arxiv.org/abs/2605.17983}{{\tt 2605.17983}}].

\bibitem{Yu:2026rba}
J.~Yu, Z.~Pan, X.~Xue, and L.~Dai, {\it {Population statistics of nanohertz
  gravitational wave sources}},  \href{http://arxiv.org/abs/2607.07477}{{\tt
  2607.07477}}.

\bibitem{Falxa:2026mhh}
M.~Falxa, {\it {Approximating the statistics of a gravitational wave
  background}},  \href{http://arxiv.org/abs/2609.07686}{{\tt 2609.07686}}.

\bibitem{Falxa:2025qxr}
M.~Falxa and A.~Sesana, \href{http://dx.doi.org/10.1103/9jdd-6ct8}{{\it
  {Modeling non-Gaussianities in pulsar timing array data analysis using
  Gaussian mixture models}}, } {\em Phys. Rev. D} {\bf 113} (2026), no.~4
  043047, [\href{http://arxiv.org/abs/2508.08365}{{\tt 2508.08365}}].

\bibitem{Foreman-Mackey:2012ig}
D.~Foreman-Mackey, D.~W. Hogg, D.~Lang, and J.~Goodman,
  \href{http://dx.doi.org/10.1086/670067}{{\it {emcee: The MCMC Hammer}}, }
  {\em Publ. Astron. Soc. Pac.} {\bf 125} (2013) 306--312,
  [\href{http://arxiv.org/abs/1202.3665}{{\tt 1202.3665}}].

\bibitem{Yeh:2022heq}
T.-H. Yeh, J.~Shelton, K.~A. Olive, and B.~D. Fields,
  \href{http://dx.doi.org/10.1088/1475-7516/2022/10/046}{{\it {Probing physics
  beyond the standard model: limits from BBN and the CMB independently and
  combined}}, } {\em JCAP} {\bf 10} (2022) 046,
  [\href{http://arxiv.org/abs/2207.13133}{{\tt 2207.13133}}].

\bibitem{Agazie:2026tui}
{\bf NANOGrav Collaboration}, G.~Agazie et~al.,
  \href{http://dx.doi.org/10.3847/2041-8213/ae7086}{{\it {The NANOGrav 15 yr
  Data Set: Piecewise Power-law Reconstruction of the Gravitational-wave
  Background}}, } {\em Astrophys. J. Lett.} {\bf 1004} (2026), no.~2 L40,
  [\href{http://arxiv.org/abs/2601.09481}{{\tt 2601.09481}}].

\bibitem{Caprini:2024hue}
{\bf LISA Cosmology Working Group Collaboration}, C.~Caprini, R.~Jinno,
  M.~Lewicki, E.~Madge, M.~Merchand, et~al.,
  \href{http://dx.doi.org/10.1088/1475-7516/2024/10/020}{{\it {Gravitational
  waves from first-order phase transitions in LISA: reconstruction pipeline and
  physics interpretation}}, } {\em JCAP} {\bf 10} (2024) 020,
  [\href{http://arxiv.org/abs/2403.03723}{{\tt 2403.03723}}].

\bibitem{Lewicki:2025hxg}
M.~Lewicki and V.~Vaskonen,
  \href{http://dx.doi.org/10.1016/j.dark.2026.102425}{{\it {Impact of cosmic
  expansion on gravitational wave spectra from strongly supercooled first-order
  phase transitions}}, } {\em Phys. Dark Univ.} {\bf 53} (2026) 102425,
  [\href{http://arxiv.org/abs/2511.15687}{{\tt 2511.15687}}].

\bibitem{NG15KDEv1}
{The NANOGrav Collaboration},
  \href{http://dx.doi.org/10.5281/zenodo.8060824}{{\it {KDE representations of
  the gravitational wave background free spectra present in the NANOGrav
  15-year dataset, version 1.0.0}}, } 2023.

\bibitem{Kelley:2017lek}
L.~Z. Kelley, L.~Blecha, L.~Hernquist, A.~Sesana, and S.~R. Taylor,
  \href{http://dx.doi.org/10.1093/mnras/stx1638}{{\it {The Gravitational Wave
  Background from Massive Black Hole Binaries in Illustris: spectral features
  and time to detection with pulsar timing arrays}}, } {\em Mon. Not. Roy.
  Astron. Soc.} {\bf 471} (2017), no.~4 4508--4526,
  [\href{http://arxiv.org/abs/1702.02180}{{\tt 1702.02180}}].

\bibitem{Raidal:2024odr}
J.~Raidal, J.~Urrutia, V.~Vaskonen, and H.~Veerm{\"a}e,
  \href{http://dx.doi.org/10.1051/0004-6361/202451345}{{\it {Eccentricity
  effects on the supermassive black hole gravitational wave background}}, }
  {\em Astron. Astrophys.} {\bf 691} (2024) A212,
  [\href{http://arxiv.org/abs/2406.05125}{{\tt 2406.05125}}].

\bibitem{Chen:2016zyo}
S.~Chen, A.~Sesana, and W.~Del~Pozzo,
  \href{http://dx.doi.org/10.1093/mnras/stx1093}{{\it {Efficient computation of
  the gravitational wave spectrum emitted by eccentric massive black hole
  binaries in stellar environments}}, } {\em Mon. Not. Roy. Astron. Soc.} {\bf
  470} (2017), no.~2 1738--1749, [\href{http://arxiv.org/abs/1612.00455}{{\tt
  1612.00455}}].

\bibitem{Goncharov:2026cdh}
B.~Goncharov, G.~Sato-Polito, X.~Bi, and M.~Zaldarriaga, {\it {A Joint Optimal
  Search for Gravitational Waves from Resolved and Unresolved Supermassive
  Binary Black Holes with Pulsar Timing Arrays}},
  \href{http://arxiv.org/abs/2606.18241}{{\tt 2606.18241}}.

\bibitem{Becsy:2023qul}
B.~B{\'e}csy et~al., \href{http://dx.doi.org/10.3847/1538-4357/ad09e4}{{\it
  {How to Detect an Astrophysical Nanohertz Gravitational Wave Background}}, }
  {\em Astrophys. J.} {\bf 959} (2023), no.~1 9,
  [\href{http://arxiv.org/abs/2309.04443}{{\tt 2309.04443}}].

\bibitem{Gardiner:2023zzr}
E.~C. Gardiner, L.~Z. Kelley, A.-M. Lemke, and A.~Mitridate,
  \href{http://dx.doi.org/10.3847/1538-4357/ad2be8}{{\it {Beyond the
  Background: Gravitational-wave Anisotropy and Continuous Waves from
  Supermassive Black Hole Binaries}}, } {\em Astrophys. J.} {\bf 965} (2024),
  no.~2 164, [\href{http://arxiv.org/abs/2309.07227}{{\tt 2309.07227}}].

\bibitem{Cecchini:2026cru}
C.~Cecchini, J.~El~Gammal, G.~Franciolini, and M.~Pieroni, {\it {Are PTA
  measurements sensitive to gravitational wave non-Gaussianities?}},
  \href{http://arxiv.org/abs/2605.05157}{{\tt 2605.05157}}.

\bibitem{NANOGrav:2023pdq}
{\bf NANOGrav Collaboration}, G.~Agazie et~al.,
  \href{http://dx.doi.org/10.3847/2041-8213/ace18a}{{\it {The NANOGrav 15 yr
  Data Set: Bayesian Limits on Gravitational Waves from Individual Supermassive
  Black Hole Binaries}}, } {\em Astrophys. J. Lett.} {\bf 951} (2023), no.~2
  L50, [\href{http://arxiv.org/abs/2306.16222}{{\tt 2306.16222}}].

\bibitem{Lamb:2025niq}
W.~G. Lamb, J.~M. Wachter, A.~Mitridate, S.~C. Sardesai, B.~B{\'e}csy, et~al.,
  \href{http://dx.doi.org/10.1103/96zk-qtck}{{\it {Finite populations and
  finite time: The non-Gaussianity of a gravitational wave background}}, } {\em
  Phys. Rev. D} {\bf 113} (2026), no.~12 123065,
  [\href{http://arxiv.org/abs/2511.09659}{{\tt 2511.09659}}].

\bibitem{Rodriguez-Gomez:2015aua}
V.~Rodriguez-Gomez et~al., \href{http://dx.doi.org/10.1093/mnras/stv264}{{\it
  {The merger rate of galaxies in the Illustris Simulation: a comparison with
  observations and semi-empirical models}}, } {\em Mon. Not. Roy. Astron. Soc.}
  {\bf 449} (2015), no.~1 49--64, [\href{http://arxiv.org/abs/1502.01339}{{\tt
  1502.01339}}].

\bibitem{Wright:2026lwu}
D.~C. Wright, A.~D. Johnson, J.~S. Hazboun, and W.~G. Lamb, {\it {Fast and
  Flow-rious: Gravitational Wave Background Bayesian Model Comparison using
  Normalizing Flows and Nested Sampling}},
  \href{http://arxiv.org/abs/2609.38122}{{\tt 2609.38122}}.

\bibitem{binaryfog}
C.~Tasillo, T.~Bringmann, and C.~Gordon,
  \href{http://dx.doi.org/10.5281/zenodo.24294394}{{\it {binaryfog: the
  supermassive black hole binary background as a nuisance in pulsar timing
  array searches}}, } 2026.
\newblock software, version 0.1.0, \url{https://github.com/tasicarl/binaryfog}.

\bibitem{Matuszak:2026xsz}
J.~Matuszak and C.~Tasillo, {\it {TransitionListener v2.0 -- Robust
  gravitational wave predictions for cosmological phase transitions}},
  \href{http://arxiv.org/abs/2605.15259}{{\tt 2605.15259}}.

\bibitem{Gouttenoire:2021kjv}
Y.~Gouttenoire, R.~Jinno, and F.~Sala,
  \href{http://dx.doi.org/10.1007/JHEP05(2022)004}{{\it {Friction pressure on
  relativistic bubble walls}}, } {\em JHEP} {\bf 05} (2022) 004,
  [\href{http://arxiv.org/abs/2112.07686}{{\tt 2112.07686}}].

\bibitem{Jinno:2022mie}
R.~Jinno, T.~Konstandin, H.~Rubira, and I.~Stomberg,
  \href{http://dx.doi.org/10.1088/1475-7516/2023/02/011}{{\it {Higgsless
  simulations of cosmological phase transitions and gravitational waves}}, }
  {\em JCAP} {\bf 02} (2023) 011, [\href{http://arxiv.org/abs/2209.04369}{{\tt
  2209.04369}}].

\bibitem{Espinosa:2010hh}
J.~R. Espinosa, T.~Konstandin, J.~M. No, and G.~Servant,
  \href{http://dx.doi.org/10.1088/1475-7516/2010/06/028}{{\it {Energy Budget of
  Cosmological First-order Phase Transitions}}, } {\em JCAP} {\bf 06} (2010)
  028, [\href{http://arxiv.org/abs/1004.4187}{{\tt 1004.4187}}].

\bibitem{deSalas:2015glj}
P.~F. de~Salas, M.~Lattanzi, G.~Mangano, G.~Miele, S.~Pastor, et~al.,
  \href{http://dx.doi.org/10.1103/PhysRevD.92.123534}{{\it {Bounds on very low
  reheating scenarios after Planck}}, } {\em Phys. Rev. D} {\bf 92} (2015),
  no.~12 123534, [\href{http://arxiv.org/abs/1511.00672}{{\tt 1511.00672}}].

\bibitem{Hiramatsu:2013qaa}
T.~Hiramatsu, M.~Kawasaki, and K.~Saikawa,
  \href{http://dx.doi.org/10.1088/1475-7516/2014/02/031}{{\it {On the
  estimation of gravitational wave spectrum from cosmic domain walls}}, } {\em
  JCAP} {\bf 02} (2014) 031, [\href{http://arxiv.org/abs/1309.5001}{{\tt
  1309.5001}}].

\bibitem{Ferreira:2022zzo}
R.~Z. Ferreira, A.~Notari, O.~Pujolas, and F.~Rompineve,
  \href{http://dx.doi.org/10.1088/1475-7516/2023/02/001}{{\it {Gravitational
  waves from domain walls in Pulsar Timing Array datasets}}, } {\em JCAP} {\bf
  02} (2023) 001, [\href{http://arxiv.org/abs/2204.04228}{{\tt 2204.04228}}].

\bibitem{Jackson:2004zg}
M.~G. Jackson, N.~T. Jones, and J.~Polchinski,
  \href{http://dx.doi.org/10.1088/1126-6708/2005/10/013}{{\it {Collisions of
  cosmic F and D-strings}}, } {\em JHEP} {\bf 10} (2005) 013,
  [\href{http://arxiv.org/abs/hep-th/0405229}{{\tt hep-th/0405229}}].

\bibitem{Blanco-Pillado:2017oxo}
J.~J. Blanco-Pillado and K.~D. Olum,
  \href{http://dx.doi.org/10.1103/PhysRevD.96.104046}{{\it {Stochastic
  gravitational wave background from smoothed cosmic string loops}}, } {\em
  Phys. Rev. D} {\bf 96} (2017), no.~10 104046,
  [\href{http://arxiv.org/abs/1709.02693}{{\tt 1709.02693}}].

\bibitem{Blanco-Pillado:2013qja}
J.~J. Blanco-Pillado, K.~D. Olum, and B.~Shlaer,
  \href{http://dx.doi.org/10.1103/PhysRevD.89.023512}{{\it {The number of
  cosmic string loops}}, } {\em Phys. Rev. D} {\bf 89} (2014), no.~2 023512,
  [\href{http://arxiv.org/abs/1309.6637}{{\tt 1309.6637}}].

\bibitem{Kuroyanagi:2014nba}
S.~Kuroyanagi, T.~Takahashi, and S.~Yokoyama,
  \href{http://dx.doi.org/10.1088/1475-7516/2015/02/003}{{\it {Blue-tilted
  Tensor Spectrum and Thermal History of the Universe}}, } {\em JCAP} {\bf 02}
  (2015) 003, [\href{http://arxiv.org/abs/1407.4785}{{\tt 1407.4785}}].

\bibitem{Kuroyanagi:2020sfw}
S.~Kuroyanagi, T.~Takahashi, and S.~Yokoyama,
  \href{http://dx.doi.org/10.1088/1475-7516/2021/01/071}{{\it {Blue-tilted
  inflationary tensor spectrum and reheating in the light of NANOGrav
  results}}, } {\em JCAP} {\bf 01} (2021) 071,
  [\href{http://arxiv.org/abs/2011.03323}{{\tt 2011.03323}}].

\bibitem{Gultekin:2009qn}
K.~G{\"u}ltekin et~al.,
  \href{http://dx.doi.org/10.1088/0004-637X/698/1/198}{{\it {The M-sigma and
  M-L Relations in Galactic Bulges and Determinations of their Intrinsic
  Scatter}}, } {\em Astrophys. J.} {\bf 698} (2009) 198--221,
  [\href{http://arxiv.org/abs/0903.4897}{{\tt 0903.4897}}].

\bibitem{Baldry:2011fj}
I.~K. Baldry et~al.,
  \href{http://dx.doi.org/10.1111/j.1365-2966.2012.20340.x}{{\it {Galaxy And
  Mass Assembly (GAMA): The galaxy stellar mass function at z {\ensuremath{<}}
  0.06}}, } {\em Mon. Not. Roy. Astron. Soc.} {\bf 421} (2012) 621--634,
  [\href{http://arxiv.org/abs/1111.5707}{{\tt 1111.5707}}].

\bibitem{Eddington:1913}
A.~S. Eddington, \href{http://dx.doi.org/10.1093/mnras/73.5.359}{{\it {On a
  formula for correcting statistics for the effects of a known error of
  observation}}, } {\em Mon. Not. Roy. Astron. Soc.} {\bf 73} (1913) 359--360.

\bibitem{Wright:2018gsm}
A.~H. Wright, S.~P. Driver, and A.~S.~G. Robotham,
  \href{http://dx.doi.org/10.1093/mnras/sty2136}{{\it {GAMA/G10-COSMOS/3D-HST:
  Evolution of the galaxy stellar mass function over 12.5 Gyr}}, } {\em Mon.
  Not. Roy. Astron. Soc.} {\bf 480} (2018), no.~3 3491--3502,
  [\href{http://arxiv.org/abs/1808.01754}{{\tt 1808.01754}}].

\bibitem{McLeod:2021gsm}
D.~J. McLeod, R.~J. McLure, J.~S. Dunlop, F.~Cullen, A.~C. Carnall, et~al.,
  \href{http://dx.doi.org/10.1093/mnras/stab731}{{\it {The evolution of the
  galaxy stellar-mass function over the last 12 billion years from a
  combination of ground-based and HST surveys}}, } {\em Mon. Not. Roy. Astron.
  Soc.} {\bf 503} (2021), no.~3 4413--4435,
  [\href{http://arxiv.org/abs/2009.03176}{{\tt 2009.03176}}].

\bibitem{Davidzon:2017rib}
I.~Davidzon et~al., \href{http://dx.doi.org/10.1051/0004-6361/201730419}{{\it
  {The COSMOS2015 galaxy stellar mass function - Thirteen billion years of
  stellar mass assembly in ten snapshots}}, } {\em Astron. Astrophys.} {\bf
  605} (2017) A70, [\href{http://arxiv.org/abs/1701.02734}{{\tt 1701.02734}}].

\bibitem{Weaver:2023cos}
J.~R. Weaver, I.~Davidzon, S.~Toft, O.~Ilbert, H.~J. McCracken, et~al.,
  \href{http://dx.doi.org/10.1051/0004-6361/202245581}{{\it {COSMOS2020: The
  galaxy stellar mass function. The assembly and star formation cessation of
  galaxies at $0.2< z \le 7.5$}}, } {\em Astron. Astrophys.} {\bf 677} (2023)
  A184, [\href{http://arxiv.org/abs/2212.02512}{{\tt 2212.02512}}].

\bibitem{Driver:2022vyh}
S.~P. Driver et~al., \href{http://dx.doi.org/10.1093/mnras/stac472}{{\it
  {Galaxy And Mass Assembly (GAMA): Data Release 4 and the z {\ensuremath{<}}
  0.1 total and z {\ensuremath{<}} 0.08 morphological galaxy stellar mass
  functions}}, } {\em Mon. Not. Roy. Astron. Soc.} {\bf 513} (2022), no.~1
  439--467, [\href{http://arxiv.org/abs/2203.08539}{{\tt 2203.08539}}].

\bibitem{Sahu:2019bhm}
N.~Sahu, A.~W. Graham, and B.~L. Davis,
  \href{http://dx.doi.org/10.3847/1538-4357/ab0f32}{{\it {Black Hole Mass
  Scaling Relations for Early-type Galaxies. I. $M_{\rm BH}$--$M_{*,{\rm sph}}$
  and $M_{\rm BH}$--$M_{*,{\rm gal}}$}}, } {\em Astrophys. J.} {\bf 876}
  (2019), no.~2 155, [\href{http://arxiv.org/abs/1903.04738}{{\tt
  1903.04738}}].

\bibitem{Saglia:2016bhm}
R.~P. Saglia, M.~Opitsch, P.~Erwin, J.~Thomas, A.~Beifiori, et~al.,
  \href{http://dx.doi.org/10.3847/0004-637X/818/1/47}{{\it {The SINFONI Black
  Hole Survey: The Black Hole Fundamental Plane Revisited and the Paths of
  (Co)evolution of Supermassive Black Holes and Bulges}}, } {\em Astrophys. J.}
  {\bf 818} (2016), no.~1 47, [\href{http://arxiv.org/abs/1601.00974}{{\tt
  1601.00974}}].

\end{thebibliography}\endgroup

\end{document}